\documentclass[usenatbib]{mn2e}

\usepackage{amsmath}
\usepackage{natbib}
\usepackage{epsfig}
\usepackage{txfonts}
\usepackage{pict2e}
\usepackage{color}
\usepackage{booktabs}
\usepackage{multirow}

\newcommand{\Kel}{{\rm K}}

\newcommand{\GHz}{{\rm GHz}}

\newcommand{\expf}[1]{{{\rm e}^{#1}}}

\newcommand{\nbb}{{n^{\rm pl}}}

\newcommand{\vgh}{{\hat{\boldsymbol\gamma}}}

\newcommand{\xe}{x_{\rm e}}

\newcommand{\id}{{\,\rm d}}

\newcommand{\beq}{\begin{equation}}   %

\newcommand{\eeq}{\end{equation}}   %

\newcommand{\beqa}{\begin{eqnarray}}   %

\newcommand{\eeqa}{\end{eqnarray}}   %

\newcommand{\beal}{\begin{align}}
\newcommand{\enal}{\end{align}}

\newcommand{\bspl}{\begin{split}}

\newcommand{\espl}{\end{split}}

\newcommand{\bsub}{\begin{subequations}}

\newcommand{\esub}{\end{subequations}}

\newcommand{\bmulti}{\begin{multline}}   %

\newcommand{\beqm}{\begin{mathletters}}   %

\newcommand{\eeqm}{\end{mathletters}}   %

\newcommand{\me}{m_{\rm e}}

\newcommand{\Ne}{N_{\rm e}}

\newcommand{\Te}{T_{\rm e}}

\newcommand{\The}{\theta_{\rm e}}

\newcommand{\sigT}{\sigma_{\rm T}}

\newcommand{\vek} [1]{\mbox{\boldmath${#1}$\unboldmath}}

\newcommand{\pot}[2]{#1 \times 10^{#2}}

\usepackage{hyperref}
\usepackage{grffile}
\usepackage{graphics}
\usepackage{booktabs}

\title[Rethinking CMB foregrounds]
{Rethinking CMB foregrounds: systematic extension of foreground parameterizations}

\author[Chluba et al.]{Jens~Chluba$^{1}$\thanks{E-mail:Jens.Chluba@manchester.ac.uk}, J.~Colin~Hill$^2$\thanks{E-mail:jch@astro.columbia.edu}, Maximilian~H.~Abitbol$^3$\thanks{E-mail:mha2125@columbia.edu}
\\
$^{1}$ Jodrell Bank Centre for Astrophysics, School of Physics and Astronomy, University of Manchester, Oxford Road, Manchester M13 9PL, UK
\\
$^{2}$ Department of Astronomy, Columbia University, Pupin Hall, New York, New York 10027, USA
\\
$^{3}$ Department of Physics, Columbia University, New York, NY, 10027, USA
}

\begin{document}

\date{\vspace{-5.9mm}{Accepted 2017 July 31. Received 2017 January 1}}

\maketitle

\begin{abstract}
Future high-sensitivity measurements of the cosmic microwave background (CMB) anisotropies and energy spectrum will be limited by our understanding and modeling of foregrounds. Not only does more information need to be gathered and combined, but also novel approaches for the modeling of foregrounds, commensurate with the vast improvements in sensitivity, have to be explored. Here, we study the inevitable effects of {\it spatial averaging} on the spectral shapes of typical foreground components, introducing a moment approach, which naturally extends the list of foreground parameters that have to be determined through measurements or constrained by theoretical models. Foregrounds are thought of as a superposition of individual emitting volume elements along the line of sight and across the sky, which then are observed through an instrumental beam. The beam and line of sight averages are inevitable. Instead of assuming a specific model for the distributions of physical parameters, our method identifies natural new spectral shapes for each foreground component that can be used to extract parameter moments (e.g., mean, dispersion, cross-terms, etc.). The method is illustrated for the superposition of power-laws, free-free spectra, gray-body and modified blackbody spectra, but can be applied to more complicated fundamental spectral energy distributions. Here, we focus on intensity signals but the method can be extended to the case of polarized emission. The averaging process automatically produces scale-dependent spectral shapes and the moment method can be used to propagate the required information across scales in power spectrum estimates. The approach is not limited to applications to CMB foregrounds but could also be useful for the modeling of X-ray emission in clusters of galaxies.
\end{abstract}

\begin{keywords}
Cosmology: cosmic microwave background -- theory -- observations
\end{keywords}

\section{Introduction}
\label{sec:Intro}
Decades of observations of the cosmic microwave background (CMB) temperature and polarization anisotropies have clearly helped to establish the cosmological concordance model \citep[e.g.,][]{WMAP_params, Planck2015params}. By now, cosmologists are close to exhausting all the primordial information from the CMB temperature anisotropies. The next steps are to tease all the information out of the $E$-mode polarization signal and the quest for the even smaller primordial polarization $B$-modes \citep{Kogut2011, PRISM2013WPII, Matsumura2014, Abazajian2015}, as well as to extract information about the late-time universe from secondary temperature anisotropies \citep[e.g., CMB lensing and the Sunyaev-Zeldovich effect;][]{Abazajian2016S4SB}. In addition, CMB spectral distortion measurements, e.g., with {\it PIXIE} \citep{Kogut2011} or {\it APSERa} \citep{Mayuri2015}, could provide a wealth of new complementary information about the Universe we live in \citep[e.g.,][]{Chluba2011therm, Sunyaev2013, Chluba2013fore, Tashiro2014, Chluba2016}.

But the aforementioned signals are (very) small and buried under (much) larger foregrounds. The hope is to be able to separate the cosmological signals from foregrounds using {\it multi-frequency observations}. In particular for CMB spectral distortions, mainly expected to be present uniformly across the sky, this is completely uncharted territory and the anticipated improvements over previous measurements \citep{Mather1994, Fixsen1996} in terms of the raw spectral sensitivity could be dramatic \citep{Kogut2011, Kogut2016SPIE}. Some prior information on spatially varying foreground signals is available, e.g., from measurements of {\it Planck} \citep{Planck2013components, PlanckSM2015}; however, at the required level of future foreground modeling we cannot {\it reliably predict} what awaits us and restrictive assumptions about the foreground model can cause biases in the deduced parameters \citep[e.g.,][]{Errard2016, Remazeilles2016, Kogut2016}. This calls for more {\it flexible} methods, which use a minimal set of prior assumptions, relevant to both CMB spectral distortions and B-mode searches. Some forecasts with explicit foreground modeling for primordial CMB distortions were presented in \citet{abitbol_pixie}.

In this work, we approach the foreground modeling problem from a more general point of view, directly considering the averaging process of spatially varying foregrounds i) along the line of sight, ii) within the experimental beam and iii) in spherical harmonic decompositions or analyses of sky-maps (e.g., when degrading the angular resolution). The goal is to identify generalized parameterizations for the foregrounds that allow us to incorporate additional features of the underlying distributions of physical parameters (e.g., spectral indices, dust temperatures, and {\it all} their statistical properties) at a level that is sufficient to robustly isolate the much smaller primordial signals.

The first problem is that spatially varying foregrounds are averaged within the beam and along the line of sight. These averages cannot be avoided and lead to modifications of the spectral response even if in every {\it infinitesimal volume element} the signal is given by a very simple known energy distribution. For example, in every direction the CMB (to very high precision) is given by a blackbody with spatially varying temperature. For a low-resolution beam, the average over these blackbodies introduces a beam $y$-distortion with $y$-parameter $y\simeq \frac{1}{2}\,\langle\Delta T^2/T^2_0\rangle$, where $\langle\Delta T^2/T^2_0\rangle$ is the beam-averaged variance of the blackbody temperatures \citep{Chluba2004}. Since the CMB temperature fluctuations are small, this usually only causes a very small beam spectral distortion ($y\simeq 10^{-9}$), which in principle can be accurately modeled using prior information from {\it Planck}. Still, this example already illustrates how averaging processes can affect the required parameterization of the signal (blackbody with varying temperature versus blackbody with $y$-type distortion), when high accuracy is reached.

Another example is the new spectral shapes introduced to the thermal Sunyaev-Zeldovich (SZ) effect \citep{Zeldovich1969} by line of sight variations of the electron temperature inside hot galaxy clusters \citep{Chluba2012moments}. In every infinitesimal volume element, the thermal SZ distortion is determined by the optical depth and electron temperature \citep{Sazonov1998, Itoh98}; however, the average of contributions from different volume elements is no longer accurately described by one mean temperature and optical depth, but depends on higher order moments of the temperature field \citep{Chluba2012moments}. The precise shape of the distortion can thus in principle be used to determine those moments, $\langle(k\Te)^k\rangle$, of the electron temperature distribution. Conversely, maps of these new observables contain all the information gained from multi-frequency SZ measurements and can be used to directly constrain the detailed structure (average temperature and density profiles) of the cluster atmosphere. Similar statements hold for the all-sky SZ distortion signal \citep{Hill2015}.

From detailed CMB measurements, we also know that the standard foreground components (e.g., synchrotron and dust) vary {\it spectrally} across the sky \citep[e.g.,][]{Planck2013components, Fuskeland2014, PlanckSM2015} and similarly we expect the spectral shapes from individual contributions to vary along the line of sight. While the effect of spatial variations across the sky can be minimized using higher angular resolution ($\rightarrow$ pencil beams), the line of sight average cannot be avoided. Thus, even the superposition of extremely simple fundamental spectral energy distributions (SEDs), e.g., power-laws, will in general not be represented by a single power-law to arbitrary precision. As we explain here, with the moment method we can parametrize the effects on the SED caused by these averages. In a similar manner, we present a moment expansion for free-free emission, gray-body and modified blackbody spectra, as examples. 
We focus on developing and illustrating the moment method for intensity signals, but it can be readily extended to polarized foregrounds, which could be relevant to future $B$-mode searches, as we investigate in a separate paper. Similarly, we plan to use the moment method to forecast the spectral distortion sensitivity of {\it PIXIE} in a future work.

While the direct moment representation for a specified fundamental SED is exact, a truncated moment hierarchy (finite number of moments) may or may not allow representing all possible spectral shapes to sufficient precision (i.e., leading to residuals that are consistent or below the noise) or may describe the signal with highly correlated parameters. In the latter case, orthogonalization schemes (e.g., Gram-Schmidt or principal component decompositions) and different choices for the weighting can provide an alternative approach, introducing new spectral amplitude parameters related to linear combinations of moments. We illustrate these aspects for gray-body spectra (Sect.~\ref{sec:blackbody}); however, we do not find orthogonalization schemes to be as beneficial and in general cases the set of orthogonal functions also depends on the average parameters and experimental settings, rendering it a less attractive procedure.

While the developed method and discussion envisions applications to CMB observations, the approach is more general and should be applicable in other regimes (e.g., X-ray modeling). One can furthermore include more complicated fundamental spectra [e.g., to parametrize the anomalous microwave emission \citep{Draine1998, Yacine2009, Hoang2016} or for improved models of the spectral shapes of dust, synchrotron, or even line emission]. However, we leave these investigations to future work. Overall, we find that the moment expansion can be used to parametrize some of the fundamental effects caused by spatial averaging to high precision, providing a first step towards generalized foreground separation methods relevant to future CMB anisotropy and spectroscopic studies.

\vspace{-4mm}
\section{Details of the moment method for spatially varying components}
\label{sec:moment_meth_gen}
In this section, we provide details about how to construct the moment expansion for a given underlying physical emission process. We separate the effects of line of sight and beam averages on the one hand (Sec.~\ref{sec:beam_average}) from the effect of spatial averages carried out on measured sky-signals (Sec.~\ref{sec:harmonic_exp_average}). Even for a perfect pencil beam, the line of sight averaging is {\it inevitable}, but we can simply associate this averaging process with the beam average itself.

\vspace{-2mm}
\subsection{Average signal within a beam}
\label{sec:beam_average}
Let us first try to understand how the average over spatially varying spectral signals along the line of sight and within the beam of an experiment affects the measured signal. We shall consider the infinitesimal SED contribution, $\delta I_\nu(\vgh, \vek{p})$, in a direction $\vgh$, which varies with frequency $\nu$ and depends on a list of spatially varying parameters $\vek{p}(\vek{r})$ at different locations, $\vek{r}$, along the line of sight. The line of sight can be parametrized by an affine parameter $s(\vgh)$ in the direction $\vgh$, so that the line of sight averaged SED is 
\beal
I_{\nu}(\vgh)\equiv\left<I_\nu(\vgh, \vek{p})\right>_{\rm l.o.s.}=\int \frac{\id I_\nu(\vgh, \vek{p})}{\id s(\vgh)} \id s(\vgh),
\end{align}
where $\vek{p}=\vek{p}(s)$ is used for the integration. We can also introduce the beam average of some quantity $X(\vgh, \nu)\equiv X_\nu(\vgh)$ by
\beal
X (\vgh_{\rm c}, \nu_{\rm c})\equiv\left< X (\vgh, \nu)\right>_{\Omega}=\int \id^2 \vgh\int \id \nu \, W(\vgh, \nu)  \, X(\vgh, \nu),
\end{align}
where $W(\vgh, \nu)$ describes the frequency-dependent beam of different channels centered at frequency, $\nu_{\rm c}=\left< \nu \right>_{\Omega}\equiv \int \nu \, W(\vgh, \nu) \id \nu\id^2 \vgh$, and in the average direction $\vgh_{\rm c}=\left< \vgh \right>_{\Omega}\equiv \int \vgh \, W(\vgh, \nu) \id \nu\id^2 \vgh$.
The specific shape of $W(\vgh, \nu)$ depends on the experiment, as discussed below. We normalize the beam as $\int W(\vgh, \nu) \id^2 \vgh\id \nu\equiv\left<1\right>_\Omega\equiv ~1$. 

We reiterate, the guiding picture is that for any observation in an average direction $\vgh_{\rm c}$, the signal from different emitters\footnote{Absorption can be thought of as negative emission unless the shape of the fundamental SED is affected. Otherwise, the SED parametrization has to be modified or both absorption and emission have to be treated independently.} along the line of sight is picked up and then averaged in different directions within the beam. Physically, the line of sight average has the same effect as the average in different directions, leading to a mixing of the fundamental SEDs of emitters with different spectral parameters, $\vek{p}$. It therefore does not have to be distinguished in the computation, and we will use $\left<X\right>$ to denote both the line of sight and beam average as a single beam average.
The effect of redshifting along the line of sight can be captured by transforming the frequency variable and adding appropriate volume weighting factors. This affects the relative contributions of emitters in the modeling, but is not explicitly discussed here without loss of generality. The beam can also have different shapes at different frequencies {\it within} the band, which furthermore can vary from channel to channel.

It is now useful to perform a multi-dimensional Taylor expansion of the underlying SED in the free model parameters
\beal
\label{eq:I_expansion}
I_\nu(\vek{p})&=I_\nu(\bar{\vek{p}})
+\sum_i (p_i-\bar{p}_i) \,\partial_{\bar{p}_i} I_\nu(\bar{\vek{p}})
\nonumber\\[-0.5mm]
&\!\!\!\!
+\frac{1}{2!}\sum_i \sum_j (p_i-\bar{p}_i)(p_j-\bar{p}_j) \,\partial_{\bar{p}_i}\partial_{\bar{p}_j} I_\nu(\bar{\vek{p}})
\nonumber\\[-0.5mm]
&+\frac{1}{3!}\sum_i \sum_j\sum_k (p_i-\bar{p}_i)(p_j-\bar{p}_j)(p_k-\bar{p}_k) \,\partial_{\bar{p}_i}\partial_{\bar{p}_j}\partial_{\bar{p}_k} I_\nu(\bar{\vek{p}})
\nonumber\\[0.5mm]
&\quad+ \ldots
\end{align}
where $\bar{\vek{p}}$ (only depending on $\vgh_{\rm c}$) describes the list of beam average- parameters that need to be specified. In the notation, we suppressed the spatial dependence of $p_i=p_i(\vek{r})$ for convenience. We also use the notation $\partial_{\bar{p}_i}\ldots\partial_{\bar{p}_j} I_\nu(\bar{\vek{p}})=\left. \partial_{p_i}\ldots\partial_{p_j} I_\nu(\vek{p})\right|_{\vek{p}=\bar{\vek{p}}}$ for brevity.

To fix the average parameters, $\bar{\vek{p}}$, for a given spectral distribution function, we require the first order terms in the Taylor expansion to vanish, $\left<\sum_i [p_i(\vek{r})-\bar{p}_i] \,\partial_{\bar{p}_i}\,I_\nu(\bar{\vek{p}})\right>=0$. This has the solution 
\beal
\label{eq:def_pbar}
\bar{p}_i = \frac{\left<p_i(\vek{r})\,\partial_{\bar{p}_i}\,I_\nu(\bar{\vek{p}})\right>}{\left<\partial_{\bar{p}_i}\,I_\nu(\bar{\vek{p}})\right>}
\end{align}
for $\left<\partial_{\bar{p}_i}\,I_\nu(\bar{\vek{p}})\right>\neq 0$. Generally, this does {\it not} imply $\bar{p}_i=\left<p_i(\vek{r})\right>$ since the frequency dependence of the beam can couple different spatial regions, where the frequency-weighting function is modified by $\partial_{\bar{p}_i}\,I_\nu(\bar{\vek{p}})$. If $W(\vgh, \nu)= B(\vgh) F(\nu)$, one can separate the frequency and spatial averages, i.e., $\left<p_i(\vek{r})\,\partial_{\bar{p}_i}\,I_\nu(\bar{\vek{p}})\right>=\left<p_i(\vek{r})\right>\left<\partial_{\bar{p}_i}\,I_\nu(\bar{\vek{p}})\right>$, such that we recover $\bar{p}_i\equiv\left<p_i(\vek{r})\right>$. Here, $B(\vgh)$ determines the spatial shape of the beam, which does not vary at different frequencies within each channel, and $F(\nu)$ describes the shape of the bandpass. Both can still vary from channel to channel.

We note immediately that the chosen definition for $\bar{p}_i$ is not always the recovered best-fitting value from the analysis. This is because the spectral functions (see below) related to higher order moments are not necessarily linearly independent from the first order derivative spectra. Thus, the best-fitting values for a truncated moment expansion can receive contributions from higher order terms. In some cases, this is not a severe problem (superposition of power-laws and free-free spectra), but, for instance, in superpositions of gray-body and modified blackbody spectra, these values can differ noteably in the non-perturbative regime. Alternative weighting schemes to determine the best estimate for the obtained value of $\bar{p}_i$ from given parameter distribution functions are discussed below.

\vspace{-0mm}
\subsubsection{Definition of Taylor-moments}
Formally, the above expressions can be simplified by introducing the Taylor-moments
\beal
\label{eq:moments_def}
\omega_{i\ldots j} = \frac{\left<[p_i(\vek{r})-\bar{p}_i]\ldots[p_j(\vek{r})-\bar{p}_j]\,
\partial_{\bar{p}_i}\ldots \partial_{\bar{p}_j}\,I_\nu(\bar{\vek{p}})\right>}
{\left<\partial_{\bar{p}_i}\ldots \partial_{\bar{p}_j}\,I_\nu(\bar{\vek{p}})\right>}
\end{align}
and spectral functions
\beal
I_{i\ldots j}(\nu_{\rm c}, \bar{\vek{p}}) = 
\left<\partial_{\bar{p}_i}\ldots \partial_{\bar{p}_j}\,I_\nu(\bar{\vek{p}})\right>,
\end{align}
where $\nu_{\rm c}$ is the central frequency of the considered channel. Here the `$\ldots$' can contain any number and permutation of the parameter indices. The total number of parameters that appears defines the order of the moment, and usually it is implicitly assumed that a perturbative expansion is possible, suggesting that a finite number of moments, ranked by their (derivative) order, can be used. The spectral shapes, $I_{i\ldots j}(\nu_{\rm c}, \bar{\vek{p}})$, provide a (generally non-orthogonal and non-complete) basis for a vector space that spans the range of possible signals that can be captured by the moment expansion.
With these definitions, we then find
\beal
\label{eq:I_av}
\left<I_\nu(\vek{p})\right>&=I(\nu_{\rm c}, \bar{\vek{p}})
+\sum_i \omega_i\,I_i(\nu_{\rm c}, \bar{\vek{p}})
+\frac{1}{2}\sum_i\sum_j \omega_{ij}\,I_{ij}(\nu_{\rm c}, \bar{\vek{p}})
\nonumber\\
&\quad+\frac{1}{6}\sum_i\sum_j\sum_k \omega_{ijk}\,I_{ijk}(\nu_{\rm c}, \bar{\vek{p}})+ \ldots
\end{align}
where $\omega_i\equiv 0$ with the definition of $\bar{\vek{p}}$ given in Eq.~\eqref{eq:def_pbar}. With this reformulation, we have identified a set of new parameters, $\omega_{i\ldots j}$, and spectral functions, $I_{i\ldots j}(\nu_{\rm c}, \bar{\vek{p}})$, which can be used to describe the beam-averaged signal. For a given theoretical model, specified by the fundamental SED, $I_\nu(\vek{p})$, one can simply compute these quantities and use them in the modeling of the foregrounds. Generally the moments have to be individually determined (i.e., measured), while the spectral functions can be computed using the precise knowledge of the beam's angle-averaged frequency dependence and the underlying spectral dependence of $I_\nu$. This provides a powerful separation of spatial and frequency-dependent functions, which we will exploit below. In particular, by studying the dependence of the spectral functions and using high-resolution measurements of average foreground parameters, one can identify and limit the contributions of different terms at varying physical scales.

There are, however, several  caveats. First, the moments, $\omega_{i\ldots j}$, in general {\it depend} on the band in a non-trivial way. This means that within each band different sets of moments (with different weighting functions) contribute, unless the beam factorization $W(\vgh, \nu)= B(\vgh) F(\nu)$ is possible, such that 
\beal
\omega_{i\ldots j} \equiv\left<[p_i(\vek{r})-\bar{p}_i]\ldots[p_j(\vek{r})-\bar{p}_j]\right>.
\end{align}
If this simplification is insufficient, one can in principle introduce additional parameters which are directly related to the beam properties and its coupling to different spectral shapes. By characterizing the beam carefully, this additional dependence can be modeled but it inevitably introduces additional weighting of the underlying spatially varying parameters and a steep growth in the number of independent moments. Thus, unless the effective list of parameters remains smaller than the number of channels, there is no way to separate the different components by using multi-frequency observations. This is indeed one of the big worries for future CMB observations, and the answer will depend on the foreground complexity and specifications of the experiments. It also renders a comparison and combination of data from different (ground-based and space-based) experiments more challenging.

A more detailed account of this problem is beyond the scope of this paper, but it has also been recognized in studies related to 21cm cosmology \citep{Mozdzen2016}. For simplicity we shall assume that the factorization $W(\vgh, \nu)= B(\vgh) F(\nu)$ is possible, but the more general case in principle can also be parametrized using a moment expansion of the beam properties.

Second, a finite expansion in terms of the moments does not necessarily converge rapidly, especially in the non-perturbative regime. Furthermore, higher moments can be directly related to the lower moments (e.g., for a Gaussian the second moment fixes all higher moments), which implies that the effective number of degrees of freedom can be much smaller, while a large number of moments might be necessary to describe the averaged distribution. The moment method is agnostic, making very few {\it a priori} assumptions. This allows us to identify shortcomings of certain parameterizations, which assume a fixed form, as we illustrate here. Alternatively, we can include simplifying assumptions for the relations among the moments to return to a specific lower dimensional parameterization or reduce the parameter space (using priors) that is spanned by the spectral basis functions. This illustrates the range of possibilities when applying the moment method.

In the above, we assumed that a list of parameters (e.g., dust temperature, spectral indices, etc.) has been identified for the moment expansion. This is by no means a trivial statement and the choice of the right variable can be tricky. In addition, the weighting of different contributions has to be considered carefully, as we illustrate below for gray-body and dust spectra.

\vspace{-4mm}
\subsection{Spherical harmonic expansion and signal processing}
\label{sec:harmonic_exp_average}
The formalism described in the previous section can be extended to include the effect of weighted averages due to spherical harmonic expansions of sky signals or when processing maps (e.g., when degrading the angular resolution). We explicitly discuss the spherical harmonic transform of the sky. Formally, the spherical harmonic coefficients of $\left<X\right>$ are given by 
\beal
\label{eq:def_almX}
a^X_{\ell m}(\nu_{\rm c})=\int \id^2 \vgh_{\rm c} \,Y^*_{\ell m}(\vgh_{\rm c}) \,\bar{X}(\nu_{\rm c}, \vgh_{\rm c}),
\end{align}
where $\bar{X}(\nu_{\rm c}, \vgh_{\rm c})=\left<X\right>$ is the beam averaged quantity in the average direction, $\vgh_{\rm c}$, and channel frequency $\nu_{\rm c}$. We can see from Eq.~\eqref{eq:I_av} that two sources of spatial variations appear when carrying out the spherical harmonic expansion, i) those from the moments, $\omega_{i\ldots j}(\vgh_{\rm c})$, and ii) those from the variation of the mean parameters, $\bar{\vek{p}}(\vgh_{\rm c})$ in different directions on the sky. 

By simply following through all the steps that lead to the moment expansion, Eq.~\eqref{eq:I_av}, we readily have the generalization
\beal
\label{eq:I_av_lm}
\bar{I}_{\nu_{\rm c}}\!(\vgh_{\rm c})&=I(\nu_{\rm c}, \bar{\vek{p}}_0)
+\!\sum_i \omega'_i(\vgh_{\rm c})\,I_i(\nu_{\rm c}, \bar{\vek{p}}_0)
+\!\frac{1}{2}\!\sum_i\sum_j \omega'_{ij}(\vgh_{\rm c})\,I_{ij}(\nu_{\rm c}, \bar{\vek{p}}_0)
\nonumber\\
&\qquad+\frac{1}{6}\sum_i\sum_j\sum_k \omega'_{ijk}(\vgh_{\rm c})\,I_{ijk}(\nu_{\rm c}, \bar{\vek{p}}_0)+ \ldots
\end{align}
where $\bar{\vek{p}}_0$ is the sky-averaged parameter vector, which can be obtained by averaging Eq.~\eqref{eq:def_pbar} over all directions. Similarly, the moments follow from Eq.~\eqref{eq:moments_def}, such that overall we find
\beal
\label{eq:moments_def_II}
\bar{p}_{i, 0} &= \int \frac{\left<p_i(r, \vgh)\,\partial_{\bar{p}_{i,0}}\,I_\nu(\bar{\vek{p}}_0)\right>}{\left<\partial_{\bar{p}_{i,0}}\,I_\nu(\bar{\vek{p}}_0)\right>}\frac{\id^2\vgh_{\rm c}}{4\pi}
\\\nonumber
\omega'_{i\ldots j}(\vgh_{\rm c}) &= \frac{\left<[p_i(r, \vgh)-\bar{p}_{i,0}]\ldots[p_j(r, \vgh)-\bar{p}_{j, 0}]\,
\partial_{\bar{p}_{i, 0}}\ldots \partial_{\bar{p}_{j, 0}}\,I_\nu(\bar{\vek{p}}_0)\right>}
{\left<\partial_{\bar{p}_{i, 0}}\ldots \partial_{\bar{p}_{j, 0}}\,I_\nu(\bar{\vek{p}}_0)\right>},
\end{align}
where the average $\left<...\right>$ connects $\vgh$ and $\vgh_{\rm c}$ through the beam shape.
The spherical harmonic coefficients of the moments are obtained using Eq.~\eqref{eq:def_almX}. 
The most important difference is that while by construction $\int \omega'_{i}(\vgh_{\rm c})\id^2\vgh_{\rm c}=0$, in general $\omega'_{i}(\vgh_{\rm c})$ no longer vanishes but fluctuates across the sky. Usually, only this term is considered, but higher order moments lead to new scale-dependent SED variations as we illustrate below. 
Also, in general $\omega'_{i\ldots j}(\vgh_{\rm c})\neq \omega_{i\ldots j}(\vgh_{\rm c})$, as the reference parameters were chosen using the sky average rather than the local quantity. 
With $W(\vgh, \nu)= B(\vgh) F(\nu)$, we have 
\beal
\label{eq:moments_def_II_simp}
\bar{p}_{i, 0} &= \int \bar{p}_i (\vgh_{\rm c})\,\frac{\id^2\vgh_{\rm c}}{4\pi}
\\\nonumber
\omega'_{i\ldots j}(\vgh_{\rm c}) &= \left<[p_i(r, \vgh)-\bar{p}_{i,0}]\ldots[p_j(r, \vgh)-\bar{p}_{j, 0}]\right>.
\end{align}
For the first and second moments, we then find
\beal
\label{eq:moments_connection}
\omega'_{i}(\vgh_{\rm c}) &= \left<[p_i(r, \vgh)-\bar{p}_{i,0}]\right>=\bar{p}_i(\vgh_{\rm c})-\bar{p}_{i,0}
\nonumber
\\[1mm]
\nonumber
\omega'_{ij}(\vgh_{\rm c}) &= \left<[p_i(r, \vgh)-\bar{p}_{i,0}][p_j(r, \vgh)-\bar{p}_{j, 0}]\right>
\nonumber\\
&=\omega_{ij}(\vgh_{\rm c})+[\bar{p}_i(\vgh_{\rm c})-\bar{p}_{i,0}][\bar{p}_j(\vgh_{\rm c})-\bar{p}_{j,0}],
\end{align}
which directly shows how the fluctuations in $\bar{p}_i$ enter as a new contribution to the moments. This can introduce new spectral shapes at different multipoles, even if in any direction one specific shape is present. We briefly highlight some of the effects here (Sect.~\ref{sec:variations}), but discuss their relevance to the analysis of future $B$-mode polarization and distortion measurements elsewhere.

\vspace{-3mm}
\section{Superposition of power-law spectra}
\label{sec:power-law}
Let us start with the simple example of a power-law spectral distribution, $I_\nu=A_0 \, (\nu/\nu_0)^\alpha$. The free parameters $\vek{p}=\{A_0, \alpha\}$ are assumed to vary spatially. This description is typically used to model the foreground caused by synchrotron emission at low frequencies \citep{Planck2013components, PlanckSM2015}. One common extension is to introduce curvature to the spectral index, $I^\ast_\nu=A_0 \, (\nu/\nu_0)^{\alpha+\frac{1}{2}\beta \ln(\nu/\nu_0)}$. Let us try to see under which conditions this approximation works by using the moment method. For this, we first compute the derivatives of $I_\nu$ with respect to $A_0$ and $\alpha$:
\beal
\partial^k_{A_0}\,I_\nu(A_0, \alpha)&=\delta_{k1}\,(\nu/\nu_0)^\alpha=\delta_{k1}\,\frac{I_\nu(A_0, \alpha)}{A_0}
\\[1mm]
\partial^k_{\alpha}\,I_\nu(A_0, \alpha)&=I_\nu(A_0, \alpha)\,\ln^k(\nu/\nu_0).
\end{align}
Since $I_\nu$ depends linearly on $A_0$, only first order derivatives with respect to $A_0$ appear. Since the derivatives $\partial_{A_0}$ and $\partial_\alpha$ commute, using Eq.~\eqref{eq:I_av}, we thus find the expansion
\beal
\left<I_\nu(\vek{p})\right>&=I(\nu_{\rm c}, \bar{\vek{p}})
+\omega_{12}I_{12}(\nu_{\rm c}, \bar{\vek{p}})+\frac{1}{2}\omega_{22}I_{22}(\nu_{\rm c}, \bar{\vek{p}})
\\[1mm] \nonumber
&\quad+\frac{1}{2}\,\omega_{122}\,I_{122}(\nu_{\rm c}, \bar{\vek{p}})+\frac{1}{6}\,\omega_{222}\,I_{222}(\nu_{\rm c}, \bar{\vek{p}})
\\[1mm] \nonumber
&\quad\quad+\frac{1}{6}\,\omega_{1222}\,I_{1222}(\nu_{\rm c}, \bar{\vek{p}})+\frac{1}{24}\,\omega_{2222}\,I_{2222}(\nu_{\rm c}, \bar{\vek{p}}) + \ldots
\end{align}
For the spectral functions, we then have 
\beal
\bar{A}_0 I_{12\ldots 2}(\nu_{\rm c}, \bar{\vek{p}})\equiv I_{2\ldots 2}(\nu_{\rm c}, \bar{\vek{p}}) = 
\left<\partial^k_{\bar{\alpha}}\,I_\nu(\bar{\vek{p}})\right>=\left<I_\nu(\bar{\vek{p}})\ln^k(\nu/\nu_0)\right>.
\end{align}
Here, we assumed that $k$ derivatives (indicated by the `$\ldots$' in the subscripts) with respect to the second parameter, $\alpha$, were taken. For the moments, we then find
\beal
\omega_{12\ldots 2} &= \frac{\left<[A_0(\vek{r})-\bar{A}_0][\alpha(\vek{r})-\bar{\alpha}]^k  \partial^k_{\bar{\alpha}}\,I_\nu(\bar{\vek{p}})\right>}
{\left<\partial^k_{\bar{\alpha}}\,I_\nu(\bar{\vek{p}})\right>}
\approx\left<[A_0-\bar{A}_0][\alpha-\bar{\alpha}]^k\right>
\nonumber\\[1mm]
\omega_{2\ldots 2} &= \frac{\left<[\alpha(\vek{r})-\bar{\alpha}]^k  \partial^k_{\bar{\alpha}}\,I_\nu(\bar{\vek{p}})\right>}
{\left<\partial^k_{\bar{\alpha}}\,I_\nu(\bar{\vek{p}})\right>}
\approx\left<[\alpha-\bar{\alpha}]^k\right>,
\end{align}
where in the second step we assumed that $W(\vgh, \nu)\approx B(\vgh) F(\nu)$. In the considered example, one hence obtains
\beal
\label{eq:gen_power_law}
\left<I_\nu(\vek{p})\right>&=I(\nu_{\rm c}, \bar{\vek{p}})
+\omega_{12}I_{2}(\nu_{\rm c}, \bar{\vek{p}})+\frac{1}{2}\left[\omega_{22}+\frac{\omega_{122}}{\bar{A}_0}\right]I_{22}(\nu_{\rm c}, \bar{\vek{p}})
\nonumber\\[1mm]
&\quad+\frac{1}{6}\left[\omega_{222}+\frac{\omega_{1222}}{\bar{A}_0}\right]\,I_{222}(\nu_{\rm c}, \bar{\vek{p}})
\nonumber\\[1mm]
&\quad\quad+\frac{1}{24}\,\left[\omega_{2222}+\frac{\omega_{12222}}{\bar{A}_0}\right]\,I_{2222}(\nu_{\rm c}, \bar{\vek{p}}) + \ldots
\end{align}
We therefore only need to consider the spectral functions $I_{2\ldots2}(\nu_{\rm c}, \bar{\vek{p}})$, and the new set of moments
\beal
\label{eq:eff_mom_power}
\omega^\ast_{2\ldots 2} &= \omega_{2\ldots 2}+\frac{\omega_{12\ldots 2}}{\bar{A}_0}
= \frac{\left<A_0(\vek{r})[\alpha(\vek{r})-\bar{\alpha}]^k\,
\partial^k_{\bar{\alpha}}\,I_\nu(\bar{A}, \bar{\alpha})\right>}
{\bar{A}_0\left<\partial^k_{\bar{\alpha}}\,I_\nu(\bar{A}, \bar{\alpha})\right>}.
\end{align}
This shows that the amplitude $A_0(\vek{r})$ enters the problem as a spatial weighting function, which leaves the shape of the SED completely unchanged. With a similar redefinition one can absorb the correction term, $\propto\omega_{12}I_{2}(\nu_{\rm c}, \bar{\vek{p}})$, leading to $\bar{\alpha}^\ast=\bar{\alpha}+\omega_{12}/\bar{A}_0$, which we have to use instead of $\alpha$ in the moments, $\omega^\ast_{2\ldots 2}$. If $\omega_{12\ldots 2}=0$ [$\leftrightarrow A_0(\vek{r})$ and $\alpha(\vek{r})$ uncorrelated], one finds $\omega^\ast_{2\ldots 2}\equiv \omega_{2\ldots 2}$. This case is illustrative for constructing the moment expansion with {\it any} overall weighting factor that does not alter the shape of the SED.

For very narrow bands, i.e, $W(\vgh, \nu)= B(\vgh) \delta(\nu-\nu_{\rm c})$, we have $I_{2\ldots 2}(\nu_{\rm c}, \bar{\vek{p}})= I(\nu_{\rm c}, \bar{\vek{p}})\ln^k(\nu_{\rm c}/\nu_0)$, such that finally 
\beal
\label{eq:I_nu_gen_exp_power_law}
\left<I_\nu(\vek{p})\right>&=\bar{A}_0 \, (\nu_{\rm c}/\nu_0)^{\bar{\alpha}^\ast}
\left[1 +\frac{1}{2}\,\omega^\ast_{22} \ln^2(\nu_{\rm c}/\nu_0) +\frac{1}{6}\,\omega^\ast_{222} \ln^3(\nu_{\rm c}/\nu_0)\right.  
\nonumber\\
&\!\!\left.+\frac{1}{24}\,\omega^\ast_{2222} \ln^4(\nu_{\rm c}/\nu_0)+\frac{1}{120}\,\omega^\ast_{22222} \ln^5(\nu_{\rm c}/\nu_0) + \ldots \right].
\end{align}
We can now compare this expression with $I^\ast_\nu=A_0 \, (\nu/\nu_0)^{\alpha+\frac{1}{2}\beta \ln(\nu/\nu_0)}$ by performing a Taylor expansion in $\beta$, which yields
\beal
I^\ast_{\nu_{\rm c}}&\!\approx \!A_0 \, (\nu_{\rm c}/\nu_0)^\alpha
\left[1 +\frac{1}{2}\,\beta \ln^2(\nu_{\rm c}/\nu_0) +\frac{1}{8}\,\beta^2 \ln^4(\nu_{\rm c}/\nu_0)+ \ldots \right].
\end{align}
This shows that $\beta\equiv \omega^\ast_{22}$ if with respect to the average $\left<\ldots\right>$ the distribution of $\alpha(\vek{r})$ is Gaussian, such that $\omega^\ast_{2\ldots 2}=(k-1)!!\,(\omega^\ast_{22})^{k/2}$ for even $k$ and zero otherwise\footnote{Indeed $I_\nu=A_0 \, (\nu/\nu_0)^{\alpha+\frac{1}{2}\beta \ln(\nu/\nu_0)}$ is obtained when averaging power-laws, $I_\nu=A \, (\nu/\nu_0)^{\alpha}$, for a Gaussian distribution of $\alpha$ with variance $\beta$.}. This is quite restrictive and even in the limit of many power-law contributions hard to achieve in physical situations.

More generally, it is better to use Eq.~\eqref{eq:I_nu_gen_exp_power_law} to capture the additional degrees of freedom introduced by the averaging. The free parameters can then be determined using different spectral bands. In a compact form, this can also be expressed as
\beal
\label{eq:power_sum_exp_power}
I^{\rm p}_{\nu}&\approx A_0 \, (\nu/\nu_0)^{\alpha+\frac{1}{2}\beta \ln(\nu/\nu_0)+\frac{1}{6}\gamma \ln^2(\nu/\nu_0)+\frac{1}{24}\delta \ln^3(\nu/\nu_0)+\ldots}.
\end{align}
Assuming that there is no additional beam-induced frequency dependence, this expression again captures all the degrees of freedom for the superposition of power-law spectra. Here, the coefficients $\beta$, $\gamma$ and $\delta$ are related to the moments $\omega^\ast_{2\ldots 2}$. In the Rayleigh-Jeans limit, Eq.~\eqref{eq:power_sum_exp_power} is also equivalent to the commonly used parameterization, $\log T_{\rm sync}(\nu)=\sum_{i=0}^N \alpha_i \, \log(\nu/\nu_0)^i$ \citep{Pritchard2010} for the antenna temperature contribution due to Galactic synchrotron. However, we find that for our examples this parameterization does not perform as well as the explicit moment representation, Eq.~\eqref{eq:I_nu_gen_exp_power_law}, which also has a clear interpretation in terms of the physical parameters and their statistical properties.

For a Gaussian distribution of spectral indices, we can also deduce a convergence criterion for the moment expansion. If the variance, $\sigma^2=\beta$, of the spectral indices exceeds $\beta\gtrsim 2/|\log(\nu/\nu_0)|$ a moment expansion is expected to have a small convergence radius around $\nu\simeq \nu_0$. For CMB applications, this limits $\beta\lesssim 0.3$, which seems to be fulfilled in observations \citep{Fuskeland2014}.

\vspace{-0mm}
\subsection{Examples for sums of power-law spectra}
\label{sec:two-p-spectra}
To develop some intuition for how the moment expansion can be applied, let us consider the simple sum of two power-law spectra, $\vek{p}=\{A_1, \alpha_1, A_2, \alpha_2\}$. This is a low dimensional parameter case, which can be used to illustrate the limitations and strength of the moment method. It is also an extreme example for the distribution of parameters, since it clearly is far from `Gaussian'. In this case, we have the moments  
\beal
\label{eq:pow_mom_theo}
\bar{A}&=A_1+A_2
\nonumber
\\
\bar{\alpha}&=\frac{A_1\alpha_1+A_2\alpha_2}{A_1+A_2}=f\, \alpha_1+(1-f)\alpha_2
\\
\nonumber
\omega_{2\ldots2}&=\frac{A_1(\alpha_1-\bar{\alpha})^k+A_2(\alpha_2-\bar{\alpha})^k}{A_1+A_2}
=\left[f(f-1)^k+(1-f)f^k\right]\Delta\alpha^k
\end{align}
with $f=A_1/(A_1+A_2)$ and $\Delta\alpha=\alpha_2-\alpha_1$. It is easy to see that for $f=1/2$ (i.e., $A_1=A_2$) all odd moments vanish, while the even ones are $\omega_{2\ldots2}=\Delta\alpha^k/2^k$. For general $f$, we have $\omega_{22}=f(1-f)\Delta \alpha^2\geq 0$, which illustrates that for a {\it sum}\footnote{This is no longer true when {\it difference} of power-law spectra is allowed.} of power-law spectra one always obtains a {\it convex} foreground spectrum. 

The expressions in Eq.~\eqref{eq:pow_mom_theo} also illustrate why thinking of the problem in terms of moments of distribution functions makes a lot of sense. The probability of finding an SED contribution with spectral index $\alpha_1$ in the beam is $f=A_1/(A_1+A_2)$ and the one for $\alpha_2$ is $\bar{f}=1-f$. We thus have the probability distribution function (PDF), $P(\alpha)=f \delta(\alpha-\alpha_1) + \bar{f} \delta(\alpha-\alpha_2)$. Then the average spectral index is, $\bar{\alpha}=\int P(\alpha) \, \alpha \id \alpha= f \alpha_1 +\bar{f} \alpha_2$, and similarly $\omega_{2\ldots2}=\int P(\alpha) \, (\alpha-\bar{\alpha})^k \id \alpha=f (\alpha_1-\bar{\alpha})^k + \bar{f} (\alpha_2-\bar{\alpha})^2$, as also found in Eq.~\eqref{eq:pow_mom_theo}. Only that in general the PDF is a much more complicated multi-dimensional function.

\begin{figure}
\centering 
\includegraphics[width=\columnwidth]{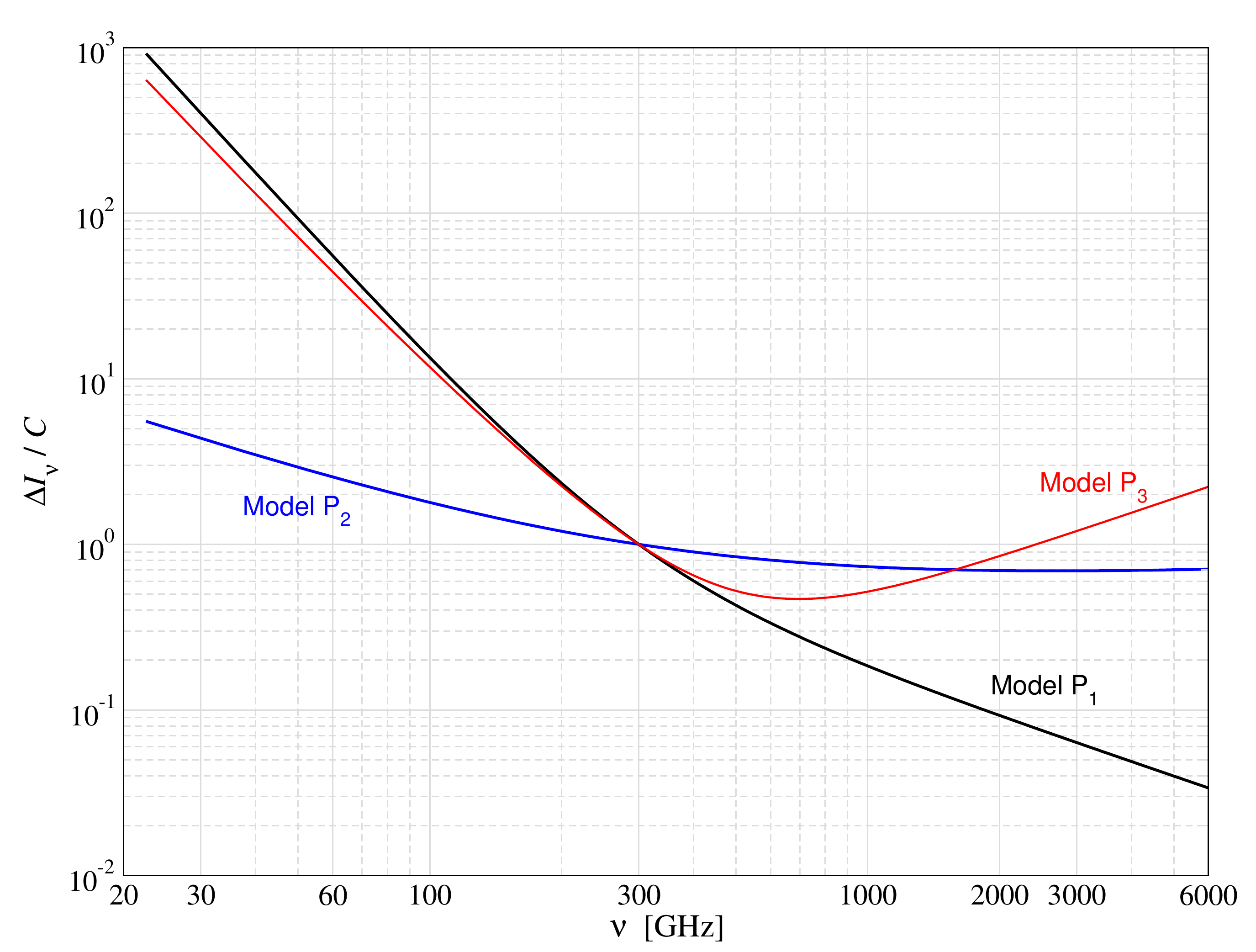}
\\[1mm]
\includegraphics[width=\columnwidth]{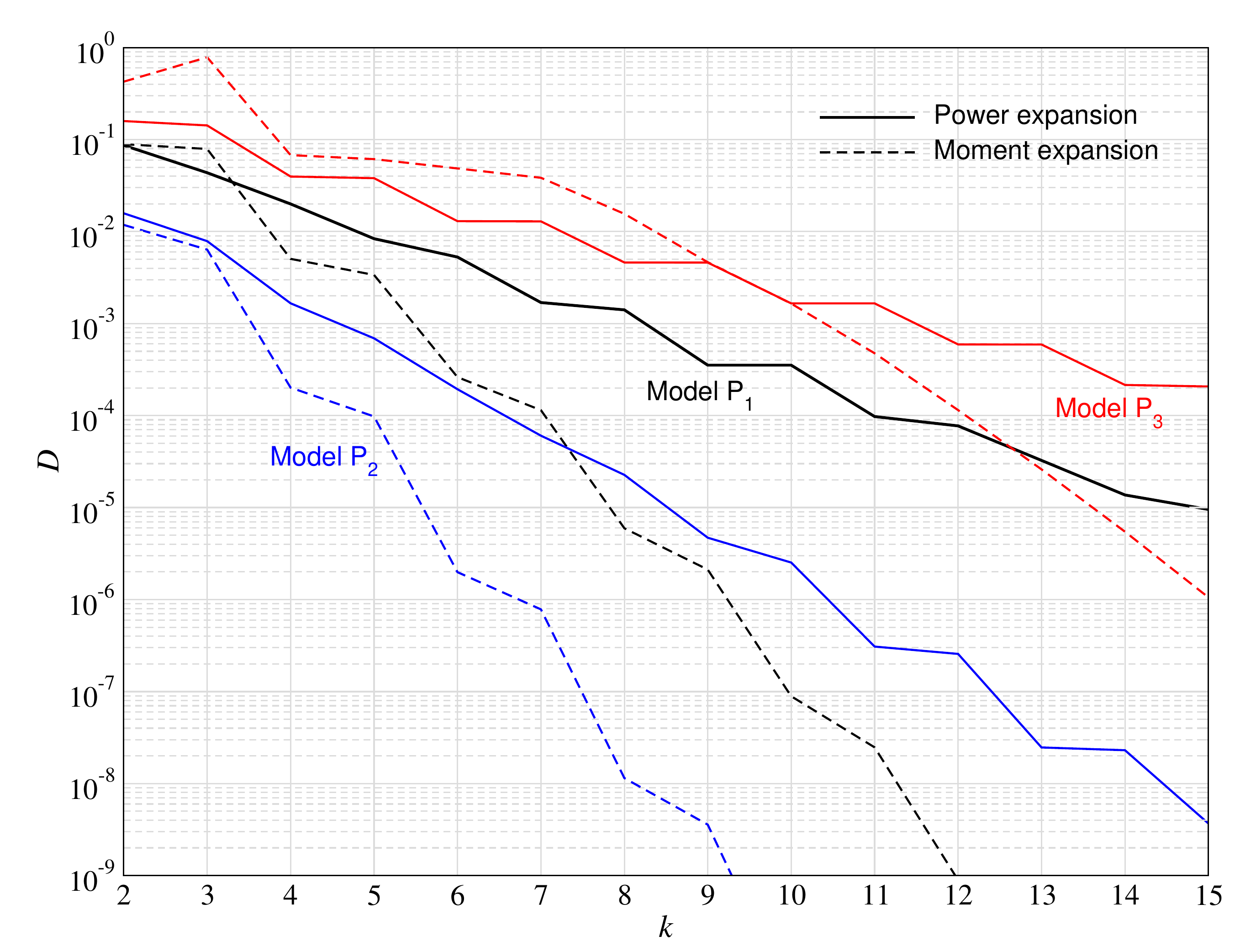}
\caption{Intensity for a superposition of power-law spectra as a function of frequency (top panel) and the convergence of the foreground parameterizations (lower panel) for different models, $P_i$. The two parameterizations Eq.~\eqref{eq:I_nu_gen_exp_power_law} and Eq.~\eqref{eq:power_sum_exp_power} are used with $k$ denoting the maximal moment index ($k=2$ for $\omega_{22}$ and $\beta$, $k=3$ for $\omega_{33}$ and $\gamma$, etc).}
\label{fig:P1_models}
\end{figure}
We now discuss the performance of different parameterizations. For a two power-law foreground, we choose the two illustrative cases $P_1=\{A_1, \alpha_1, A_2, \alpha_2\}=\{C/2, -2.9, C/2, -0.9\}$ and $P_2=\{A_1, \alpha_1, A_2, \alpha_2\}=\{C/2, -0.9, C/2, 0.1\}$ with pivot frequency $\nu_0=300\,\GHz$. Here $C$ is a normalization constant with dimension Jy/sr. The latter model has a low-frequency behavior that is similar to that of synchrotron emission in our Galaxy \citep{PlanckSM2015}. Varying the fraction $f$ is similar to changing $\nu_0$ and we tried different fractions but the conclusions did not change much.
We now simply perform a least-square fit in logarithmic coordinates\footnote{This is not a crucial choice.} (both in $\nu$ and $\Delta I_\nu$) for data points between $30\,\GHz$ and $6000\,\GHz$ with $\Delta \nu=15\,\GHz$ (this is similar to {\it PIXIE}), considering noise-less reconstructions for illustration of the main effects. We then compute the convergence of the two parameterizations, Eq.~\eqref{eq:I_nu_gen_exp_power_law} and Eq.~\eqref{eq:power_sum_exp_power}, varying the number of terms. To assess the convergence, we use the simple criterion
\beal
\label{eq:Conv_power}
\mathcal{D}=\sqrt{\frac{1}{N_{\rm ch}} \sum_i (I_{\nu_i}^{\rm model}/I_{\nu_i}^{\rm input}-1)^2},
\end{align}
which shows how large the average relative deviation of the fit from the input is. Here, $N_{\rm ch}\simeq 400$ is the number of channels. Since the CMB signals are typically several orders of magnitude smaller than the foregrounds, values for $\mathcal{D}\simeq 10^{-4}-10^{-3}$ need to be achieved. This is obtained with $\simeq 3-4$ moments for model $P_2$ and $\simeq 5-6$ moments for $P_1$. Model $P_3$, which has the largest change of the spectral index from low to high frequencies, shows the slowest convergence of the three considered cases. 

In Fig.~\ref{fig:P1_models} we show the intensity for the sum of power-laws and the convergences of the foreground parameterizations as a function of the included terms. For the models $P_1$ and $P_2$, the convergence is more rapid with the moment expansion, Eq.~\eqref{eq:I_nu_gen_exp_power_law}. We varied the combination of the spectral indices and also $f$ but found that this parameterization generally performed better than Eq.~\eqref{eq:power_sum_exp_power}. Of course by directly assuming a two power-law model the convergence is perfect already for four parameters. However, if the underlying foreground is composed of a distribution of power-laws (physically very plausible), then it should be better to use the moment expansion, Eq.~\eqref{eq:I_nu_gen_exp_power_law}, to allow more freedom. 

To further illustrate this point, we chose a 5-power-law model, $P_3=\{A_1, \alpha_1, ..., A_5, \alpha_5\}=\{0.3\,C, -2.9, 0.35\,C, -2.1, 0.15\,C,$ $ -1.5, 0.05\,C, -1.1, 0.15\,C, 0.9\}$. As Fig.~\ref{fig:P1_models} shows, even if the overall convergence is rather slow, also for this case the moment expansion performs better than Eq.~\eqref{eq:power_sum_exp_power} when including many terms ($k\gtrsim 9-10$). 
For comparison, we also performed explicit (finite number) power-law superpositions. Assuming a two power-law approximation, one is able to represent the model $P_3$ with $1\%-10\%$ precision. This does not improve much for a three or four power-law approximation, highlighting that only for a sum of five power-laws (identical to the 10 parameter input model) convergence can be achieved. Also, a simple sum of power-laws exhibits significant parameter degeneracies, which make finding a solution harder. This problem is at least partially mitigated with the moment method.

We also mention that the convergence criterion, Eq.~\eqref{eq:Conv_power}, is quite demanding, requiring an overall relative accuracy in the full frequency domain. When studying the performance for fixed channel sensitivity, we find a lower number of moments suffices in many cases. This is particularly true if the average spectrum is monotonic.

\subsubsection{Comparison of the best-fitting parameters with the theoretical moment values}
\label{sec:pow_exercise}
In the exercise for Fig.~\ref{fig:P1_models}, we simply used the expressions in Eq.~\eqref{eq:I_nu_gen_exp_power_law} and Eq.~\eqref{eq:power_sum_exp_power} to represent the different examples and determined the free parameters using a noiseless $\chi^2$ minimization. For the moment expansion, Eq.~\eqref{eq:I_nu_gen_exp_power_law}, we can explicitly compare the obtained best-fitting values with the theoretical values, Eq.~\eqref{eq:pow_mom_theo}. 
\begin{table}
\centering
\caption{Comparison of first few moments for power-law model $P_2$. In the fitting procedure (last three columns), the number of parameters was varied.}
\begin{tabular}{lcccc}
\hline
\hline
Parameter & Theory & 2 parameters & 3 parameters & 5 parameters
\\[1pt]
\hline
$\bar{A}/C$ & $1$ & $1.039$ & $0.977$ & $1.000$
\\[1pt]
$\bar{\alpha}$ & $-0.4$ & $-0.169$ & $-0.399$ & $-0.400$
\\[1pt]
$\omega_{22}$ & $0.25$ & -- & $0.301$ & $0.248$
\\[1pt]
$\omega_{222}$ & $0$ & -- & -- & $-0.001$
\\[1pt]
$\omega_{2222}$ & $0.0625$ & -- & -- & $0.070$
\\[1pt]
\hline
\hline
\label{tab:one}
\end{tabular}
\end{table}
In Table~\ref{tab:one}, we show this comparison for model $P_2$ when varying the number of parameters that are included in the fit. As alluded to in the introduction, the recovered values from the fit depend on the included number of parameters in the representation. This is because the spectral shapes of higher moments are not completely independent of the lower-moment shapes. However, by including more moments, the values converge rapidly towards the theoretical moment values. In a real analysis, this behavior can be used as a diagnostic to determine at what level the moment expansion can be truncated, a procedure that is limited by the number of channels.

\vspace{-2mm}
\section{Superposition of free-free spectra}
\label{sec:ffspectra}
Another CMB foreground, present at low-frequencies, is caused by the free-free emission of thermal electrons ($\Te \simeq 7000\,\Kel$). The plasma is assumed to be optically thin at the relevant frequencies ($\nu \geq 1\,\GHz$), so that the fundamental spectral shape is given by
\beal
\label{eq:I_nu_gen_ff_in}
\delta I^{\rm ff}_\nu&\approx
\frac{hc\delta t\,\alpha}{\sqrt{6 \pi^3}\,\The^{1/2}}\sum_i Z_i^2 N_i \, g_{\rm ff}(Z_i, \Te, \nu)
\,\frac{\expf{x-\xe}-1}{\expf{x}-1}
\end{align}
with $x=\xe\,\Te/T_0=h\nu/k T_0$, $\The=k\Te/\me c^2$ and where $g_{\rm ff}$ denotes the thermally averaged free-free Gaunt-factor. The factor, $f(x, \xe)=(\expf{x-\xe}-1)/(\expf{x}-1)$, accounts for stimulated emission\footnote{Without stimulated free-free emission due to the CMB we would have $f(x, \xe)\rightarrow \expf{-\xe}$, which decays much faster towards high frequencies, but is usually also neglected.} and absorption of ambient CMB photons. This can be directly obtained from the radiative transfer equation (see Appendix~\ref{app:ff_derivation}). At $\nu\lesssim 1\,{\rm THz}$, it can usually be neglected, but it does suppress the emission at higher frequency (see Fig.~\ref{fig:ff}). 

The Gaunt-factor can be obtained from detailed tables \citep[e.g.,][]{Itoh2000}; however, to demonstrate the important effects we shall use the simpler approximation\footnote{We treat these expressions as exact, given that the Gaunt factor can be precisely computed if necessary \citep{Karzas1961}.} \citep{Draine2011Book}
\beal
\label{eq:Gaunt_low}
g_{\rm ff}(Z_i, \Te, \nu)
&=1+\ln\left[1+\left(\frac{\nu_{\rm ff}(Z_i, \Te)}{\nu}\right)^{\frac{\sqrt{3}}{\pi}} \right]
\nonumber\\
\nu_{\rm ff}(Z_i, \Te)
&\approx \frac{255.33\,\GHz}{Z_i} \left[\frac{\Te}{10^3\,{\rm K}}\right]^{3/2},
\end{align}
where $Z_i$ is the charge of the nucleus of ion $i$. 
Neglecting the high-frequency suppression [$f(x, \xe)=1$], we can perform a simple moment expansion\footnote{We also tried an expansion in $\Te$ and $1/\Te$, but found this to converge quite slowly. Using $\ln \nu_{\rm ff}$ as a parameter is also motivated by the fact that most of the emission is found at low frequencies where one has $g_{\rm ff}(Z_i, \Te, \nu)\approx 1+\frac{\sqrt{3}}{\pi}\,\ln \left(\frac{\nu_{\rm ff}(Z_i, \Te)}{\nu}\right)$.} in $\xi=\ln \nu_{\rm ff}$. 
With $\gamma=\frac{\sqrt{3}}{\pi}$, $\eta=(\nu_{\rm ff}/\nu)^\gamma$ and
\beal
\label{eq:g_ff_k}
\partial_{\xi} g_{\rm ff}&=\frac{\gamma\,\eta}{(1 + \eta)},
&&\partial^2_{\xi} g_{\rm ff}=\frac{\gamma^2\,\eta }{(1 + \eta)^2}
\nonumber\\
\partial^3_{\xi} g_{\rm ff}&=\frac{\gamma^3\,\eta\,(1-\eta) }{(1 + \eta)^3}, 
&&\partial^4_{\xi} g_{\rm ff}=\frac{\gamma^4\,\eta\,(1-4\eta+\eta^2) }{(1 + \eta)^4}
\end{align}
for the first few derivatives of $g_{\rm ff}$ with respect to $\xi$, we then have
\beal
\label{eq:I_nu_gen_ff}
&\left<I^{\rm ff}_\nu\right> \approx \bar\epsilon \, g_{\rm ff}(\bar{\nu}_{\rm ff}, \nu)+\bar\epsilon \sum_{k=2}^\infty\frac{g^{(k)}_{\rm ff}(\bar{\nu}_{\rm ff}, \nu)}{k!}\, \omega^{(k)}_{\rm ff}
\nonumber\\
&g^{(k)}_{\rm ff}(\bar{\nu}_{\rm ff}, \nu)=\left.\partial^k_\xi g_{\rm ff}\right|_{\nu_{\rm ff}=\bar{\nu}_{\rm ff}}
\\ \nonumber
&\bar\epsilon=\left<\sum_i \epsilon_i \right>,\quad \ln\bar\nu_{\rm ff}=\frac{\left<\sum_i \epsilon_i \ln \nu_{\rm ff}\right>}{\bar\epsilon},\quad
\omega^{(k)}_{\rm ff}=\frac{\left<\sum_i \epsilon_i \ln^k(\nu_{\rm ff}/\bar\nu_{\rm ff})\right>}{\bar\epsilon}.
\end{align}
Here, we introduced the parameter $\epsilon_i=hc\delta t\,\alpha\,Z_i^2 N_i/[\sqrt{6 \pi^3}\,\The^{1/2}]$, which just modifies the relative proportion of emission from different ions and can be kept as one single weight parameter. In this case, no separate moment expansion in $\Te$ and $Z_i$ has to be performed, since variations in $Z_i$ can be modeled as variations in $\Te^*=\Te/Z^{2/3}_i$, or equivalently, variations of $\ln \nu_{\rm ff}$. This moment expansion converges very rapidly; however, it fails to accurately capture the high-frequency suppression at $\nu\gtrsim 1\,{\rm THz}$.

\vspace{-2mm}
\subsection{Inclusion of the high-frequency suppression}
To also include the high-frequency suppression caused by ambient CMB photons, we have to add the separate dependence on $Z_i$. For $f(x, \xe)$, we need $\Te$, which can be expressed as $\Te=\kappa\,\nu_{\rm ff}^{2/3}\, Z_i^{2/3}$, with $\kappa\approx 24.851\,\Kel \,\GHz^{-2/3}$. We can then define $\bar T_{\rm e}$ using $\ln \bar{\nu}_{\rm ff}$ and $\bar{Z}$ to determine $\bar x_{\rm e}=x\,T_0/\bar T_{\rm e}$. Since $\rho_{\rm e}=T_0/\bar T_{\rm e}\ll 1$, we can write
\beal
\label{eq:f_supp}
f(x, \xe)&= f(x, \bar x_{\rm e})+\sum_{k=1}^\infty \frac{J_k(x, \bar x_{\rm e})}{k!} 
\,\frac{(\rho_{\rm e}-\bar{\rho}_{\rm e})^k}{\bar{\rho}_{\rm e}^k}
\nonumber\\
J_k(x, x')&=(-1)^k\,\frac{{x'}^k\expf{x-x'}}{\expf{x}-1}.
\end{align}
For CMB applications, it should be sufficient to include the first two terms in $\Delta \rho_{\rm e}=\rho_{\rm e}-\bar{\rho}_{\rm e}$.
To perform the moment expansion in $\ln \nu_{\rm ff}$ and $Z_i$, we define $\zeta_i=Z_i^{-2/3}$. 
This yields
\beal
\label{eq:DI_nu_gen_ff}
&\left<I^{\rm ff}_\nu\right>_{\rm high} = 
\left<I^{\rm ff}_\nu\right> f(x, \bar x_{\rm e})+
\left<\Delta I^{\rm ff}_\nu\right> 
\\ \nonumber
&\bar\zeta=\frac{\left<\sum_i \epsilon_i \,Z_i^{-2/3}\right>}{\bar\epsilon}\equiv \frac{1}{\bar{Z}^{2/3}},\qquad
\bar T_{\rm e}=\kappa\,\bar{\nu}_{\rm ff}^{2/3}/\bar\zeta,
\end{align}
where the additional correction $\left<\Delta I^{\rm ff}_\nu\right>$ is obtained from a moment expansion of 
$\Delta I^{\rm ff}_\nu=\epsilon_i \, g_{\rm ff}(\nu_{\rm ff}, \nu)\,[f(x, \xe)-f(x, \bar x_{\rm e})]$. At zeroth order in $\Delta \rho_{\rm e}$, the high frequency suppression is simply captured by multiplying $\left<I^{\rm ff}_\nu\right>$ in Eq.~\eqref{eq:I_nu_gen_ff} with $f(x, \bar x_{\rm e})$. However, variations in the temperature and composition do require a more detailed treatment. The derivations are straightforward but cumbersome. Defining the frequency-dependent functions
\beal
\label{eq:Gkm_ff}
G^{(l, m)}_{k}(\bar{\nu}_{\rm ff}, \bar\zeta, \nu)&=
\frac{\bar{\zeta}^m\,\partial_\xi^l \partial_{\zeta_i}^m}{\bar{\rho}_{\rm e}^k\,k! \, l! \, m!} 
\left[g_{\rm ff}\,(\rho_{\rm e}-\bar{\rho}_{\rm e})^k\right]_{\nu_{\rm ff}=\bar{\nu}_{\rm ff}, \zeta_i=\bar{\zeta}}
\end{align}
we can write
\beal
\label{eq:DI_nu_gen_ff_cont}
&\left<\Delta I^{\rm ff}_\nu\right> =
\bar\epsilon 
 \sum_{l=0}^\infty \sum_{m=0}^\infty \left\{\sum_{k=1}^\infty J_k(x, \bar x_{\rm e})\,G^{(l, m)}_{k}(\bar{\nu}_{\rm ff}, \bar\zeta, \nu)\right\}  \, \sigma^{(l,m)}_{\rm ff}
\nonumber \\ 
&\sigma^{(k,m)}_{\rm ff}=\frac{\left<\sum_i \epsilon_i \ln^k(\nu_{\rm ff}/\bar\nu_{\rm ff})(\zeta_i-\bar\zeta)^m\right>}{\bar\epsilon}.
\end{align}
We note that $\sigma^{(k,0)}_{\rm ff}\equiv \omega^{(k)}_{\rm ff}$, $G^{(0, 0)}_{k}=0$ and $\sigma^{(1,0)}_{\rm ff}=\sigma^{(0,1)}_{\rm ff}=0$ by construction. The first few functions necessary for a fourth order expansion in $\ln \nu_{\rm ff}$ and $Z_i$ and second order in $\bar\rho_{\rm e}$ are given in Appendix~\ref{app:derivs_ff}. At low frequencies, $f(x, \xe)\rightarrow 1$. In this case, no new spectral shapes are added through $\left<\Delta I^{\rm ff}_\nu\right>$ and the number of free parameters, equivalent to the number of independent spectral functions, can be reduced to only using $\bar\epsilon$ , $\bar\nu_{\rm ff}$ and $\omega^{(k)}_{\rm ff}$. 

Consequently, only the high frequency tail ($\nu\gtrsim 1\,{\rm THz}$) can be used to obtain direct information about $\bar Z$, which specifies the chemical composition of the medium and its ionization state, through the moments $\sigma^{(k, l)}_{\rm ff}$. But how many coefficients $\sigma^{(k, l)}_{\rm ff}$ can in principle be constrained independently? 
The functions $G^{(l, m)}_{k}(\bar{\nu}_{\rm ff}, \bar\zeta, \nu)$ are all just linear combinations of $g_{\rm ff}$ and its derivatives, $g^{(j)}_{\rm ff}$, up to $j=3$ (see Appendix~\ref{app:derivs_ff}). These would be degenerate with simple variations of $\nu_{\rm ff}$, having the same spectral dependence as in the expansion, Eq.~\eqref{eq:I_nu_gen_ff}. However, the functions $J_k(x, \bar x_{\rm e})$ introduce new spectral shapes that can in principle be used to break the degeneracy.
Up to fourth order in $\ln \nu_{\rm ff}$ and $Z_i$ ($l+m\leq 4$) and second order in $\bar\rho_{\rm e}$ one thus finds $7$ linearly independent spectral shapes. The function $J_1(x, \bar x_{\rm e})\,G^{(0, 1)}_{1}(\bar{\nu}_{\rm ff}, \bar\zeta, \nu)$ fixes $\bar\zeta$.  Since $\sigma^{(k,0)}_{\rm ff}\equiv \omega^{(k)}_{\rm ff}$ (3~parameters) are determined through $\left<I^{\rm ff}_\nu\right> f(x, \bar x_{\rm e})$, only 6 additional $\sigma^{(k,m)}_{\rm ff}$ remain unspecified. This means that all $\sigma^{(k,m)}_{\rm ff}$ can in principle be extracted from detailed measurements of the free-free spectrum at low {\it and} sufficiently high frequencies. The dimensionality of the problem reduces at low frequencies, when only $J_1\simeq {\rm const}$ contributes, as mentioned above. In addition, instrumental noise and contributions from other foregrounds will reduce our ability to distinguish different chemical compositions and temperature distributions so that ultimately a much lower number of parameters can be independently inferred.

\vspace{2mm}
\subsection{Simple approximations for single-temperature plasmas}
\begin{figure}
\centering 
\includegraphics[width=\columnwidth]{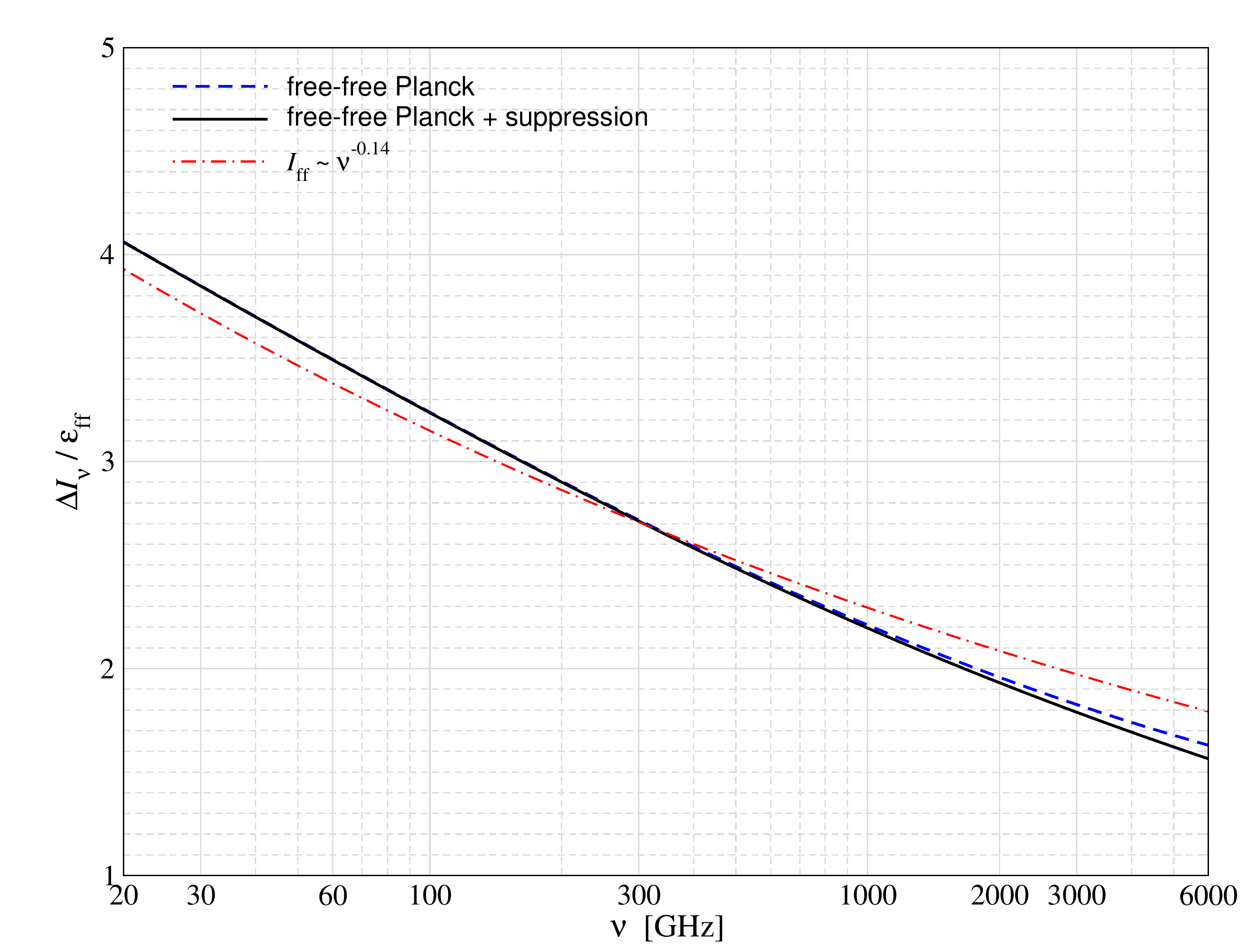}
\caption{Free-free intensity (in units of $\epsilon_{\rm ff}\simeq 150\,{\rm Jy/sr}$) for different cases.}
\label{fig:ff}
\end{figure}

For {\it Planck} measurements \citep{PlanckSM2015}, the free-free emission is modeled using a single-temperature approximation with $\nu_{\rm ff}(1, \Te)\approx 4727\,\GHz$ for $\Te=(7000\pm11)\,\Kel$. The high-frequency suppression is neglected, such that one can write\footnote{The Rayleigh-Jeans temperature is obtained as $T_{\rm b}=c^2 I^{\rm ff, P}_\nu/(2 k \nu^2)$.}
\beal
\label{eq:I_nu_Pl_ff}
&I^{\rm ff, P}_\nu \approx \epsilon_{\rm ff} \, g_{\rm ff}(\nu_{\rm ff}, \nu),
\end{align}
where $\epsilon_{\rm ff}\simeq 150 \, {\rm Jy/sr}$ \citep{PlanckSM2015}. To include the high-frequency suppression, we simply multiply by $f(x, \xe)$. These two cases are shown in Fig.~\ref{fig:ff} together with a simple power-law fit $I^{\rm ff, P}_\nu\propto \nu^{-0.14}$ $(\equiv T_{\rm b}\propto \nu^{-2.14})$, which is close to the template used by \citet{PySky}. The latter only captures the overall trend but fails to reproduce the curvature of the spectrum.

A more accurate alternative representation of the free-free emission can be obtained using a power-law moment expansion, Eq.~\eqref{eq:I_nu_gen_exp_power_law}. We find
\beal
\label{eq:I_nu_Pl_ff_2}
&I^{\rm ff, 2}_\nu \approx \frac{3.229}{\nu_{100}^{0.1518}}\,\epsilon_{\rm ff} \,\left[1 - \pot{6.129}{-3} \ln^2\nu_{100}\right]
\end{align}
with $\nu_{100}\equiv \frac{\nu}{100\,\GHz}$ to work to $\lesssim 0.5\%$ precision in the frequency range $1\,\GHz\lesssim \nu\lesssim 6\,{\rm THz}$. For better than $\simeq 0.06\%$ precision we obtain the fourth order power-law moment expansion
\beal
\label{eq:I_nu_Pl_ff_4}
I^{\rm ff, 4}_\nu &\approx \frac{3.235}{\nu_{100}^{0.1527}}\,\epsilon_{\rm ff} \left[1 - \pot{7.116}{-3} \ln^2\nu_{100}\right.
\nonumber\\
&\qquad+\left.\pot{8.701}{-5} \ln^3\nu_{100}+\pot{6.157}{-5} \ln^4\nu_{100}\right].
\end{align}
Both expressions include the high-frequency suppression effect and were obtained by noiseless fits using power-law moments as free parameters. This shows that the free-free emission for a single-temperature component can be fully represented using a superposition of power-laws. Thus, in the simplest CMB foreground scenarios it does not seem necessary to add any extra parameters when considering both synchrotron and free-free emission with spatial variations. On the full sky, they apparently can be treated as one component with slightly more complex frequency structure. The spatial-spectral variations are then captured by the spatial variations of the moment values. In CMB analysis, the main motivation for modeling free-free and synchrotron independently is thus really driven by external priors but not because the data distinguishes them easily. However, this is not expected to be sufficient in more general cases, which could be relevant to non-CMB applications.

\subsection{Free-free models with multiple components}
\label{sec:model_ff_example}
In general, the Galactic free-free emission has contributions from multiple components. In this case, a single-temperature representation of the spectrum will no longer work and the moment expansions, Eq.~\eqref{eq:I_nu_gen_ff} and \eqref{eq:DI_nu_gen_ff}, are expected to perform better, while also capturing the more general cases with distributions of temperatures and chemical composition. 

\begin{figure}
\centering 
\includegraphics[width=\columnwidth]{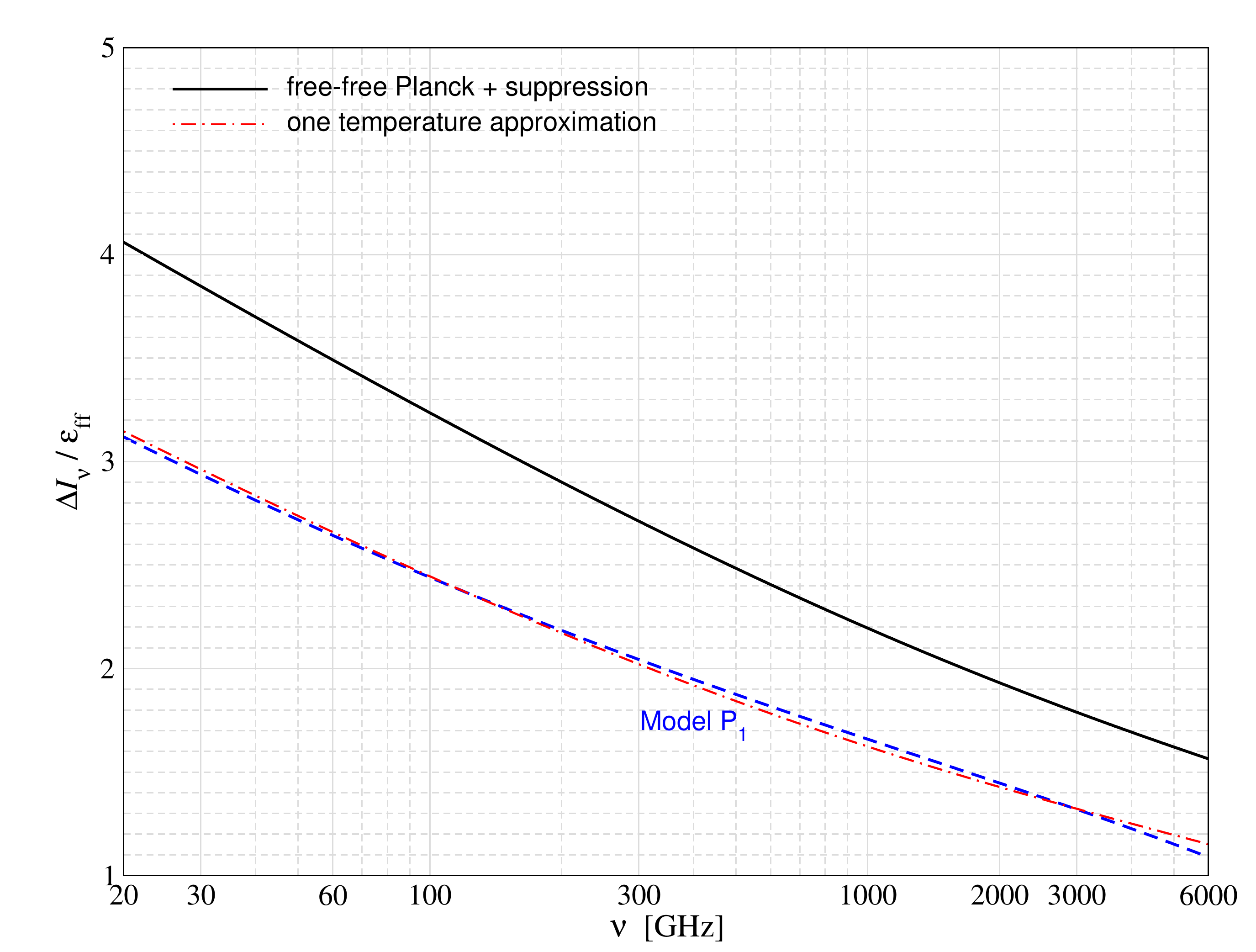}
\caption{Free-free intensity (in units of $\epsilon_{\rm ff}$) for different models. For comparison we also show the $\Te=7000\,\Kel$ {\it Planck} model, Eq.~\eqref{eq:I_nu_Pl_ff}, with the high-frequency suppression added. The change in the normalization is absorbed when determining $\epsilon_{\rm ff}$, but changes in the spectral shape remain.The red line represents a single temperature approximation. The moment approximations are not shown as they basically coincide with the input model.}
\label{fig:ff_PX}
\end{figure}

To illustrate the effects, we start with a two-temperature case, $P_1=\{\bar\epsilon_1, T_{\rm e, 1}, \bar\epsilon_2, T_{\rm e, 2}\}=\{0.5\,\epsilon_{\rm ff}, 7000\,\Kel, 0.5\,\epsilon_{\rm ff}, 500\,\Kel\}$, representing a mix of hot and cold components. For now we assume $Z_i=1$. The two components are generally not in the same region but could be situated at different locations along the line of sight. The relatively large fractional contribution from the low temperature component is physically slightly difficult to achieve, since at low temperatures the ionization degree (of hydrogen) is low. However, the effective weight scales as $\epsilon_i \simeq 1/\sqrt{\Te}$ and also there is typically more gas in the colder phase, so that we still consider this model for illustrative purposes.
Model $P_1$ is illustrated in Fig.~\ref{fig:ff_PX}. Approximating it with a single temperature spectrum\footnote{We use 100 log-spaced frequency bins in the range $1\,\GHz\lesssim \nu\lesssim 6\,{\rm THz}$.}, $I^{\rm ff, 1}_\nu \approx \bar{\epsilon} \, g_{\rm ff}(\bar{\nu}_{\rm ff}, \nu) f(x, \xe)$, we find $\bar{\nu}_{\rm ff}\approx 1600.4\,\GHz$ ($\Te\approx 3400\,\Kel$) and $\bar\epsilon\approx0.8992\,\epsilon_{\rm ff}$ (red line in Fig.~\ref{fig:ff_PX}). This representation works to $\lesssim 6\%$ precision in the considered frequency range (see Fig.~\ref{fig:ff_PX}). 
Explicitly using the free-free moment expansion, Eq.~\eqref{eq:I_nu_gen_ff} and \eqref{eq:DI_nu_gen_ff}, with $\bar{\epsilon}$, $\bar{\nu}_{\rm ff}$ and $\omega_{\rm ff}^{(2)}$ as free parameters, we obtain the values given in Table~\ref{tab:two}. This approximation already represents the free-free spectrum to $\lesssim 0.3\%$ precision at $1\,\GHz\lesssim \nu\lesssim 6\,{\rm THz}$. Adding also $\omega_{\rm ff}^{(3)}$ and $\omega_{\rm ff}^{(4)}$, we obtain the values given in the last column of Table~\ref{tab:two}. This approximation already works to better than $0.005\%$ precision. While in the latter case, the spectrum is described by one additional parameter than the input model (5 versus 4 parameters), this approximation is more general, not implicitly assuming a two-temperature model and allowing one to capture the effect of more general spatial variations in the temperature of the electrons. This again demonstrates the potential of the moment expansion.

\begin{table}
\centering
\caption{Comparison of first few moments for the free-free model $P_1$. In the fitting procedure (last three columns), the number of parameters was varied. For all cases we set $Z_i=1$.}
\begin{tabular}{lcccc}
\hline
\hline
Parameter & Theory & 2 parameters & 3 parameters & 5 parameters
\\[1pt]
\hline
$\bar{\epsilon}/\epsilon_{\rm ff}$ & $1$ & $0.8991$ & $1.001$ & $0.999$
\\[1pt]
$\bar{\nu}_{\rm ff}\, [\GHz]$ & $653.18$ & $1600.4$ & $651.88$ & $658.32$
\\[1pt]
$\omega_{\rm ff}^{(2)}$ & $3.9176$ & -- & $3.659$ & $3.835$
\\[1pt]
$\omega_{\rm ff}^{(3)}$ & $0$ & -- & -- & $-0.276$
\\[1pt]
$\omega_{\rm ff}^{(4)}$ & $15.348$ & -- & -- & $11.054$
\\[1pt]
\hline
\hline
\label{tab:two}
\end{tabular}
\end{table}

\newpage
We can again compare the obtained best-fitting values, with the theoretical predictions for the moments. For a general two-component case, $P=\{C_{\rm ff}(1-f), T_{\rm e, 1}, C_{\rm ff} f, T_{\rm e, 2}\}$, we have
\beal
\bar\epsilon&=C_{\rm ff}
\nonumber\\
\ln\bar{\nu}_{\rm ff}&=(1-f)\ln \nu_{\rm ff}(T_{\rm e, 1})+f \ln \nu_{\rm ff}(T_{\rm e, 2})
\nonumber\\
\omega^{(k)}_{\rm ff}&=(1-f)\ln^k[\nu_{\rm ff}(T_{\rm e, 1})/\bar\nu_{\rm ff}]+f \ln^k[\nu_{\rm ff}(T_{\rm e, 2})/\bar\nu_{\rm ff}]
\\ \nonumber
\bar\rho_{\rm e}&=T_0/[\kappa\,\bar{\nu}_{\rm ff}^{2/3}]\approx \pot{1.097}{-3}\,\left[\frac{\bar{\nu}_{\rm ff}}{10^3\,\GHz}\right]^{-2/3}
\\ \nonumber
\sigma^{(k, 0)}_{\rm ff}&=\omega^{(k)}_{\rm ff}, \qquad \sigma^{(k, m)}_{\rm ff}=0\quad\text{for $m>0$}
\end{align}
For model $P_1$, we find $\bar{\nu}_{\rm ff}=653.18\,\GHz$, $\bar\rho_{\rm e}=\pot{1.4571}{-3}$, $\omega^{(2)}_{\rm ff}=3.9176$, $\omega^{(4)}_{\rm ff}=15.348$, and $\omega^{(6)}_{\rm ff}=60.126$ (odd moments vanish for $f=0.5$). 
Comparing this with the values obtained from different model in the fit, we again see that with increasing order, the recovered values approach the theoretical one quite fast (see Table~\ref{tab:two}). This behavior is similar to that of the power-law moment expansion (Table~\ref{tab:one}).

We also illustrate how well the theoretical moments can be used to model the shape of the spectrum. Including only $\bar{\nu}_{\rm ff}$ and $\bar\rho_{\rm e}$ ($\equiv$ single-temperature model) in the moment expansion represents the low-frequency free-free spectrum very well, while underestimating the high-frequency spectrum by $\simeq 7\%$ at $\nu\simeq 1\,{\rm THz}$. Including $\omega^{(2)}_{\rm ff}=3.9176$ as additional parameter, the discrepancy decreases to $\lesssim 0.6\%$ at $1\,\GHz\lesssim \nu\lesssim 3\,{\rm THz}$. Adding the next non-vanishing moment, $\omega^{(4)}_{\rm ff}=15.348$, the moment expansion captures the full spectrum at the $\lesssim 0.02\%$ level in the same range.

When also varying $f$ for the two-temperature case, the approximation also requires inclusion of $\omega^{(3)}_{\rm ff}\neq 0$, which no longer vanishes. For the chosen extreme difference in temperatures, this representation is only accurate at the $\simeq 0.1\%$ level. This can be improved by adding higher order moments. However, with much smaller spread in the temperature even a second order expansion is found to converge very rapidly. Our study indicates that the free-free moment expansion allows us to accurately represent the free-free emission from more general temperature distributions with a minimal number of assumptions.

\vspace{-3mm}
\subsection{Variations in the chemical composition and ionization}
As alluded to above, when adding variations in the chemical composition, the low and high frequency spectra become more independent. This may be relevant to detailed modeling of the continuum X-ray component from clusters \citep{Ponente2011MNRAS}, where the gas is found in highly-ionized states with significant contributions from metals. For the modeling of emission in our galaxy, at temperatures $T\lesssim10^4\,\Kel$ the dominant contributions are from singly-ionized hydrogen and helium, which both have $Z=1$. Thus, variations in $Z$ are not expected to enter at a significant level to the spectral shape. 

When varying the distribution of temperatures, e.g., adding more components with different weights, we find that even a small addition of low-temperature plasma causes a single-temperature approximation to depart from the free-free spectrum at the $1\%$ level. In contrast, adding high temperature components the spectrum remains well described by a single temperature approximation. This can be understood since for high temperatures one has
\beal
g_{\rm ff}(Z_i, \Te, \nu)
&\approx 1+\ln\left(\frac{\nu_{\rm ff}(Z_i, \Te)}{\nu}\right)^{\frac{\sqrt{3}}{\pi}} =
1+\frac{\sqrt{3}}{\pi}\left[\ln\nu_{\rm ff}(Z_i, \Te) -  \ln \nu\right]
\nonumber
\end{align}
and $f(x, \xe)\approx 1$ at CMB frequencies. This means that adding multiple high temperature components, in the CMB regime [$\nu \lesssim \nu_{\rm ff}(Z_i, \Te)\simeq \mathcal{O}(10^3\,\GHz)$] only leads to an overall change in the zero offset [$\simeq 1+\frac{\sqrt{3}}{\pi}\ln\nu_{\rm ff}(Z_i, \Te)$] but no change in the spectral dependence, $\simeq -\frac{\sqrt{3}}{\pi} \ln \nu$. Thus, a sufficient approximation can indeed be obtained with a single-temperature model. The inferred effective temperature is close to 
\beal
\ln \bar{T}^*_{\rm e}= \frac{\left< \sum \epsilon_i\, \ln\Te^* \,\partial_{\ln \Te^*} \ln \nu_{\rm ff}(\Te^*) \right> }{\left< \sum \epsilon_i\,\partial_{\ln \Te^*} \ln \nu_{\rm ff}(\Te^*) \right>}
\equiv \frac{\left< \sum \epsilon_i\,\ln\Te^* \right>}{\bar\epsilon},
\nonumber
\end{align}
with $\Te^*=\Te \zeta_i$. This expression explicitly shows how the chemical composition enters into the problem and that the weighted average has to be carried out over $\epsilon_i \ln(\Te\zeta_i)$. However, at high frequencies ($\nu\gtrsim 1\,{\rm THz}$), corrections can become noticeable at $\simeq 0.1\%-1\%$ level even in this situation.

This highlights that for CMB applications the shape of the free-free spectrum is quite insensitive to the detailed temperature structure and composition. Thus, simpler parameterizations such as a power-law moment expansion should be applicable, in particular for CMB applications. We also highlight that in the presence of instrumental noise and other foregrounds (which we neglected here), physically distinct components are expected to become spectrally indistinguishable. This can lead to significant model-dependence (in form of prior choices), which can also cause biases.

\vspace{2mm}
\section{Superposition of gray-body spectra}
\label{sec:blackbody}
We can also apply the moment method to spatially varying gray-body spectra, with intensity $I_\nu(A_0, T)=A_0 \nu^3/(\expf{h\nu/kT}-1)$. This case is not directly relevant to CMB applications but illustrates how to use the moment expansion.
It is convenient to perform the expansion in terms of\footnote{We will discuss this choice in Sect.~\ref{sec:gray_single}.} $\beta=1/T$, since in this case a closed form for the derivatives of the distribution function with good convergence properties can be given (see Appendix \ref{app:derivs}). The factor $A_0$ is again a simple weight factor which does not affect the spectral shape. For the first few derivatives with respect to $\beta$, we find
\beal
\beta\, \partial_{\beta}\,I_\nu(A_0, T)&= - I_\nu(A_0, T) \,\frac{x\expf{x}}{\expf{x}-1}
\\\nonumber
\beta^2\, \partial^2_{\beta}\,I_\nu(A_0, T)&=  I_\nu(A_0, T) \,\frac{x\expf{x}}{\expf{x}-1}\,x\coth(x/2)
\\\nonumber
\beta^3\, \partial^3_{\beta}\,I_\nu(A_0, T)&= - I_\nu(A_0, T) \,\frac{x\expf{x}}{\expf{x}-1}\,x^2\frac{\cosh(x)+2}{\cosh(x)-1}
\\\nonumber
\beta^4\, \partial^4_{\beta}\,I_\nu(A_0, T)&=  I_\nu(A_0, T) \,\frac{x\expf{x}}{\expf{x}-1}\,\frac{x^3}{2}\frac{\cosh(x)+5}{\sinh^2(x/2)}\coth(x/2)
\\\nonumber
\beta^5\, \partial^5_{\beta}\,I_\nu(A_0, T)&= - I_\nu(A_0, T) \,\frac{x\expf{x}}{\expf{x}-1}\,\frac{x^4}{8}\frac{33+26\cosh(x)+\cosh(2x)}{\sinh^4(x/2)}
\end{align}
with $x=h\nu/kT$. The same arguments as in the previous section apply, so that Eq.~\eqref{eq:gen_power_law} carries over with $\alpha\rightarrow 1/T$ in the moments. For narrow spectral bands, we then find the moment representation
\beal
\label{eq:I_nu_gen_Gray}
\left<I_\nu\right>&=\frac{\bar{A}_0 \nu^3}{\expf{x}-1}
\left\{1 +\frac{1}{2}\,\omega^{\rm g}_{22} \,Y_2(x) +\frac{1}{6}\, \omega^{\rm g}_{222} \,Y_3(x) 
\right.
\nonumber
\\
\nonumber
&\qquad
\left.
+\frac{1}{24}\, \omega^{\rm g}_{2222} \,Y_4(x)
+\frac{1}{120}\, \omega^{\rm g}_{22222}  \,Y_5(x)+ \ldots\right\}
\nonumber\\
\frac{1}{\bar{T}} &= 
\frac{\left<A_0(\vek{r})/T(\vek{r})\right>}{\bar{A}_0}
\\ \nonumber
\omega^{\rm g}_{2\ldots 2} &= \bar{T}^k\omega^\ast_{2\ldots 2} 
=\frac{\left<A_0(\vek{r})[\bar{T}/T(\vek{r})-1]^k\right>}
{\bar{A}_0},
\end{align}
where now $x=h\nu/k\bar{T}$ and we defined $Y_k=[(-\beta)^k\partial^k_{\beta}\,I_\nu]/I_\nu$ (e.g., $Y_1(x)=x\,\expf{x}/(\expf{x}-1)$ and $Y_2(x)=Y_1(x)\,x\coth(x/2)$, etc.). 
The first correction term is similar to a Compton $y$-type distortion \citep{Zeldovich1969}, with effective $y$-parameter $y=\frac{1}{2}\,\omega_{22}^{\rm g}$, but generalized to gray-body spectra. The effective moments, $\omega^\ast_{2\ldots2}$, are obtained using Eq.~\eqref{eq:eff_mom_power}, with $\alpha\rightarrow 1/T$. 

Below we illustrate how to apply the gray-body moment expansion for several cases. However, Eq.~\eqref{eq:I_nu_gen_Gray} can also be used to model the correction to the average CMB spectrum caused by the temperature fluctuations of the CMB itself. In this case $A_0=(2h/c^2)$, so that the moments simply characterize the fractional temperature variations around the average CMB blackbody without any albedo variation. Since $\Delta T/T\simeq 10^{-5}-10^{-4}$, the overall effect is captured by adding only a $y$-parameter, $y\simeq 10^{-9}$ \citep{Chluba2004, Chluba2012}, which can be neglected. The variation caused by the CMB dipole gives rise to $y= \pot{(2.525\pm0.012)}{-7}$ \citep{Chluba2004, Chluba2016}, which can be accurately removed.

\subsection{Convergence at high frequencies}
\label{sec:conv_GB}
The functions $Y_k$ are all positive. At low frequencies, they scale like $Y_k\simeq 1$ [so that $\left<I_\nu(\vek{p}^\ast)\right>\simeq {\rm const} \times \nu^2 $], while for $x\gg 1$, they all approach $Y_k\simeq x^k$. Higher order moments thus become distinguishable mainly at high frequencies. This leads to rather slow convergence of the expansion in the non-perturbative regime, implying that without additional prior knowledge about the relations between the moments, alternative treatments, e.g., directly based on the temperature distribution function, may be needed. This problem reappears for the superposition of spatially varying thermal dust emission, and is also known in connection with relativistic corrections to the Sunyaev-Zeldovich effect \citep{Sazonov1998, Itoh98, Nozawa2006, Chluba2012SZpack}. 
This also implies that it becomes difficult to use the high-frequency spectrum to improve the modeling of the low-frequency spectrum.


\subsection{Examples for sums of two gray-body spectra} 
Let us consider the simple case of two gray-body spectra, with parameters $A_i, T_i$. We then have 
\beal
\label{eq:GB_2_M}
\bar{A}&=A_1+A_2,
\qquad \bar{\beta}=\frac{f}{T_1}+\frac{1-f}{T_2}
\\
\nonumber
\omega_{2\ldots2}&=\left[f(f-1)^k+(1-f)f^k\right]\frac{(T_1-T_2)^k}{[(1-f) T_1+f T_2]^k}
\end{align}
with $f=A_1/(A_1+A_2)$. For $A_1=A_2=A$ one has the moments $\omega_{2\ldots 2}=(T_1-T_2)^k/(T_1+T_2)^k$ for even $k$ and zero otherwise. At $x\gg 1$, with $Y_k(x)\approx x^k$ as expected we thus find $\left<I_\nu\right>\approx 2A\,\nu^3\expf{-x}\cosh[x\,(T_1-T_2)/(T_1+T_2)]\equiv A\,\nu^3[\expf{-x_1}+\expf{-x_2}]$ for $x_i=h\nu/kT_i$. To achieve full convergence at high frequencies, an infinite number of moments is required, while at low frequencies, the convergence is much more rapid. This demonstrates that in the non-perturbative regime ($\leftrightarrow T_1$ and $T_2$ differ strongly or $x$ become very large) it is difficult to use the high-frequency spectrum to constrain the low-frequency model (see also Sect.~\ref{sec:conv_GB}).

\begin{figure}
\centering 
\includegraphics[width=\columnwidth]{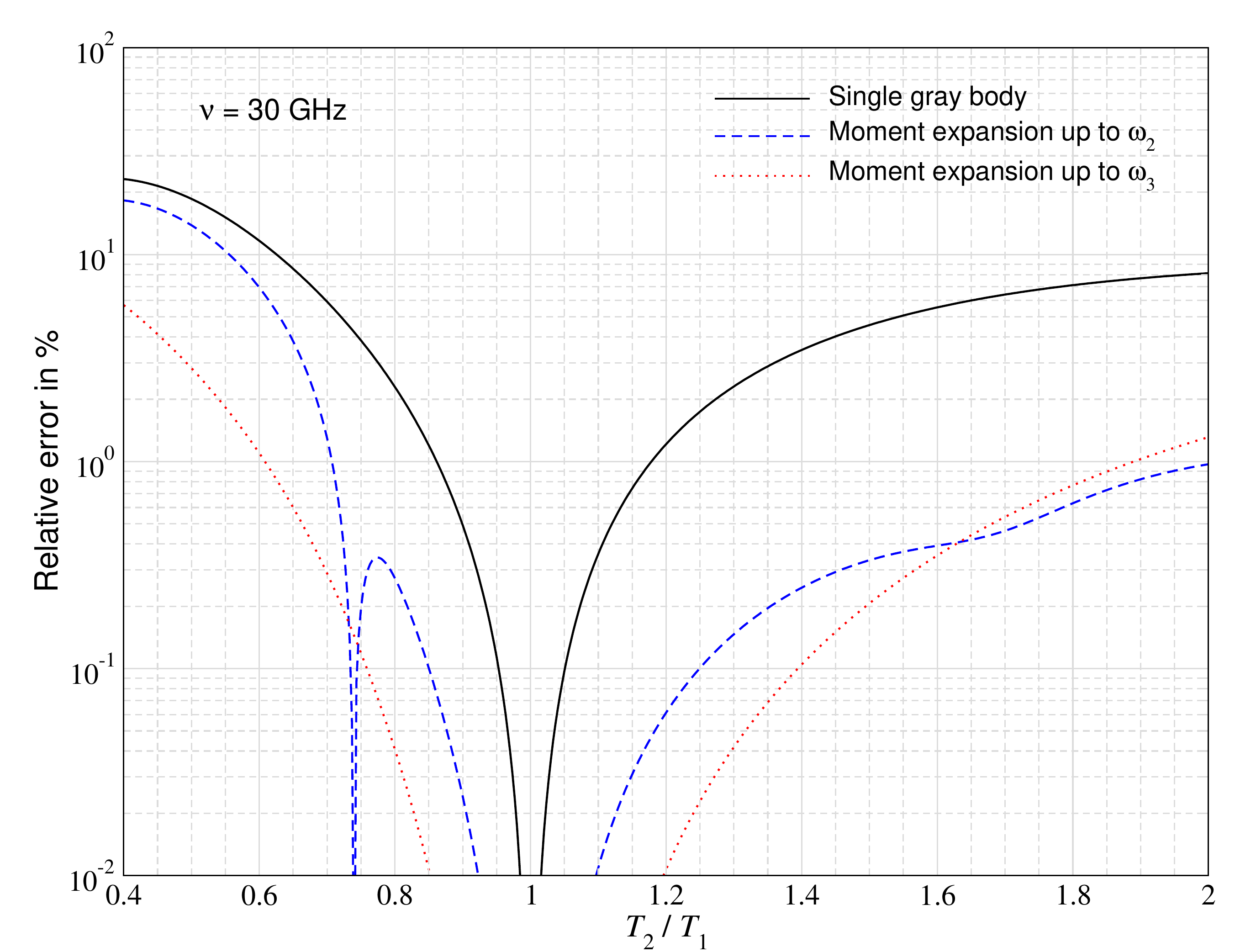}
\\
\includegraphics[width=\columnwidth]{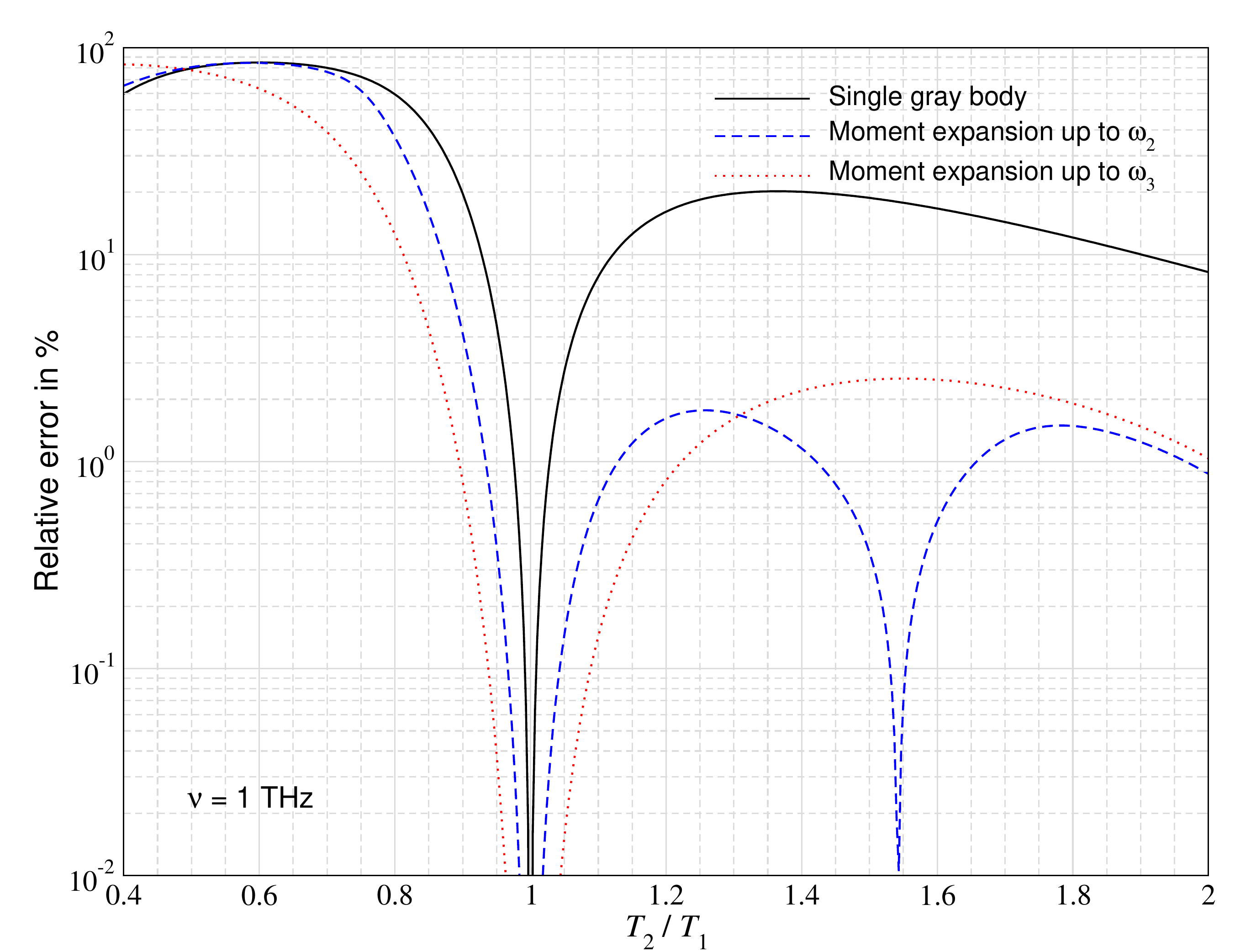}
\caption{Relative departure from the input model for the sum of two gray-body spectra with $A_1=0.3 A$, $T_1=3\,\Kel$, $A_2=0.7 A$ and varying $T_2$ at two representative frequencies.}
\label{fig:R_error_gray}
\end{figure}
We start with an example in the perturbative regime, choosing $P_1=\{A_1, T_1, A_2, T_2\}=\{0.3 A, 3\,\Kel, 0.7 A, 4\,\Kel$\}. A single gray-body spectrum with $T^{\rm fit}=3.8336\,\Kel$ and amplitude $A^{\rm fit}=0.9329\,A$ represents the average SED at the level of $\lesssim 3\%-4\%$ at $\nu\lesssim 550\,\GHz$. The given best-fitting values are not far away from what is obtained with the moment expressions in Eq.~\eqref{eq:GB_2_M}, which imply $\bar{T}=3.6363\,\Kel$ and amplitude\footnote{We will address the cause of the small difference in Sect.~\ref{sec:gray_single}.} $\bar{A}=A$. This approximation fails, however, at the level of $\gtrsim 10\%$ at $\nu\gtrsim  550\,\GHz$, exceeding $\simeq 50\%$ at about $2\,{\rm THz}$. Adding the first moment, $\omega_2$, as a free parameter, the spectrum is represented at a level better than $\simeq 1.5\%$ at all frequencies $\nu\lesssim 1\,{\rm THz}$ and to better than $\simeq 0.25\%$ at all frequencies $\nu\lesssim 600\,\GHz$. Although not perfect, this level of precision is reached with {\it fewer} parameters than the input model, without assuming a specific number of components, which again illustrates the potential of the method. 

Here we immediately mention a caveat of the moment representation for gray-body spectra. The approximation was obtained by fitting the average SED assuming zero noise. The best-fitting parameters were $R_1=\{A^{\rm fit}, T^{\rm fit}, \omega^{\rm fit}_2\}=\{0.9932\,A, 3.6667\,\Kel, 0.01341\}$. Comparing this to what is expected from the expressions in Eq.~\eqref{eq:GB_2_M}, $M_1=\{A, 3.6363\,\Kel, 0.01928\}$, shows that the best-fitting parameters do not match perfectly. This is most obvious for the value of $\omega_2$, which deviates by $\simeq 68\%$. We have seen a similar behavior for the superposition of power-laws and free-free spectra (Sect.~\ref{sec:pow_exercise} and \ref{sec:model_ff_example}).
The source of this difference is that the spectral function $\propto\bar{A}_0 \nu^3\,Y_2(x)/(\expf{x}-1)$ is not linearly independent of $\propto\bar{A}_0 \nu^3/(\expf{x}-1)$. The recovered values of a truncated moment expansion thus are expected to differ from the theoretical values and only approach these when a sufficiently large number of moments is evaluated. This can be used as a {\it diagnostic} for how many moments may be needed to describe a given spectral shape in real experimental situations, as closer to convergence the moment values will not vary anymore when adding more components. A cure for this issue can be obtained using orthogonalization schemes, as we briefly discuss below (Sect.~\ref{sec:gray_ortho}). However, in this case the interpretation of the obtained parameters in terms of the underlying temperature moments $\omega_{2\ldots2}$, becomes more cumbersome.  

While we found the moment representation to work very well for the model chosen above, let us investigate how the convergence changes with increasing temperature difference and varied fractional contribution. For this, we determined the best-fitting parameters for a given input model using varying representations in the analysis. In Fig.~\ref{fig:R_error_gray}, we show the results for $A_1=0.3 A$, $T_1=3\,\Kel$ and $A_2=0.7 A$ varying $T_2$. We illustrate the relative difference with respect to the input spectrum at $\nu=30\,\GHz$, representative of the behavior in the Rayleigh-Jeans part of the spectrum, and $\nu=1\,{\rm THz}$, for the Wien part of the spectrum. For $T_1=T_2$, all approximations naturally show no departure from the input spectrum. The convergence radius around $T_1\simeq T_2$ is larger at low frequencies than at high frequencies, owing to the asymptotic behavior of the moment expansion in this regime. 
For the chosen example, the convergence is faster at low than at high frequencies. Similarly, we find that for $T_2>T_1$, the departure is smaller than for $T_2<T_1$. In the latter case one thus needs more moments to reach the same level of precision. This observation depends on the chosen value for $f$, which up-weights $T_2$ here. A small contribution of a colder component in the Rayleigh-Jeans part of the average spectrum thus seems easier to incorporate than a small contribution of a hotter component in the Wien-tail. We find that the curves become relatively symmetric around $T_1= T_2$ when fixing the height of the maxima for each component to be equal. All these cases are simply to illustrate the  behavior of the moment expansion for gray-body spectra.

\vspace{-0mm}
\subsubsection{Best-fitting single gray-body parameters}
\label{sec:gray_single}
We already saw that the best-fitting values for the parameters in a moment expansion depend on the number of terms that are included. The same is true when we simply approximate the superposition of gray-body spectra with a single gray-body spectrum. The best-fitting temperature and amplitude are not simply given by the average parameters $\left<A_i\right>$ or $\left<T_i\right>$. In our moment expansion we used $\beta_i=T^{-1}_i$ as temperature parameter, and again, the best-fitting value is usually not determined by $T^*=\left<\beta_i\right>^{-1}$ or $T^*=\left<A_i\right>/\left<A_i\beta_i\right>$ [see Eq.~\eqref{eq:I_nu_gen_Gray}]. How can we estimate the best-fitting parameters for a single gray-body approximation? The answer is: by using moments of the frequency-integrated spectrum. 

Like for a blackbody spectrum, we can define the number and energy density of the fundamental spectral shape. These are simply $N_\gamma=A_i\,c_N\,T_i^3$ and  $\rho_\gamma=A_i\,c_\rho\,T_i^4$, where the values of two constants are not important. We then have two equations $A^*\,(T^*)^3=\left<A_i\,T_i^3\right>$ and $A^*\,(T^*)^4=\left<A_i\,T_i^4\right>$ when assuming a single gray-body approximation for the superposition. This yields
\beal
\label{eq:Sol_sing_T_G}
A^*&=\left<A_i\,T_i^3\right>^4 / \left<A_i\,T_i^4\right>^3, \qquad
T^*=\left<A_i\,T_i^4\right> / \left<A_i\,T_i^3\right>.
\end{align}
For our example, $P_1=\{0.3 A, 3\,\Kel, 0.7 A, 4\,\Kel\}$, we then find $A^*=0.9292\,A$ and $T^*=3.8469$, which is in very good agreement with the values of a single gray-body approximation recovered from an explicit fit ($T^{\rm fit}=3.8336\,\Kel$ and amplitude $A^{\rm fit}=0.9329\,A$). As mentioned above, this departs notably from the moment values ($\bar{T}=3.6363\,\Kel$ and $\bar{A}=A$) that are obtained using Eq.~\eqref{eq:I_nu_gen_Gray}.

The difference between the best-fitting values and those obtained with Eq.~\eqref{eq:I_nu_gen_Gray} can become much more dramatic. Choosing $P_2=\{0.5 A, T_0, 0.7 A, 1.5\,T_0, 0.4 A, T_0/1.5, 0.2 A, T_0/3.5\}$ with $T_0=2.726\,\Kel$, we find $\bar{A}=1.8\,A$ and $\bar{T}=2.165\,\Kel$ using the moment definitions, Eq.~\eqref{eq:I_nu_gen_Gray}. 
From Eq.~\eqref{eq:Sol_sing_T_G}, we have $A^*=1.1329\,A$ and $T^*=3.765\,\Kel$ and a fit gives $A^{\rm fit}=1.1245\,A$ and $T^{\rm fit}=3.743\,\Kel$, which clearly illustrates the point.\footnote{In the presence of noise, other foreground components, and differences in the frequency coverage, the answers can furthermore differ significantly.} 
This example shows that the distribution of temperatures becomes so wide that a perturbative expansion requires many moments to converge. The weighting ($A_i\,T_i^3$ instead of $A_i$) and choice of variables ($T_i$ instead of $1/T_i$) furthermore can lead to improved convergence properties (higher level of independence between moments). In spite of all this, the moment representation, Eq.~\eqref{eq:I_nu_gen_Gray}, can still be used to approximate the average SED in a meaningful way, even without explicitly rearranging the moments and spectral functions. This is because the moment expansion naturally captures the dominant new degrees of freedom (higher order derivatives) of each moment order. A different weighting scheme or change of variables (which ultimately is equivalent to a different weighting scheme) does not alter this statement, but only changes the interpretation of the moment values in terms of the underlying temperature distribution function.

For completeness, we also give the moment expansion using weight $B=A\,T^3$ and $p=T$. In this case, the fundamental SED reads $I_\nu(B, T)=B\,(\bar{T}/T)^3\,x^3/(\expf{x\bar{T}/T}-1)$, with the derivatives
\beal
T\, \partial_{T}\,I_\nu(B, T)&= I_\nu(B, T) \,\left[Y_1(x)-3\right]
\\\nonumber
T^2\, \partial^2_{T}\,I_\nu(B, T)&= I_\nu(B, T) \,\left[Y_2(x)-8Y_1(x)+12\right]
\\\nonumber
T^3\, \partial^3_{T}\,I_\nu(B, T)&= I_\nu(B, T) \,\left[Y_3(x)-15Y_2(x)+60Y_1(x)-60\right]
\\\nonumber
T^4\, \partial^4_{T}\,I_\nu(B, T)&= I_\nu(B, T) \,\left[Y_4(x)-24Y_3(x)+180Y_2(x)\right.
\nonumber\\ \nonumber
&\qquad\qquad\qquad\qquad\qquad\left.-480Y_1(x)+360\right].
\end{align}
Defining the functions $Z_k=[T^k\, \partial^k_{T}\,I_\nu(B, T)]/I_\nu(B, T)$, we find the alternative moment expansion
\beal
\label{eq:I_nu_gen_Gray_mod}
\left<I_\nu\right>&=\frac{\bar{B} \,x^3}{\expf{x}-1}
\left\{1 +\frac{1}{2}\,\kappa^{\rm g}_{22} \,Z_2(x) +\frac{1}{6}\, \kappa^{\rm g}_{222} \,Z_3(x) 
\right.
\\
\nonumber
&\qquad\qquad\qquad\qquad\qquad
\left.
+\frac{1}{24}\, \kappa^{\rm g}_{2222} \,Z_4(x) + \ldots\right\}
\\ \nonumber
\bar{B} &= 
\left<B(\vek{r})\right>\equiv \left<A(\vek{r}) T^3(\vek{r})\right>,
\;\;
\bar{T} = 
\frac{\left<B(\vek{r}) T(\vek{r})\right>}{\left<B(\vek{r})\right>}\equiv \frac{\left<A(\vek{r}) T^4(\vek{r})\right>}{\left<A(\vek{r}) T^3(\vek{r})\right>}
\\ \nonumber
\kappa^{\rm g}_{2\ldots 2} 
&=\frac{\left<B(\vek{r})[T(\vek{r})/\bar{T}-1]^k\right>}
{\bar{B}}\equiv \frac{\left<A(\vek{r})T^3(\vek{r})[T(\vek{r})/\bar{T}-1]^k\right>}{\left<A(\vek{r}) T^3(\vek{r})\right>}.
\end{align}
This expression is close to optimal regarding the independence of individual moment coefficients (see Sect.~\ref{sec:gray_ortho}).

\vspace{-0mm}
\subsection{Orthonormal basis functions for gray-body spectra}
\label{sec:gray_ortho}
\begin{figure}
\centering 
\includegraphics[width=\columnwidth]{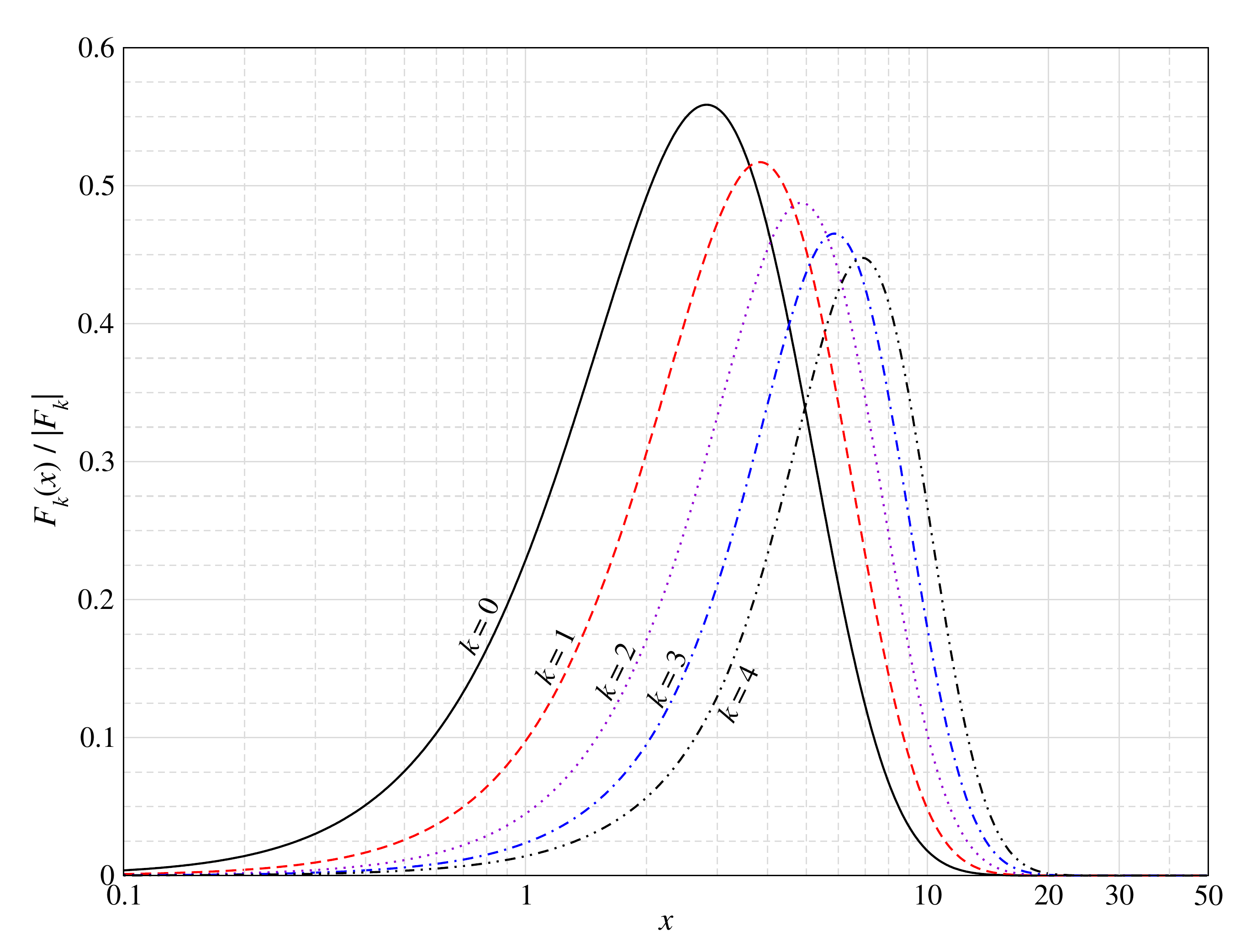}
\\[2mm]
\includegraphics[width=\columnwidth]{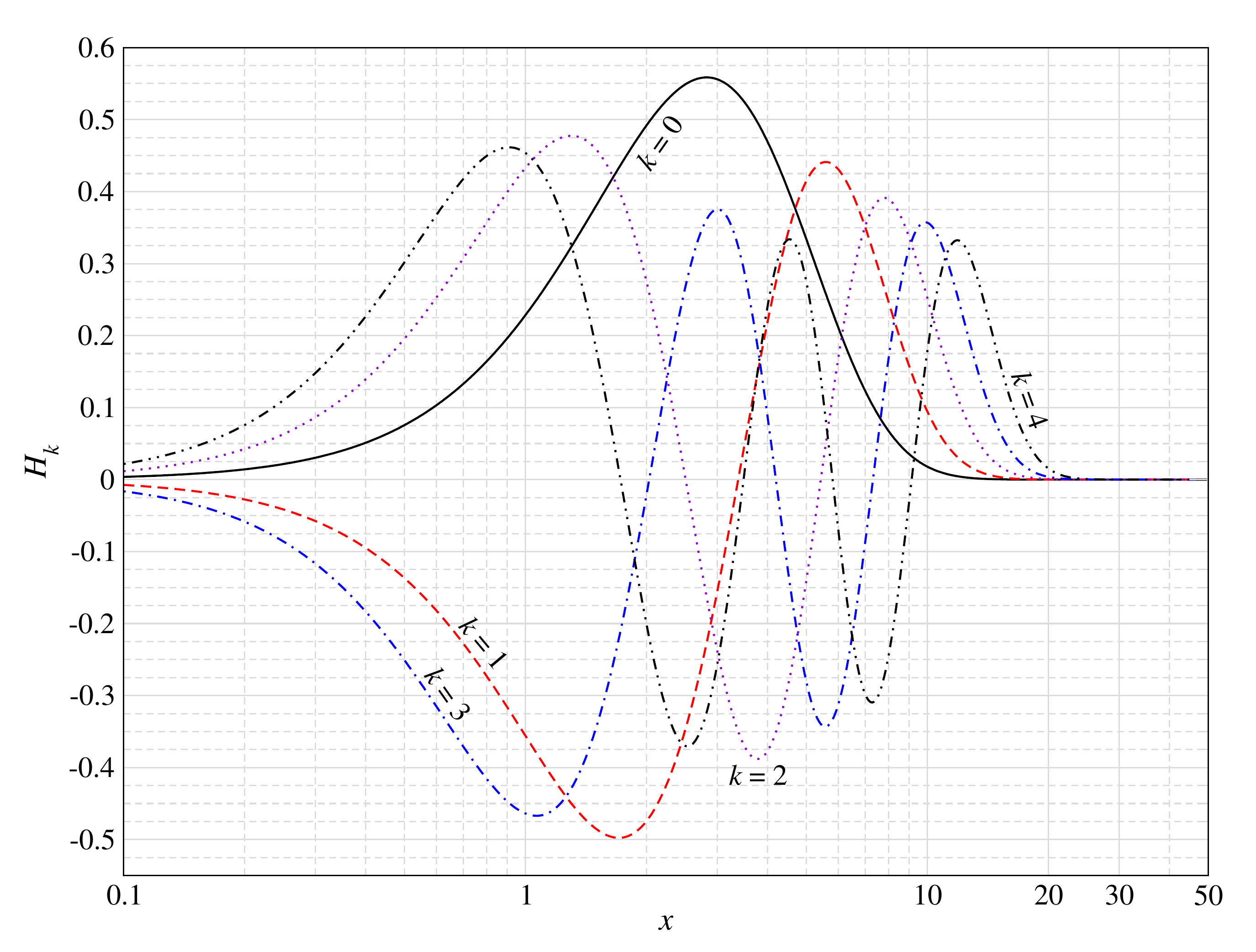}
\caption{Gray-body spectral functions, $F_k(x)$ and $H_k(x)$ for $k\leq 4$.}
\label{fig:F_k_gray}
\end{figure}

Since the functions $Y_k\simeq 1$ at low frequencies, all moments become degenerate in this regime. It is thus useful to think about a new set of functions that minimize the correlations between different moments. This can be achieved using a Gram-Schmidt orthogonalization scheme, ordering the functions with respect to their moment order, which provides a perturbative ordering. By defining $x=h\nu/k \bar{T}$, we can remove the explicit dependence on $\bar{T}$. Before the orthonormalization we then have the basic spectral functions
\beal
F_k(x)=\frac{x^3}{\expf{x}-1} \,Y_{k}(x)
\end{align}
with $Y_0(x)=1$. These functions are shown in Fig.~\ref{fig:F_k_gray} for $k\leq 4$ and fully determine the moment expansion for the superposition of gray-body spectra. With increasing $k$, the maximum of $F_k$ moves towards higher frequencies and smaller contributions are visible at low frequencies. 
We now define the dot-product
\beal
\left<F_k(x)\,F_m(x)\right>=\int_0^\infty F_k(x)\,F_m(x) \id x.
\end{align}
This could be modified by adjusting the range of integration to match some experimental setting, but we do not explore this possibility any further. For the norms of the first few functions we then have $|F_0|=2.5447$, $|F_1|=9.4417$, $|F_2|=44.469$, $|F_3|=253.91$, $|F_4|=1703.81$. Within a limited space (including fourth order moment terms only) we then find the orthonormal basis
\beal
\label{eq:H_k_examples}
H_0(x)&=
\left(0.392 97, 0, 0, 0, 0\right)^{T}\cdot\vek{F}(x)
\nonumber\\
H_1(x)&=
\left(-1.116 87, 0.31911, 0, 0, 0\right)^{T}\cdot\vek{F}(x)
\\
H_2(x)&=
\left(2.259 44, -1.31301, 0.16413, 0, 0\right)^{T}\cdot\vek{F}(x)
\nonumber\\
H_3(x)&=
\left(-3.885 31, 3.387 61, -0.854 07, 0.063 26, 0\right)^{T}\cdot\vek{F}(x)
\nonumber\\ \nonumber
H_4(x)&= 
\left(6.077 88, -7.038 48, 2.658 12, -0.395 18, 0.019 76\right)^{T}\cdot\vek{F}(x),
\end{align}
where we defined $\vek{F}=\left(F_0, F_1, F_2, F_3, F_4\right)$. These functions are also illustrated in Fig.~\ref{fig:F_k_gray} and show an increasing number of oscillations with $k$ at quasi-constant overall amplitude. This nicely illustrates why the moment expansion can exhibit asymptotic convergence, with both high and low frequencies being affected. 

We can also compare Eq.~\eqref{eq:H_k_examples} with Eq.~\eqref{eq:I_nu_gen_Gray_mod}. From the coefficients, we have $|F_0|\,H_0(x)=F_0(x)$ and $k!|F_0|\,H_k(x)\approx F_0(x)\,Z_k(x)$. This shows that the moment expansion in Eq.~\eqref{eq:I_nu_gen_Gray_mod}, with modified weighting, $B=A\,T^3$, results in a quasi-orthogonal ($\leftrightarrow$ optimal) re-summation of terms. We will discuss a similar approach for the superposition of modified blackbody spectra (Sect.~\ref{sec:alt_mBB}).

With the basis functions, $H_k(x)$, we can also write 
\beal
\label{eq:I_nu_gen_Gray_orth}
\left<I_\nu\right>&=A^*_0\left\{ H_0(x)+\sum_{k=2}^{k_{\rm max}} \sigma^{\rm g}_{k}\,H_k(x)\right\}
\end{align}
for the superposition of gray-body spectra, with overall amplitude $A_0^*=|F_0|\,A_0\,(k\,\bar{T}/h)^3$. In an idealized case, the coefficients $\sigma^{\rm g}_{k}$ are uncorrelated and directly determine the underlying moments $\omega^{\rm g}_{2\ldots 2}$ up to a given order. However, the presence of other signals (which we did not project out) and truncation in the frequency domain or bandpass effects can re-introduce correlations. This is not necessarily a severe limitation and applying the idealized expansion renders a comparison of the results from different experiments easier.

While for a superposition of gray-body spectra a Gram-Schmidt orthogonalization leads to an efficient re-summation of higher order moments, more generally this approach does not improve matters as much. In particular, generally it will be impossible to define a basis that is independent of the leading order average spectral parameter values. This was avoided for the gray-body spectra by using the transformation $\nu\rightarrow x$. However, for a superposition of power-law spectra this already is not possible and the set of basis functions would depend on the average spectral index, $\bar\alpha$. This limits the benefits of orthogonalization schemes.

\vspace{-0mm}
\section{Superposition of thermal dust spectra}
\label{sec:dust}
We now investigate how a superposition of thermal dust spectra can be modeled. Here, we first develop a general expansion in terms of Taylor moments. Our examples will then be guided by models that may be relevant to future CMB analyses. 
We start with the commonly used parameterization for a single-temperature dust spectrum, $I_\nu(A_0, \alpha, T)=A_0\,(\nu/\nu_0)^\alpha\, \nu^3/(\expf{h\nu/kT}-1)$, which is referred to as a modified blackbody spectrum. This model for the fundamental spectral energy distribution can be motivated by considering the physical properties of dust grains in the vicinity of stars \citep{Draine1998, Finkbeiner1999, Draine2003, Planck2013components}. Physically speaking, this shape is certainly not expected to be exact, as many different types of dust grains can contribute. Especially at high frequencies, the spectral shape can be much more rich \cite[e.g.,][]{Draine2003}. Nevertheless, as a first step we can extend the widely used dust models using the moment expansion to incorporate the effect of spatial variations parametrically. 

The superposition of modified blackbody spectra shares many of the features found for the superposition of gray-body spectra discussed in Sect.~\ref{sec:blackbody}. In particular, at high frequencies the convergence of a moment expansion is again expected to be slow in the non-perturbative regime. This highlights the fact that the high-frequency spectrum of the dust SED is quite sensitive to the detailed shape of the underlying parameter distribution functions and thus hard to use to constrain the low-frequency spectrum.  

To obtain the moment expansion, we will use the parameters $\vek{p}=\{A_0, \alpha, \beta=1/T\}$. Alternative parameterizations and weighting schemes are briefly discussed below (Sect.~\ref{sec:alt_mBB}). In the model, $I_\nu(A_0, \alpha, T)$, no second derivatives with respect to $A_0$ appear, however, many additional spectral shapes are created due to mixed derivatives with respect to $\alpha$ and $\beta=1/T$. Since all the derivatives commute, we only have to deal with moments of the form $\omega_{12\ldots 2}$, $\omega_{13\ldots 3}$, $\omega_{12\ldots 23\ldots 3}$, $\omega_{2\ldots 23\ldots 3}$, $\omega_{2\ldots 2}$ and $\omega_{3\ldots 3}$. The relevant spectral functions are $I_{1i\ldots j}(\nu_{\rm c}, \bar{\vek{p}})=I_{i\ldots j}(\nu_{\rm c}, \bar{\vek{p}})/\bar{A}_0$ with $i,j\in\{2, 3\} \equiv \{\alpha,\beta\}$, so that the moments $\omega_{12\ldots 2}$, $\omega_{13\ldots 3}$ and $\omega_{12\ldots 23\ldots 3}$ can again be absorbed by defining $\omega^{\rm d}_{i\ldots i j\ldots j}=\omega_{i\ldots i j\ldots j}+\omega_{1 i\ldots i j\ldots j}/\bar{A}_0$. Assuming narrow bands and $W(\vgh, \nu)= B(\vgh) F(\nu)$, we then find
\beal
\label{eq:I_nu_gen_BB}
\left<I_\nu\right>&=\frac{\bar{A}_0 \, (\nu/\nu_0)^{\bar{\alpha}}\nu^3}{\expf{x}-1}
\left\{1 
+\frac{1}{2}\,\omega^{\rm d}_{22} \ln^2(\nu/\nu_0) 
\right.
\nonumber\\[-0.5mm]
&\quad\left.
+ \omega^{\rm d}_{23} \ln(\nu/\nu_0) \, Y_1(x) 
+\frac{1}{2}\,\omega^{\rm d}_{33} Y_2(x) 
\right.
\nonumber\\[-0.5mm]
&\qquad\left.
+\frac{1}{6}\,\omega^{\rm d}_{222} \ln^3(\nu/\nu_0)
+\frac{1}{2}\, \omega^{\rm d}_{223} \ln^2(\nu/\nu_0)Y_1(x)
\right.
\nonumber\\[-0.5mm]
&\quad\qquad\left.
+\frac{1}{2}\, \omega^{\rm d}_{233} \ln(\nu/\nu_0)Y_2(x)
+\frac{1}{6}\, \omega^{\rm d}_{333} Y_3(x)
+ \ldots\right\}
\nonumber\\[-0.5mm]
\bar{\alpha} 
&= \frac{\left<A_0(\vek{r})\,\alpha(\vek{r})\right>}{\bar{A}_0},
\qquad 
\frac{1}{\bar{T}} 
= \frac{\left<A_0(\vek{r})/T(\vek{r})\right>}{\bar{A}_0}
\\[-0.5mm] \nonumber
\omega^{\rm d}_{2\ldots 2 3\ldots3} 
&= 
\frac{\left<A_0(\vek{r})[(\alpha(\vek{r})-\bar{\alpha}]^k\,[(\bar{T}/T(\vek{r})-1]^m\right>}{\bar{A}_0}.
\end{align}
Higher order terms can be easily added in a similar way. This expression again captures all the degrees of freedom introduced by the averaging inside the beam and along the line of sight. Note that the simple product assuming individual superpositions of power-law and gray-body spectra would be {\it insufficient}, since the cross-terms $\propto \ln^k(\nu/\nu_0)\,Y_m(x)$, would not have independent coefficients to take into account possible correlations between $\alpha$ and $T$. For example, the term for $k=1$ and $m=1$ would be absent.

\vspace{-2mm} 
\subsection{Behavior in the Rayleigh-Jeans limit} 
\label{sec:RJ}
In the Rayleigh-Jeans limit ($h\nu \ll k T$), we can re-express the modified blackbody spectrum as
\beal
I^{\rm RJ}_\nu(A_0, \alpha, T)\approx A_0\,(\nu/\nu_0)^{\alpha+2}\,\left[1-\frac{x}{2}+\frac{x^2}{12}-\frac{x^4}{720}+\ldots \right].
\end{align}
This represents a superposition of power-law spectra, which in principle can be approximated with Eq.~\eqref{eq:I_nu_gen_exp_power_law}. Since $x=h\nu/kT$, the temperature variations of the dust itself enter as a weight-factor for the power-law moment amplitudes, introducing specific correlations among the moments. The Taylor series of $1/(\expf{x}-1)$ furthermore can only be recovered when allowing {\it differences} between power-laws instead of just sums as considered previously. We will discuss the applicability of this approximation below (Sect.~\ref{sec:RJ_test}).

\vspace{-2mm} 
\subsection{Alternative parameterizations and weighting schemes} 
\label{sec:alt_mBB}
In Sect.~\ref{sec:gray_single}, we illustrated how the choice of parameters and weighting affects the moment expansion for  gray-body spectra. In particular, we found that the recovered best-fitting parameters for truncated moment expansions are sensitive to these choices. In a similar manner as for the gray-body spectra, by setting $\nu_0=k T / h$ and rewriting $I_\nu(A_0, \alpha, T)=A_0\,(\nu/\nu_0)^\alpha\, \nu^3/(\expf{h\nu/kT}-1)$ as
\beal
\label{eq:I_nu_gen_BB}
I_\nu&=A_0 \left(\frac{kT}{h}\right)^3  \frac{x^{3+\alpha}}{\expf{x}-1}
=B(T, \alpha) \frac{\bar{x}^{3+\alpha}\,\left(\bar{T}/T\right)^{3+\alpha}}{\expf{\bar{x}\,\bar{T}/T}-1}
\end{align}
we can derive an alternative moment expansion for the modified blackbody spectra with new weighting $B(T, \alpha)=A_0\,(kT/h)^{3+\alpha}$. However, we find that this approach does not significantly change the capabilities of the moment expansion to represent different SEDs. It only leads to a re-summation of terms up to a given moment order. We therefore did not consider this approach any further.

\vspace{-0mm} 
\subsection{Two-temperature dust models} 
To illustrate the application of the moment expansion, let us consider a simple two-temperature dust model, with SED parameters $\vek{p}=\{A_1, A_2, \alpha_1, \alpha_2, T_1, T_2\}$. We then have $\bar{A}=A_1+A_2$, $\bar{\alpha}=(A_1\alpha_1+A_2\alpha_2)/\bar{A}$, $\bar{T}=\bar{A}/(A_1/T_1+A_2/T_2)$ and the moments
\beal
\omega^{\rm d}_{2\ldots 2 3\ldots 3}&=\sum^2_{a=1} \frac{A_a}{\bar{A}} (\alpha_a-\bar{\alpha})^k  (\bar{T}/T_a-1)^m.
\end{align}
The general behavior of the moment expansion for different cases is similar to that of gray-body spectra, as mentioned above. The additional parameter, $\alpha$, causes the number of moments per perturbation order to increase strongly. For example, adding all second order moment terms means $3$ extra parameters instead of 1 for the gray-body spectrum; at third order we need $7$ instead of 2 and so on. The required number of parameters can thus grow quickly for complicated dust distribution functions.

\begin{figure}
\centering 
\includegraphics[width=\columnwidth]{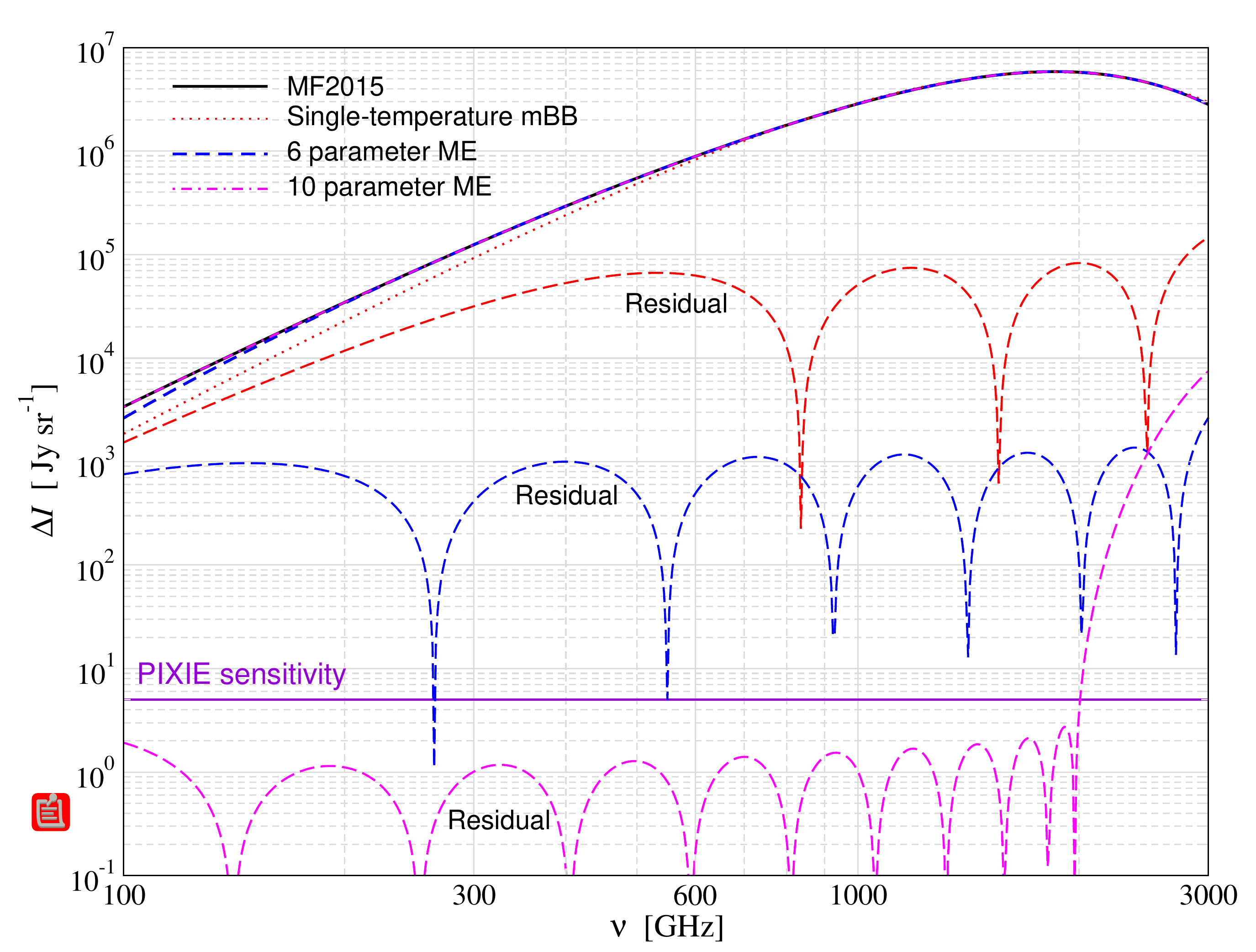}
\caption{Representation of a two-temperature modified blackbody spectrum from \citet{Meisner2015} using a varying number of moments (see text for details). The long-dashed lines show the residuals in comparison with the estimated absolute sensitivity of {\it PIXIE}.}
\label{fig:mBB_2}
\end{figure}
To illustrate the method, let us consider a concrete example, setting the model parameters to $P_1=\{A_1, A_2, \alpha_1, \alpha_2, T_1, T_2\}=\{A\,f_1, A\,(1-f_1), 1.63, 2.82, 9.75\,\Kel, 15.7\,\Kel\}$ with $f_1=0.34188$ and $\nu_0=3\,{\rm THz}$, based on recent modeling of CMB data \citep{Meisner2015}. We fix $A$ such that $\left<I_\nu\right>\simeq 3388\,{\rm Jy\,sr^{-1}}$ at $\nu=100\,\GHz$, but for illustrations of the method, the absolute value does not matter, as the spectral shape is fixed without the amplitude. 
We assume 200 frequency bins in log-$\nu$ of different ranges and perform a simple $\chi^2$-fit using Eq.~\eqref{eq:I_nu_gen_BB} setting $\nu_0=k\bar{T}/h$. This choice of the pivot frequency leads to faster convergence.
Representing the resultant spectrum (see Fig.~\ref{fig:mBB_2}) with a single-temperature modified blackbody in the range $100\,\GHz\lesssim \nu\lesssim 3\,{\rm THz}$ yields $\alpha=1.819$ and $T=18.45\,\Kel$. 
We find that at low frequencies this approximation shows a large departure from the true SED, underestimating the emission by a factor of two (see Fig.~\ref{fig:mBB_2}). Adding the first order moment corrections, we find $\alpha=1.270$, $T=15.27\,\Kel$, $\omega^{\rm d}_{22}=-0.4438$, $\omega^{\rm d}_{23}=0.1695$ and $\omega^{\rm d}_{33}=0.2094$, which provides an approximation that is better than $\simeq 0.4\%$ at $300\,\GHz\lesssim \nu\lesssim 3\,{\rm THz}$, departing only notably at low frequencies, reaching $\simeq 20\%$ at $\nu\simeq 100\,\GHz$ (see Fig.~\ref{fig:mBB_2}). This representation uses the same number of parameters (six in total) as the input model, but without assuming a two-temperature case, and can in principle already capture a broader range of distributions in temperature and spectral indices.

The convergence of the moment representation also depends on the frequency range that is assumed in the fitting process. For a second order moment representation at $100\,\GHz\lesssim \nu\lesssim 2\,{\rm THz}$, we obtain $\alpha=1.412$, $T=14.84\,\Kel$, $\omega^{\rm d}_{22}=-0.2793$, $\omega^{\rm d}_{23}=0.05132$ and $\omega^{\rm d}_{33}=0.2191$. This representation improves the fit at low frequencies, with the departure decreasing to $\lesssim 8\%$ at $\nu\simeq 100\,\GHz$. This illustrates that in precision measurements over a wide range of frequencies, the high frequency part of the dust spectrum can drive the solution away from the optimal solution at low frequencies. This could affect the ability to recover primordial distortion signals in high precision CMB spectroscopy, but also is relevant to the modeling of foregrounds for primordial $B$-modes. Incorrect and incomplete (truncated) parameterizations for the foregrounds can thus yield biased results. Convergence properties of the moment expansion can again be tested by varying the moment order, which provides a powerful diagnostic, but is limited by the available number of channels.

If we now include all moments up to third order (3+3+4=10 parameters), we find $\alpha=2.057$, $T=13.33\,\Kel$, $\omega^{\rm d}_{22}=0.002363$, $\omega^{\rm d}_{23}=-0.4189$, $\omega^{\rm d}_{33}=0.3789$, $\omega^{\rm d}_{222}=-0.004326$, $\omega^{\rm d}_{223}=0.5326$, $\omega^{\rm d}_{233}=-0.2657$ and $\omega^{\rm d}_{333}=0.08050$ at $100\,\GHz\lesssim \nu\lesssim 2\,{\rm THz}$. We find that this approximation represents the dust spectrum at $100\,\GHz\lesssim \nu\lesssim 2\,{\rm THz}$ to better than $\simeq 0.06\%$ precision (see Fig.~\ref{fig:mBB_2}). We can see that the higher order moments start decreasing rapidly, showing that the chosen example is in the perturbative regime and converges fairly quickly. The achieved level of precision would be about $4-5$ times better than the expected sensitivity of {\it PIXIE} for absolute CMB spectroscopy \citep[$\Delta I_\nu\simeq 5\,{\rm Jy/sr}$;][]{Kogut2011}. Again, the parameterization is more general than simply assuming a two-temperature model, capturing more general distributions of dust temperatures and spectral indices. 

The overall representation of the average spectrum with the third order moment expansion degrades significantly (by more than one order of magnitude in terms of absolute precision) when increasing the upper frequency to $\nu=3\,{\rm THz}$. As expected from our discussion of gray-body spectra, in this case higher order moments are required to compensate for the asymptotic behavior of the basis functions. For a given finite order and distribution of temperatures, this behavior is inevitable with the moment expansion. In this case, alternative approaches that directly assume representations of the temperature distribution functions could be used; however, a precise description of the underlying probability distribution functions (with a significant number of parameters) is still required. Similarly, one could use two hierarchies of moment expansions, forming a two-temperature basis to improve the local convergence. We leave a more detailed discussion of these ideas to future work.

\vspace{-0mm}
\subsubsection{Rayleigh-Jeans approximation}
\label{sec:RJ_test}
We saw in Sect.~\ref{sec:RJ}, at low frequencies the superposition of dust spectra can be thought of as a superposition of power-laws. 
It turns out that using a simple power-law expansion to represent the average dust spectrum for the model discussed above works quite well at $30\,\GHz\lesssim \nu\lesssim 500\,\GHz$, when including only $N\simeq 4$ moments (6 parameters in total). A sum of synchrotron and dust spectra, relevant to CMB applications, also works well with $N\simeq 6$ moments (8 parameters in total, which is one less than the input model) at $30\,\GHz\lesssim\nu\lesssim 200\,\GHz$. However, at higher frequencies corrections related to the Taylor series of $1/(\expf{x}-1)$ become too large so that many power-law moments ($N\gg 6$) have to be included for a sufficient representation of the dust spectrum. In this case, the general dust moment expansion should be preferred.

\begin{figure}
\centering 
\includegraphics[width=\columnwidth]{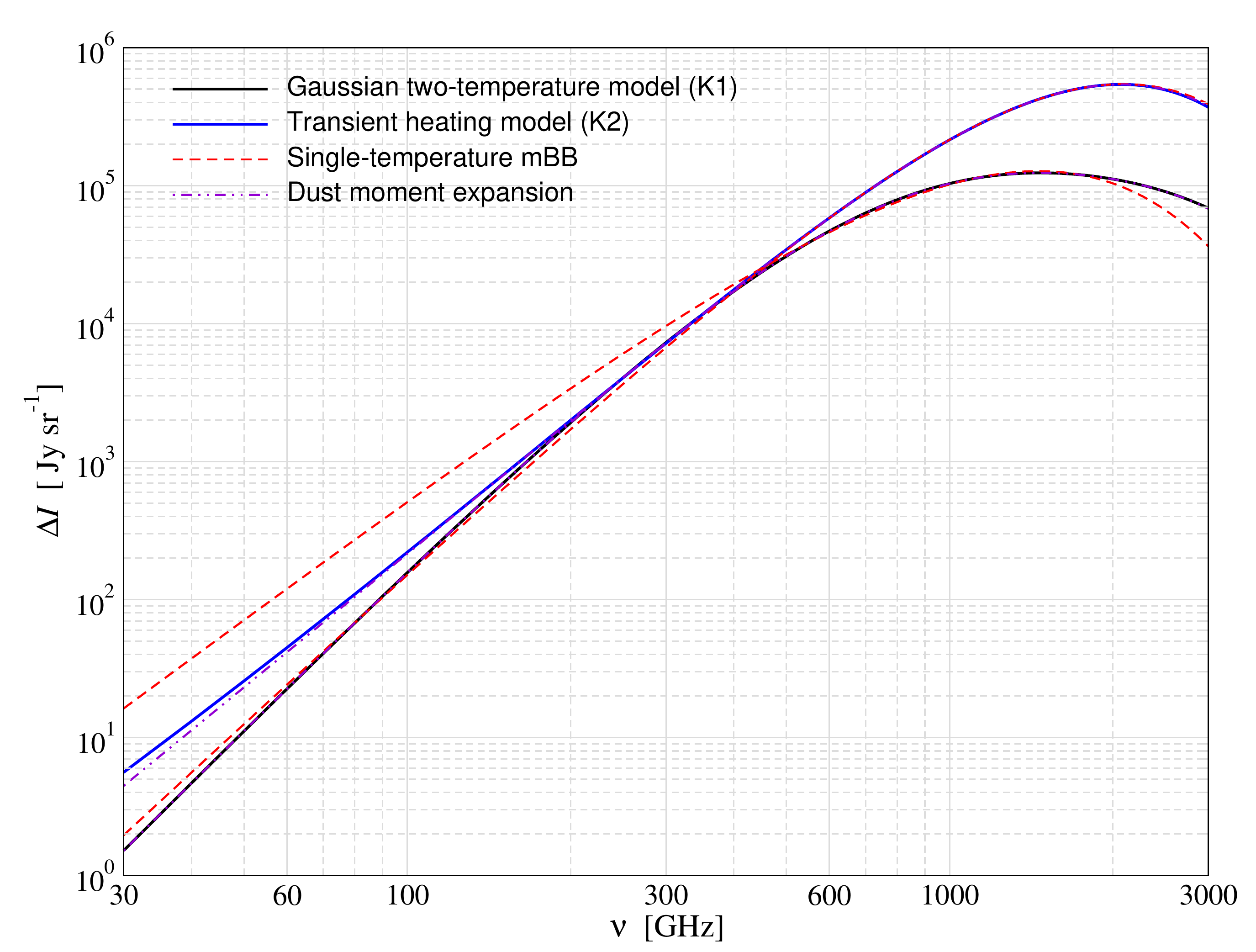}
\caption{Moment representation of dust models recently discussed by \citet{Kogut2016}. The fits were obtained in the frequency range $100\,\GHz\lesssim \nu\lesssim 2\,{\rm THz}$. The red lines assume a single-temperature dust model. The violet (double-dotted dashed) lines assume a 10/6 parameter moment expansion for K1/K2. Notice that the 10 parameter moment expansion for K1 covers the input model curve even outside the fit domain.}
\label{fig:mBB_dist}
\end{figure}
\subsection{Distributions of temperatures}
\label{sec:T_distributions}
To demonstrate the potential of the dust moment representation, we analyze two more general dust models considered by \citet{Kogut2016}, one assuming a sum of two Gaussians for the dust temperature (K1), the other using a transient heating model\footnote{We cordially thank Alan Kogut for providing the average SEDs to us.}  (K2). The resultant SEDs are shown in Fig.~\ref{fig:mBB_dist}. 

Without knowing the details of the temperature model, we can represent these spectra using the dust moment expansion, Eq.~\eqref{eq:I_nu_gen_BB}. Fitting in the range $100\,\GHz\lesssim \nu\lesssim 2\,{\rm THz}$, we find $\alpha=0.9379$ and $T=18.15\,\Kel$ (K1) and $\alpha=1.671$ and $T=21.69\,\Kel$ (K2) assuming a single-temperature modified blackbody spectrum. The overall amplitude, $A_0$, is be determined using $I_\nu=156.61\,{\rm Jy/sr}$ (K1) and $I_\nu=220.11\,{\rm Jy/sr}$ (K2) at $\nu=100\,\GHz$. Clearly, this representation fails to approximate the SED for model K1 but already provides a good approximation for K2 (see Fig.~\ref{fig:mBB_dist}).

A second order moment expansion (6 parameters) significantly improves the fit in both cases, leaving residuals at the level of $|\Delta I_\nu|\lesssim 50-200\,{\rm Jy/sr}$ (K1) and $|\Delta I_\nu|\lesssim 5\,{\rm Jy/sr}$ (K2) in the considered frequency range. For K2, the 6 parameter moment expansion is shown in Fig.~\ref{fig:mBB_dist}. We find that at $100\,\GHz\lesssim \nu\lesssim 2\,{\rm THz}$, the transient heating model can be fully represented down to the {\it PIXIE} sensitivity using only 6 parameters without a priori assumptions about the temperature model, while outside this range higher order moments are required. We also find that for both models, a two-temperature dust model leads to a similar performance, albeit being less general.

Using a third order moment representation (10 parameters), we obtain residuals that are more than one order of magnitude below the sensitivity of {\it PIXIE} in both cases. For K1, we have $\alpha=2.327$, $T=14.56\,\Kel$, $\omega^{\rm d}_{22}=-0.2983$, $\omega^{\rm d}_{23}=-0.1718$, $\omega^{\rm d}_{33}=-0.1396$, $\omega^{\rm d}_{222}=0.03299$, $\omega^{\rm d}_{223}=0.5685$, $\omega^{\rm d}_{233}=-0.07951$ and $\omega^{\rm d}_{333}=0.03106$ (see Fig.~\ref{fig:mBB_dist}). This shows that the dust moment expansion is able to handle more complicated dust spectra with few a priori assumptions about the underlying distribution functions. 
It also indicates that a significantly larger number of foreground parameters has to be considered. This means that the overall dust foreground exhibits a richer spectral-spatial morphology related to each new parameter. With approaches based on probability distribution function modeling \citep[e.g., see][for a recent example]{Jacques2017_3Ddust}, this aspect cannot be easily incorporated, which highlights another benefit of the moment approach.

\vspace{-4mm}
\section{Effect of spatial variations on the energy distribution at different angular scales}
\label{sec:variations} 
As mentioned in Sect.~\ref{sec:harmonic_exp_average}, the moment method could provide a simple way to propagate scale-dependent changes to the SED from small to large scales. This point can be illustrated using the spatially varying CMB temperature and the effect on the SED of different multipoles. A similar effect occurs due to the spatial variations of the dust temperature and spectral index, as discussed below and in more detail in a future paper (Hill et al., in preparation).

\vspace{-3mm}
\subsection{CMB blackbody superposition}
\label{sec:CMB_superposition}

A spherical harmonic transformation of the (ideal CMB-only) sky intensity leads to distortions ($\leftrightarrow$ modified frequency dependence) of the multipole coefficients due to the weighted average ($\leftrightarrow$ superposition) of blackbodies with different temperatures. This is an {\it artefact} of the map making procedure, which (under idealized assumptions) could be completely avoided by using maps of the thermodynamic temperature instead. To illustrate this point, let us consider the CMB blackbody intensity in different directions
\beal
\label{eq:CMB_blackbody}
I^{\rm CMB}_{\nu}(\vgh)&= \frac{2h}{c^2} \frac{\nu^3}{\expf{h\nu/kT(\vgh)}-1}
\\ \nonumber 
&\approx \frac{2h}{c^2} \frac{\nu^3}{\expf{x}-1}\left[1+G(x) (\Delta +\Delta^2) + \frac{1}{2} Y(x) \,\Delta^2 + \mathcal{O}(\Delta^3)\right],
\end{align}
where $x=h\nu/k\bar{T}$, with $\bar{T}$ denoting the full-sky average temperature and $\Delta(\vgh)=T(\vgh)/\bar{T}-1$. In the second step we assumed $\Delta \ll 1$ and introduced $G(x)\!=\!x\expf{x}/(\expf{x}-1)$ and $Y(x)\!=\!G(x)[x\coth(x/2)-4]$. We neglected higher order terms; however, these can become important at very high frequencies \citep{Chluba2004}.

We started with a blackbody at thermodynamic temperature $T(\vgh)$ in every direction, $\vgh$; thus, even the Taylor series, Eq.~\eqref{eq:CMB_blackbody}, still simply represents a pure blackbody spectrum in every direction. If we compute the all-sky average intensity, we find
\beal
\label{eq:CMB_blackbody_av}
\bar{I}^{\rm CMB}_{\nu}&\approx \frac{2h}{c^2} \frac{\nu^3}{\expf{x}-1}\left[1+ 2 y_{\rm sup}\,G(x)  + y_{\rm sup}\, Y(x) \right],
\end{align}
where $y_{\rm sup}=\int \frac{\Delta^2(\vgh)}{8\pi}\id^2\vgh $, which in general does not vanish. The frequency dependence of the average intensity spectrum is no longer a pure blackbody, but has a $y$-type distortion \citep{Chluba2004}. One can absorb the temperature shift term $\propto 2 y_{\rm sup}\,G(x)$ by redefining the average temperature to $T'=\bar{T}(1+2y_{\rm sup})$, but the last term cannot be removed in this way. This already demonstrates how fluctuations at the smallest scales propagate to the largest, changing the SED from a pure blackbody in each direction to a distorted blackbody \citep{Chluba2004}.

Similarly, for the multipole coefficients with $\ell>0$, a distortion is created when carrying out the spherical harmonic transformation of the intensity. Defining the antenna temperature fluctuation as
\beal
\label{eq:CMB_blackbody_DT}
\Delta T^{\rm CMB}(\nu, \vgh)&= \frac{c^2}{2h}\frac{(\expf{x}-1)}{\nu^3 G(x)}[I^{\rm CMB}_{\nu}-\bar{I}^{\rm CMB}_{\nu}],
\end{align}
the multipoles for $\ell>0$ are given by
\beal
\label{eq:CMB_blackbody_DT_lm}
\Delta T^{\rm CMB}_{\ell m}(\nu)&\approx \Delta_{\ell m} +[\Delta^2]_{\ell m} + \frac{1}{2} \frac{Y(x)}{G(x)} \,[\Delta^2]_{\ell m},
\end{align}
where $[X]_{\ell m}=\int Y^*_{\ell m} (\vgh)\,X(\vgh)\id^2\vgh$ is the multipole coefficient of $X(\vgh)$. This shows two features of the spherical harmonic transformation of the intensity map: i) the derived antenna temperature fluctuations are frequency dependent, even if the thermodynamic temperature fluctuations, $\Delta_{\ell m}$, per definition are  independent of frequency; ii) independent thermodynamic temperature multipoles couple at second order due to the weighted average (mixing). Both of these effects are small for the primordial CMB temperature fluctuations at multipoles $\ell>1$ \citep[e.g.,][]{Chluba2004}; however, our motion with respect to the CMB rest frame causes a relative leakage of power $\mathcal{O}(\varv/c) \simeq 10^{-3}$ between adjacent multipoles \citep[e.g.,][]{Challinor2002, Amendola2010, Chluba2011ab}, which causes a larger effect that needs to be taken into account \citep{Planck2013abber}. With the moment expansion, here defined by $\Delta_{\ell m}$ and $2 \, y_{\ell m}=[\Delta^2]_{\ell m}$, this effect can be trivially included and propagated. The CMB sky is thus described by maps of $\Delta_{\ell m}$ and $y_{\ell m}$, which have slightly different spatial morphology and statistical properties.

We reiterate that these issues could be completely avoided by converting to thermodynamic temperature {\it before} the spherical harmonic transformation, requiring a different map making procedure. In the presence of foregrounds, this step is non-trivial, but the total SEDs and the power spectra of different moment values can be used to handle these issues. We also mention that a similar distortion is created due to the superposition of blackbodies of different temperatures for polarized contributions \citep{Chluba2015}, which could be relevant to high precision $E$- and $B$-mode searches. A more detailed consideration of these problems is left to future work. 

\vspace{0mm}
\subsection{Thermal dust superposition}
\label{sec:dust_superposition}

As a second illustration, we consider a simple example in which only the spectral index of a thermal dust SED is allowed to vary over the sky, asking how a spherical harmonic transformation affects the SED at different multipoles.  First, we generate a reference template dust map at $353$ GHz, consisting of a Gaussian random field with a power spectrum described by a pure power-law, $C_{\ell} \propto \ell^{-2.7}$, matching recent \emph{Planck} measurements~\citep{Planck2016GNILC}.  Second, we assume that this template can be rescaled to other frequencies according to a modified blackbody SED.  The dust temperature of this SED is fixed to $T_{\rm d} = 19$ K~\citep{Planck2016GNILC} and does not vary over the sky.  The spectral index $\alpha$, however, is allowed to vary on the sky.  We generate maps of $\alpha$ in which the value in each pixel is drawn from a Gaussian of mean $\alpha_{\rm fid} = 1.6$ and standard deviation $0.1$, matching recent \emph{Planck} full-sky results~\citep{Planck2016GNILC}.  The amplitude in each direction on the sky is set by the reference 353 GHz template.  We generate the reference template at HEALPix resolution $N_{\rm side} = 256$ (pixel size $\approx 0.052$ deg${}^2$), while the spectral index maps are generated at $N_{\rm side} = 64$ (pixel size $\approx 0.84$ deg${}^2$).

Using these ingredients, we construct maps at ten linearly spaced frequencies from $20$ GHz to $353$ GHz, generating $100$ independent realizations (i.e., $100$ spectral index maps and $100$ reference template maps).  We do not add noise or beam smoothing for this simple illustration.  We compute the angular power spectrum for each frequency map and average the results in seven multipole bins linearly spaced between $2 < \ell < 400$.  We average over all 100 realizations at each frequency, and compute error bars from the scatter among these 100 realizations.

\begin{figure}
\centering 
\includegraphics[width=\columnwidth]{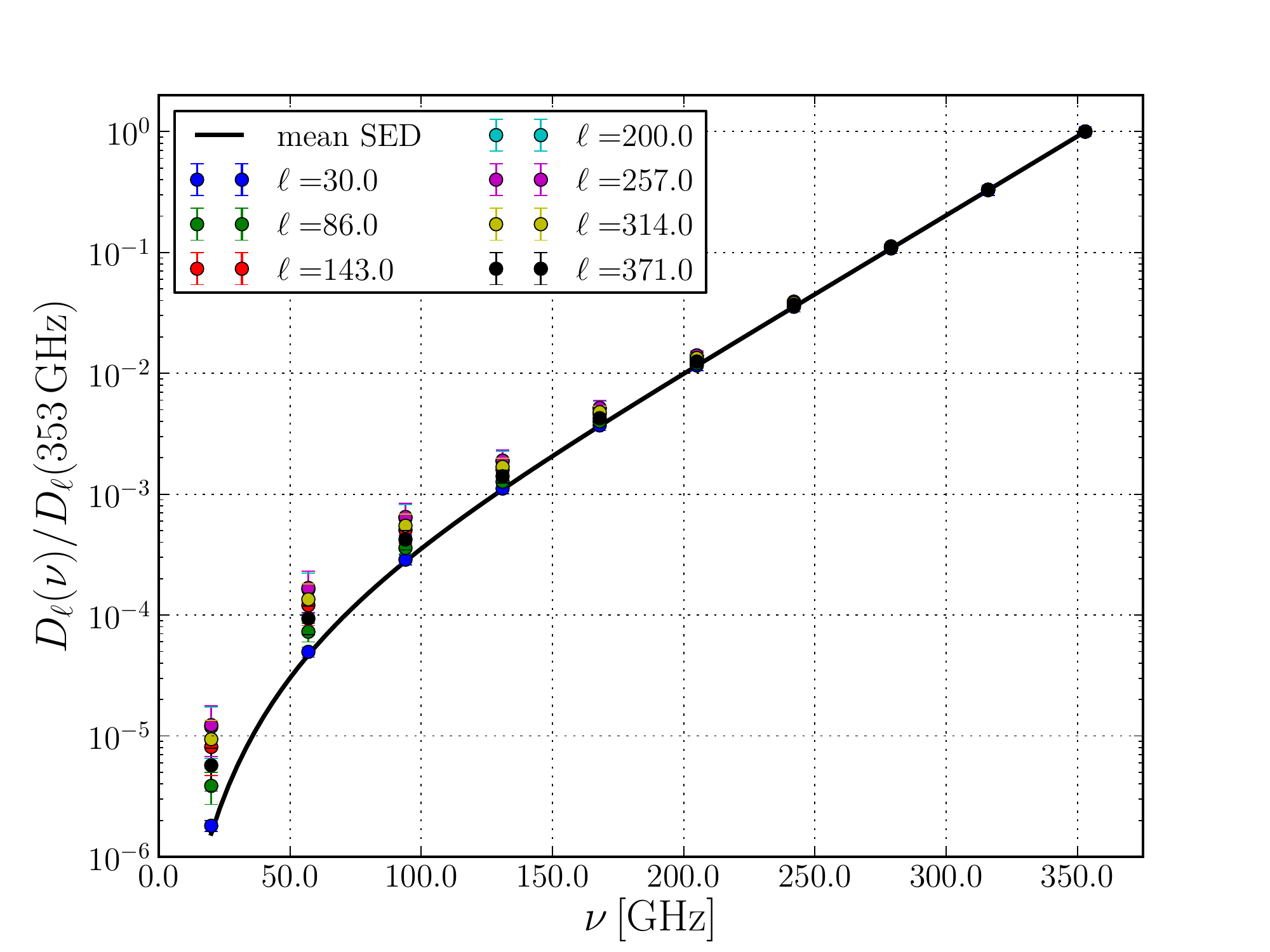}
\\[1mm]
\includegraphics[width=\columnwidth]{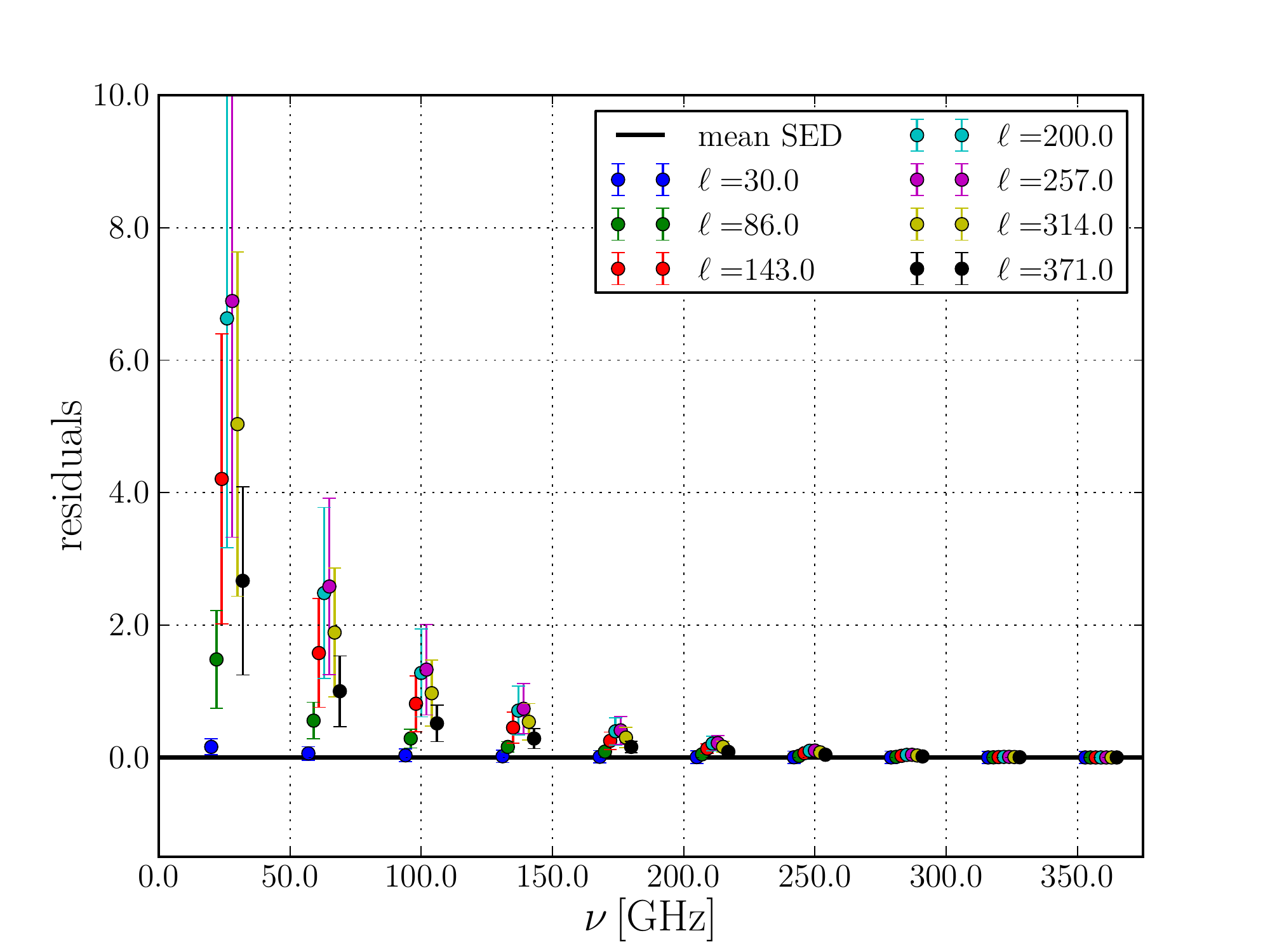}
\caption{Effective SEDs of angular power spectra of modified blackbody simulations (top panel) and fractional residuals with respect to the input mean SED (bottom panel).  The solid black curve shows the input mean SED, a modified blackbody with spectral index $\alpha_{\rm fid} = 1.6$ and temperature $T_d = 19$ K.  The amplitude is set by normalizing all results to the 353 GHz reference template.  In the bottom panel, the points are slightly offset in the horizontal direction for additional clarity.  The effective SED shows the largest departure from the assumed form on the scales where the spectral index variations are largest ($\ell \approx 200$--$300$, corresponding to the pixel size of the generated spectral index map).  In reality, the SED properties will vary over a range of scales and can include both temperature and spectral index variations, yielding more complex behavior than in this simple example.}
\label{fig:dust_PS}
\end{figure}

The results are shown in Fig.~\ref{fig:dust_PS}, plotted as the effective SED for each bandpower in terms of $D_{\ell} \equiv \ell(\ell+1) C_{\ell}/(2\pi)$ at each frequency, normalized to the 353 GHz result.  The effective SED deviates most significantly from the input mean SED (i.e., the square of the modified blackbody described above) on the scales at which the spectral index has been specified to vary, i.e., $\ell \approx 200$--$300$ (corresponding to the pixel size of the spectral index map, as given above).  On much larger or much smaller scales, the effective SED tends back toward the input mean SED (e.g., for the lowest bandpower centered at $\ell = 30$, the behavior is nearly indistinguishable from the input mean SED). Of course, in reality the dust properties likely vary over a range of scales, which would lead to more complex behavior than seen here.  Nonetheless, this simple example already illustrates the key point: averaging over populations with differing SEDs (even if drawn from the same fundamental SED shape, in this case a modified blackbody) will generically produce a multipole-dependent SED in the power spectrum.  The moment method provides a way to efficiently capture this behavior, by introducing new moment maps with varying statistical properties and spatial morphology which combine through new spectral functions related to higher derivatives with respect to the fundamental SED parameters.

\vspace{0mm}
\section{The modeling of other spectral components in CMB applications} 
\label{sec:other}
In this paper, we focused on a few common examples appearing in CMB foreground analysis. There are several additional known and less known components \citep[see][for some overview]{deZotti2015, PlanckSM2015}. 
At low frequencies, anomalous microwave emission (AME), possibly produced by spinning dust grains in the intergalactic medium, contributes significantly, showing a broad maximum around $\nu\simeq 10\,\GHz-30\,\GHz$ \citep[e.g.,][]{Draine1998, Yacine2009, Hoang2016}. This component exhibits a rich phenomenology of relatively broad spectral shapes, depending on the grain sizes and optical properties, temperature and ambient stellar light, to name a few of the ingredients. It is usually treated using spectral templates with a free amplitude and position \citep{PlanckSM2015}. Significant new information about this component is now becoming available [e.g., with {\it C-BASS} \citep{Irfan2015} and {\it QUIJOTE} \citep{GSantos2015}]. This will help in understanding the complexity of the AME and a moment expansion may provide one avenue towards a refined treatment in component separation analyses, capturing variations in the amount of spinning dust and typical emissivity but also other physical properties. In this case, an orthogonalization scheme may be beneficial to reduce the number of common modes for the different variables. 

At intermediate CMB frequencies ($30\,\GHz\lesssim \nu \lesssim 200\,\GHz$), the cumulative CO rotational line emission from star-forming galaxies throughout the Universe becomes important \citep{Righi2008b, deZotti2015, Mashian2016}. Recent measurements with {\it ALMA} \citep{Carilli2016} show that in agreement with previous model estimates, this component will have to be considered carefully to reach the full potential of future CMB spectral distortion measurements.
The uncertainties in the modeling of this component are significant and require refined theoretical studies; however, also in this case, a moment expansion in the main parameters may provide a viable approach for incorporating it in future CMB analyses. Future large galaxy surveys may provide additional constraints to inform the theoretical modeling of the CO emission. 

We also mention that at high frequencies additional dust-like components are present. One is due to the cosmic infrared background (CIB) which can be described using a halo-model approach \citep[e.g.,][]{Vince2015}. At frequencies above the dust peak ($\nu \simeq 1.5-2\,{\rm THz}$), the SED in this case is typically much flatter than that of a modified blackbody spectrum, while at lower frequencies a dust moment expansion is expected to provide an efficient representation. This further highlights that measurements at THz frequencies will not necessarily help in constraining the low frequency tails of the dusty foregrounds. 

Another dusty component is caused by the cumulative emission of intergalactic dust \citep{Imara2016}. Since the fundamental SED in these models is assumed to be given by a modified blackbody spectrum, a dust moment expansion is expected to efficiently represent this contribution. \footnote{For some of the models presented in \citet{Imara2016}, we already confirmed this statement, finding only $N=6$ parameters to suffice.}

\vspace{-0mm}
\section{Conclusions} 
We developed and illustrated a new generalized parameterization for the modeling of spatially varying foreground components relevant to CMB observations. The method relies on a minimal set of assumptions about the underlying distributions of physical parameters that determine the SED of the considered component. The parameter list is extended by moments of the underlying distribution functions, which fix the amplitude of new spectral shapes (related to derivatives of the fundamental SED) that are caused by line of sight and beam averages. We applied the method to superpositions of power-law (Sect.~\ref{sec:power-law}), free-free (Sect.~\ref{sec:ffspectra}), gray-body (Sect.~\ref{sec:blackbody}), and modified blackbody (Sect.~\ref{sec:dust}) spectra, illustrating how these can be used to describe different foreground models. 

All our examples assume noiseless reconstructions for a single type of component, meant to illustrate the main features and limitations of different parameterizations. Our analysis clearly suggests that even simply incorporating the inevitable effects of spatial variations of SED parameters calls for more general methods with a larger number of degrees of freedom being required. This provides an indication for the true challenges that are awaiting us in future CMB B-mode and spectral distortion searches.

We find power-law moment expansions [Eq.~\eqref{eq:I_nu_gen_exp_power_law}] to converge quite rapidly. For the modeling of synchrotron emission, we expect $N\simeq 3-4$ moments to provide sufficient spectral freedom to reach sensitivity levels required for absolute CMB spectroscopy. However, for power-law superpositions with stronger variation of the spectral index, the convergence of the power-law moment expansion can become quite slow (see Sect.~\ref{sec:two-p-spectra}).

For free-free we obtained two moment expansions [Eq.~\eqref{eq:I_nu_gen_ff} and Eq.~\eqref{eq:DI_nu_gen_ff}]. The first represents the commonly used model for the optically thin free-free spectrum; however, it neglects free-free absorption of CMB photons and stimulated free-free emission, which leads to a small suppression at high frequencies ($\nu\geq 1\,{\rm THz}$) and is captured by the second representation, Eq.~\eqref{eq:DI_nu_gen_ff}. Variations in the chemical compositions and ionization degree are not expected to be as relevant to CMB foreground modeling, but applications of the moment method to the continuum X-ray emission from clusters could require a more detailed treatment.

We also find that the free-free SED can be described using a simple power-law moment expansion with $N\simeq 3-4$ moments. This suggests that the combined emission from synchrotron and free-free could be modeled using one parametrization. Only at very high frequencies ($\nu\geq 1\,{\rm THz}$) do these two components become spectrally distinguishable once the effects of spatial variations are included. In a similar manner, we expect a power-law moment expansion to capture the contributions from radio-point sources in low-angular resolution observations, when masking cannot be applied. This could greatly reduce the number of independent parameters that are required to model these low-frequency foregrounds, with the different components mainly being distinguished via physical priors (e.g., by assuming that free-free has a typical spectral index of $\alpha \simeq -0.14$ rather than $\alpha\simeq -0.9$ for synchrotron).

For the superposition of gray-body spectra, we developed two moment expansions with individual weighting [Eq.~\eqref{eq:I_nu_gen_Gray} and Eq.~\eqref{eq:I_nu_gen_Gray_mod}]. We showed that the convergence of the moment expansion is slow at high frequencies (in the Wien tail), which can lead to biases at low frequencies. The two considered weighting schemes are physically equivalent; however, an interpretation of the obtained best-fitting moments in truncated moment expansions in terms of the underlying parameter distribution functions can be affected significantly (Sect.~\ref{sec:gray_single}). We also derived an orthonormal set of basis functions for the gray-body moment expansion (Sect.~\ref{sec:gray_ortho}), however, our discussion does not suggest that this will greatly improve the applicability of the moment parametrization in this case, as the number of free parameters is not reduced.

Finally, we provide a moment expansion for modified blackbody spectra [Eq.~\eqref{eq:I_nu_gen_BB}]. We blindly applied this parametrization to recent models for the dust emission discussed by \citet{Kogut2016}, showing that without many a priori assumptions about the temperature distribution functions we obtain accurate representations of the average SEDs (Fig.~\ref{fig:mBB_dist}). Our modeling suggested that once general distributions of temperature and dust opacities are allowed, about $\simeq 6-10$ parameters are required to model the individual dust components down to the level of sensitivity required in absolute CMB spectroscopy. We expect a similar number to be necessary for future CMB polarization studies; with more restrictive models biases in the deduced cosmological parameters (in particular the tensor-to-scalar ratio) could be produced. The capabilities of future CMB anisotropy experiments have to be explicitly tailored towards this challenge.

In Sect.~\ref{sec:variations}, we also briefly discussed the effect of spatial variations on the SED at different scales. Specifically, the dust SED can show significant scale-dependent modifications caused by spatial variations of the spectral index (see Fig.~\ref{fig:dust_PS}). The moment expansion provides a simple way for parameterizing and propagating these effects in real maps. This is achieved by separating spatial from spectral variations though the moment expansion, where each moment map can have a new spatial morphology. For SZ clusters, this was illustrated in \citet{Chluba2012moments} and for CMB foreground modeling we plan to study this thoroughly in the future.

We close by saying that one of the important assumptions of the moment method is that a detailed form for the fundamental SED is known. Nature may not comply with the spectral forms used here (and in standard CMB foreground analyses); however, improvements of the fundamental SEDs can be easily implemented. We also expect the moment method to be extendable to other foreground components (see Sect.~\ref{sec:other}) and to CMB polarization. Given the physically motivated extensions of foreground parameterizations, the moment expansion can furthermore be used as a diagnostic, testing the robustness of the foreground modeling and derived cosmological parameters. External data sets can also be used to place priors on the moment values and the relations of different moments. 
Overall, the moment method provides a much more general framework for the treatment of CMB foregrounds.
  
\vspace{-5mm}
\small
\section*{Acknowledgments}
JC is supported by the Royal Society as a Royal Society University Research Fellow at the University of Manchester, UK.  This work was partially supported by a Junior Fellow award from the Simons Foundation to JCH.

\small 

\vspace{-0mm}
\bibliographystyle{mn2e}
\bibliography{Lit}

\begin{thebibliography}{62}
\expandafter\ifx\csname natexlab\endcsname\relax\def\natexlab#1{#1}\fi

\bibitem[{{Abazajian} {et~al}\mbox{.}(2016){Abazajian}, {Adshead}, {Ahmed},
  {Allen}, {Alonso}, {Arnold}, {Baccigalupi}, {Bartlett}, {Battaglia},
  {Benson}, {Bischoff}, {Borrill}, {Buza}, {Calabrese}, {Caldwell},
  {Carlstrom}, {Chang}, {Crawford}, {Cyr-Racine}, {De Bernardis}, {de Haan},
  {di Serego Alighieri}, {Dunkley}, {Dvorkin}, {Errard}, {Fabbian}, {Feeney},
  {Ferraro}, {Filippini}, {Flauger}, {Fuller}, {Gluscevic}, {Green}, {Grin},
  {Grohs}, {Henning}, {Hill}, {Hlozek}, {Holder}, {Holzapfel}, {Hu},
  {Huffenberger}, {Keskitalo}, {Knox}, {Kosowsky}, {Kovac}, {Kovetz}, {Kuo},
  {Kusaka}, {Le Jeune}, {Lee}, {Lilley}, {Loverde}, {Madhavacheril}, {Mantz},
  {Marsh}, {McMahon}, {Meerburg}, {Meyers}, {Miller}, {Munoz}, {Nguyen},
  {Niemack}, {Peloso}, {Peloton}, {Pogosian}, {Pryke}, {Raveri}, {Reichardt},
  {Rocha}, {Rotti}, {Schaan}, {Schmittfull}, {Scott}, {Sehgal}, {Shandera},
  {Sherwin}, {Smith}, {Sorbo}, {Starkman}, {Story}, {van Engelen}, {Vieira},
  {Watson}, {Whitehorn}, \& {Kimmy Wu}}]{Abazajian2016S4SB}
{Abazajian} K.~N. {et~al.}, 2016, ArXiv:1610.02743

\bibitem[{{Abazajian} {et~al}\mbox{.}(2015){Abazajian}, {Arnold}, {Austermann},
  {Benson}, {Bischoff}, {Bock}, {Bond}, {Borrill}, {Calabrese}, {Carlstrom},
  {Carvalho}, {Chang}, {Chiang}, {Church}, {Cooray}, {Crawford}, {Dawson},
  {Das}, {Devlin}, {Dobbs}, {Dodelson}, {Dor{\'e}}, {Dunkley}, {Errard},
  {Fraisse}, {Gallicchio}, {Halverson}, {Hanany}, {Hildebrandt}, {Hincks},
  {Hlozek}, {Holder}, {Holzapfel}, {Honscheid}, {Hu}, {Hubmayr}, {Irwin},
  {Jones}, {Kamionkowski}, {Keating}, {Keisler}, {Knox}, {Komatsu}, {Kovac},
  {Kuo}, {Lawrence}, {Lee}, {Leitch}, {Linder}, {Lubin}, {McMahon}, {Miller},
  {Newburgh}, {Niemack}, {Nguyen}, {Nguyen}, {Page}, {Pryke}, {Reichardt},
  {Ruhl}, {Sehgal}, {Seljak}, {Sievers}, {Silverstein}, {Slosar}, {Smith},
  {Spergel}, {Staggs}, {Stark}, {Stompor}, {Vieregg}, {Wang}, {Watson},
  {Wollack}, {Wu}, {Yoon}, \& {Zahn}}]{Abazajian2015}
{Abazajian} K.~N. {et~al.}, 2015, Astroparticle Physics, 63, 66

\bibitem[{{Abitbol} {et~al}\mbox{.}(2017){Abitbol}, {Chluba}, {Hill}, \&
  {Johnson}}]{abitbol_pixie}
{Abitbol} M.~H., {Chluba} J., {Hill} J.~C., {Johnson} B.~R., 2017, \mnras

\bibitem[{{Ali-Ha{\"i}moud} {et~al}\mbox{.}(2009){Ali-Ha{\"i}moud}, {Hirata},
  \& {Dickinson}}]{Yacine2009}
{Ali-Ha{\"i}moud} Y., {Hirata} C.~M., {Dickinson} C., 2009, \mnras, 395, 1055

\bibitem[{{Amendola} {et~al}\mbox{.}(2011){Amendola}, {Catena}, {Masina}, \&
  {et al.}}]{Amendola2010}
{Amendola} L., {Catena} R., {Masina} I., {et al.}, 2011, \jcap, 7, 27

\bibitem[{{Andr{\'e}} {et~al}\mbox{.}(2014){Andr{\'e}}, {Baccigalupi},
  {Banday}, {Barbosa}, {Barreiro}, {Bartlett}, {Bartolo}, {Battistelli},
  {Battye}, {Bendo}, {Beno{\#523}t}, {Bernard}, {Bersanelli}, {B{\'e}thermin},
  {Bielewicz}, {Bonaldi}, {Bouchet}, {Boulanger}, {Brand}, {Bucher},
  {Burigana}, {Cai}, {Camus}, {Casas}, {Casasola}, {Castex}, {Challinor},
  {Chluba}, {Chon}, {Colafrancesco}, {Comis}, {Cuttaia}, {D'Alessandro}, {Da
  Silva}, {Davis}, {de Avillez}, {de Bernardis}, {de Petris}, {de Rosa}, {de
  Zotti}, {Delabrouille}, {D{\'e}sert}, {Dickinson}, {Diego}, {Dunkley},
  {En{\ss}lin}, {Errard}, {Falgarone}, {Ferreira}, {Ferri{\`e}re}, {Finelli},
  {Fletcher}, {Fosalba}, {Fuller}, {Galli}, {Ganga}, {Garc{\'{\i}}a-Bellido},
  {Ghribi}, {Giard}, {Giraud-H{\'e}raud}, {Gonzalez-Nuevo}, {Grainge},
  {Gruppuso}, {Hall}, {Hamilton}, {Haverkorn}, {Hernandez-Monteagudo},
  {Herranz}, {Jackson}, {Jaffe}, {Khatri}, {Kunz}, {Lamagna}, {Lattanzi},
  {Leahy}, {Lesgourgues}, {Liguori}, {Liuzzo}, {Lopez-Caniego}, {Macias-Perez},
  {Maffei}, {Maino}, {Mangilli}, {Martinez-Gonzalez}, {Martins}, {Masi},
  {Massardi}, {Matarrese}, {Melchiorri}, {Melin}, {Mennella}, {Mignano},
  {Miville-Desch{\^e}nes}, {Monfardini}, {Murphy}, {Naselsky}, {Nati},
  {Natoli}, {Negrello}, {Noviello}, {O'Sullivan}, {Paci}, {Pagano}, {Paladino},
  {Palanque-Delabrouille}, {Paoletti}, {Peiris}, {Perrotta}, {Piacentini},
  {Piat}, {Piccirillo}, {Pisano}, {Polenta}, {Pollo}, {Ponthieu},
  {Remazeilles}, {Ricciardi}, {Roman}, {Rosset}, {Rubino-Martin}, {Salatino},
  {Schillaci}, {Shellard}, {Silk}, {Starobinsky}, {Stompor}, {Sunyaev},
  {Tartari}, {Terenzi}, {Toffolatti}, {Tomasi}, {Trappe}, {Tristram},
  {Trombetti}, {Tucci}, {Van de Weijgaert}, {Van Tent}, {Verde}, {Vielva},
  {Wandelt}, {Watson}, \& {Withington}}]{PRISM2013WPII}
{Andr{\'e}} P. {et~al.}, 2014, \jcap, 2, 6

\bibitem[{{Bennett} {et~al}\mbox{.}(2003){Bennett}, {Halpern}, {Hinshaw},
  {Jarosik}, {Kogut}, {Limon}, {Meyer}, {Page}, {Spergel}, {Tucker}, {Wollack},
  {Wright}, {Barnes}, {Greason}, {Hill}, {Komatsu}, {Nolta}, {Odegard},
  {Peiris}, \& {Verde}}]{WMAP_params}
{Bennett} C.~L. {et~al.}, 2003, \apjs, 148, 1

\bibitem[{{Carilli} {et~al}\mbox{.}(2016){Carilli}, {Chluba}, {Decarli},
  {Walter}, {Aravena}, {Wagg}, {Popping}, {Cortes}, {Hodge}, {Weiss},
  {Bertoldi}, \& {Riechers}}]{Carilli2016}
{Carilli} C.~L. {et~al.}, 2016, \apj, 833, 73

\bibitem[{{Challinor} \& {van Leeuwen}(2002)}]{Challinor2002}
{Challinor} A., {van Leeuwen} F., 2002, \prd, 65, 103001

\bibitem[{{Chluba}(2011)}]{Chluba2011ab}
{Chluba} J., 2011, \mnras, 415, 3227

\bibitem[{{Chluba}(2013)}]{Chluba2013fore}
{Chluba} J., 2013, \mnras, 436, 2232

\bibitem[{{Chluba}(2016)}]{Chluba2016}
{Chluba} J., 2016, \mnras, 460, 227

\bibitem[{{Chluba} {et~al}\mbox{.}(2015){Chluba}, {Dai}, {Grin}, {Amin}, \&
  {Kamionkowski}}]{Chluba2015}
{Chluba} J., {Dai} L., {Grin} D., {Amin} M.~A., {Kamionkowski} M., 2015,
  \mnras, 446, 2871

\bibitem[{{Chluba} {et~al}\mbox{.}(2012{\natexlab{a}}){Chluba}, {Khatri}, \&
  {Sunyaev}}]{Chluba2012}
{Chluba} J., {Khatri} R., {Sunyaev} R.~A., 2012{\natexlab{a}}, \mnras, 425,
  1129

\bibitem[{{Chluba} {et~al}\mbox{.}(2012{\natexlab{b}}){Chluba}, {Nagai},
  {Sazonov}, \& {Nelson}}]{Chluba2012SZpack}
{Chluba} J., {Nagai} D., {Sazonov} S., {Nelson} K., 2012{\natexlab{b}}, \mnras,
  426, 510

\bibitem[{{Chluba} \& {Sunyaev}(2004)}]{Chluba2004}
{Chluba} J., {Sunyaev} R.~A., 2004, \aap, 424, 389

\bibitem[{{Chluba} \& {Sunyaev}(2012)}]{Chluba2011therm}
{Chluba} J., {Sunyaev} R.~A., 2012, \mnras, 419, 1294

\bibitem[{{Chluba} {et~al}\mbox{.}(2013){Chluba}, {Switzer}, {Nelson}, \&
  {Nagai}}]{Chluba2012moments}
{Chluba} J., {Switzer} E., {Nelson} K., {Nagai} D., 2013, \mnras, 430, 3054

\bibitem[{{De Zotti} {et~al}\mbox{.}(2016){De Zotti}, {Negrello}, {Castex},
  {Lapi}, \& {Bonato}}]{deZotti2015}
{De Zotti} G., {Negrello} M., {Castex} G., {Lapi} A., {Bonato} M., 2016, \jcap,
  3, 047

\bibitem[{{Desjacques} {et~al}\mbox{.}(2015){Desjacques}, {Chluba}, {Silk}, {de
  Bernardis}, \& {Dor{\'e}}}]{Vince2015}
{Desjacques} V., {Chluba} J., {Silk} J., {de Bernardis} F., {Dor{\'e}} O.,
  2015, \mnras, 451, 4460

\bibitem[{{Draine}(2003)}]{Draine2003}
{Draine} B.~T., 2003, \araa, 41, 241

\bibitem[{{Draine}(2011)}]{Draine2011Book}
{Draine} B.~T., 2011, {Physics of the Interstellar and Intergalactic Medium}.
  Princeton University Press

\bibitem[{{Draine} \& {Lazarian}(1998)}]{Draine1998}
{Draine} B.~T., {Lazarian} A., 1998, \apjl, 494, L19

\bibitem[{{Errard} {et~al}\mbox{.}(2016){Errard}, {Feeney}, {Peiris}, \&
  {Jaffe}}]{Errard2016}
{Errard} J., {Feeney} S.~M., {Peiris} H.~V., {Jaffe} A.~H., 2016, \jcap, 3, 052

\bibitem[{{Finkbeiner} {et~al}\mbox{.}(1999){Finkbeiner}, {Davis}, \&
  {Schlegel}}]{Finkbeiner1999}
{Finkbeiner} D.~P., {Davis} M., {Schlegel} D.~J., 1999, \apj, 524, 867

\bibitem[{{Fixsen} {et~al}\mbox{.}(1996){Fixsen}, {Cheng}, {Gales}, {Mather},
  {Shafer}, \& {Wright}}]{Fixsen1996}
{Fixsen} D.~J., {Cheng} E.~S., {Gales} J.~M., {Mather} J.~C., {Shafer} R.~A.,
  {Wright} E.~L., 1996, \apj, 473, 576

\bibitem[{{Fuskeland} \& {et al.}(2014)}]{Fuskeland2014}
{Fuskeland} U., {et al.}, 2014, \apj, 790, 104

\bibitem[{{G{\'e}nova-Santos} {et~al}\mbox{.}(2015){G{\'e}nova-Santos},
  {Rubi{\~n}o-Mart{\'{\i}}n}, {Rebolo}, {Pel{\'a}ez-Santos},
  {L{\'o}pez-Caraballo}, {Harper}, {Watson}, {Ashdown}, {Barreiro},
  {Casaponsa}, {Dickinson}, {Diego}, {Fern{\'a}ndez-Cobos}, {Grainge},
  {Guti{\'e}rrez}, {Herranz}, {Hoyland}, {Lasenby}, {L{\'o}pez-Caniego},
  {Mart{\'{\i}}nez-Gonz{\'a}lez}, {McCulloch}, {Melhuish}, {Piccirillo},
  {Perrott}, {Poidevin}, {Razavi-Ghods}, {Scott}, {Titterington}, {Tramonte},
  {Vielva}, \& {Vignaga}}]{GSantos2015}
{G{\'e}nova-Santos} R. {et~al.}, 2015, \mnras, 452, 4169

\bibitem[{{Hill} {et~al}\mbox{.}(2015){Hill}, {Battaglia}, {Chluba}, {Ferraro},
  {Schaan}, \& {Spergel}}]{Hill2015}
{Hill} J.~C., {Battaglia} N., {Chluba} J., {Ferraro} S., {Schaan} E., {Spergel}
  D.~N., 2015, Physical Review Letters, 115, 261301

\bibitem[{{Hoang} {et~al}\mbox{.}(2016){Hoang}, {Vinh}, \& {Quynh
  Lan}}]{Hoang2016}
{Hoang} T., {Vinh} N.-A., {Quynh Lan} N., 2016, \apj, 824, 18

\bibitem[{{Hu}(1995)}]{Hu1995PhD}
{Hu} W., 1995, arXiv:astro-ph/9508126

\bibitem[{{Imara} \& {Loeb}(2016)}]{Imara2016}
{Imara} N., {Loeb} A., 2016, \apj, 825, 130

\bibitem[{{Irfan} {et~al}\mbox{.}(2015){Irfan}, {Dickinson}, {Davies},
  {Copley}, {Davis}, {Ferreira}, {Holler}, {Jonas}, {Jones}, {King}, {Leahy},
  {Leech}, {Leitch}, {Muchovej}, {Pearson}, {Peel}, {Readhead}, {Stevenson},
  {Sutton}, {Taylor}, \& {Zuntz}}]{Irfan2015}
{Irfan} M.~O. {et~al.}, 2015, \mnras, 448, 3572

\bibitem[{{Itoh} \& et. al.(2000)}]{Itoh2000}
{Itoh} N., et. al., 2000, \apjs, 128, 125

\bibitem[{{Itoh} {et~al}\mbox{.}(1998){Itoh}, {Kohyama}, \& {Nozawa}}]{Itoh98}
{Itoh} N., {Kohyama} Y., {Nozawa} S., 1998, \apj, 502, 7

\bibitem[{{Karzas} \& {Latter}(1961)}]{Karzas1961}
{Karzas} W.~J., {Latter} R., 1961, \apjs, 6, 167

\bibitem[{{Kogut} {et~al}\mbox{.}(2016){Kogut}, {Chluba}, {Fixsen}, {Meyer}, \&
  {Spergel}}]{Kogut2016SPIE}
{Kogut} A., {Chluba} J., {Fixsen} D.~J., {Meyer} S., {Spergel} D., 2016, in
  Proc.SPIE, Vol. 9904, SPIE Conference Series, p. 99040W

\bibitem[{{Kogut} \& {Fixsen}(2016)}]{Kogut2016}
{Kogut} A., {Fixsen} D.~J., 2016, \apj, 826, 101

\bibitem[{{Kogut} {et~al}\mbox{.}(2011){Kogut}, {Fixsen}, {Levin}, {Limon},
  {Lubin}, {Mirel}, {Seiffert}, {Singal}, {Villela}, {Wollack}, \&
  {Wuensche}}]{Kogut2011}
{Kogut} A. {et~al.}, 2011, \apj, 734, 4

\bibitem[{{Mart{\'{\i}}nez-Solaeche}
  {et~al}\mbox{.}(2017){Mart{\'{\i}}nez-Solaeche}, {Karakci}, \&
  {Delabrouille}}]{Jacques2017_3Ddust}
{Mart{\'{\i}}nez-Solaeche} G., {Karakci} A., {Delabrouille} J., 2017,
  ArXiv:1706.04162

\bibitem[{{Mashian} {et~al}\mbox{.}(2016){Mashian}, {Loeb}, \&
  {Sternberg}}]{Mashian2016}
{Mashian} N., {Loeb} A., {Sternberg} A., 2016, \mnras, 458, L99

\bibitem[{{Mather} {et~al}\mbox{.}(1994){Mather}, {Cheng}, {Cottingham},
  {Eplee}, {Fixsen}, {Hewagama}, {Isaacman}, {Jensen}, {Meyer}, {Noerdlinger},
  {Read}, \& {Rosen}}]{Mather1994}
{Mather} J.~C. {et~al.}, 1994, \apj, 420, 439

\bibitem[{{Matsumura} {et~al}\mbox{.}(2014){Matsumura}, {Akiba}, {Borrill},
  {Chinone}, {Dobbs}, {Fuke}, {Ghribi}, {Hasegawa}, {Hattori}, {Hattori},
  {Hazumi}, {Holzapfel}, {Inoue}, {Ishidoshiro}, {Ishino}, {Ishitsuka},
  {Karatsu}, {Katayama}, {Kawano}, {Kibayashi}, {Kibe}, {Kimura}, {Kimura},
  {Koga}, {Kozu}, {Komatsu}, {Lee}, {Matsuhara}, {Mima}, {Mitsuda}, {Mizukami},
  {Morii}, {Morishima}, {Murayama}, {Nagai}, {Nagata}, {Nakamura}, {Naruse},
  {Natsume}, {Nishibori}, {Nishino}, {Noda}, {Noguchi}, {Ogawa}, {Oguri},
  {Ohta}, {Otani}, {Richards}, {Sakai}, {Sato}, {Sato}, {Sekimoto}, {Shimizu},
  {Shinozaki}, {Sugita}, {Suzuki}, {Suzuki}, {Tajima}, {Takada}, {Takakura},
  {Takei}, {Tomaru}, {Uzawa}, {Wada}, {Watanabe}, {Yoshida}, {Yamasaki},
  {Yoshida}, \& {Yotsumoto}}]{Matsumura2014}
{Matsumura} T. {et~al.}, 2014, Journal of Low Temperature Physics, 176, 733

\bibitem[{{Meisner} \& {Finkbeiner}(2015)}]{Meisner2015}
{Meisner} A.~M., {Finkbeiner} D.~P., 2015, \apj, 798, 88

\bibitem[{{Mozdzen} {et~al}\mbox{.}(2016){Mozdzen}, {Bowman}, {Monsalve}, \&
  {Rogers}}]{Mozdzen2016}
{Mozdzen} T.~J., {Bowman} J.~D., {Monsalve} R.~A., {Rogers} A.~E.~E., 2016,
  \mnras, 455, 3890

\bibitem[{{Nozawa} \& et~al.(2006)}]{Nozawa2006}
{Nozawa} S., et~al., 2006, Nuovo Cimento B Serie, 121, 487

\bibitem[{{Planck Collaboration} {et~al}\mbox{.}(2016{\natexlab{a}}){Planck
  Collaboration}, {Adam}, {Ade}, {Aghanim}, {Arnaud}, {Ashdown}, {Aumont},
  {Baccigalupi}, {Banday}, {Barreiro}, \& et~al.}]{PlanckSM2015}
{Planck Collaboration} {et~al.}, 2016{\natexlab{a}}, \aap, 594, A9

\bibitem[{{Planck Collaboration} {et~al}\mbox{.}(2014{\natexlab{a}}){Planck
  Collaboration}, {Ade}, {Aghanim}, {Armitage-Caplan}, {Arnaud}, {Ashdown},
  {Atrio-Barandela}, {Aumont}, {Baccigalupi}, {Banday}, \&
  et~al.}]{Planck2013components}
{Planck Collaboration} {et~al.}, 2014{\natexlab{a}}, \aap, 571, A12

\bibitem[{{Planck Collaboration} {et~al}\mbox{.}(2016{\natexlab{b}}){Planck
  Collaboration}, {Ade}, {Aghanim}, {Arnaud}, {Ashdown}, {Aumont},
  {Baccigalupi}, {Banday}, {Barreiro}, {Bartlett}, \&
  et~al.}]{Planck2015params}
{Planck Collaboration} {et~al.}, 2016{\natexlab{b}}, \aap, 594, A13

\bibitem[{{Planck Collaboration} {et~al}\mbox{.}(2014{\natexlab{b}}){Planck
  Collaboration}, {Aghanim}, {Armitage-Caplan}, {Arnaud}, {Ashdown},
  {Atrio-Barandela}, {Aumont}, {Baccigalupi}, {Banday}, {Barreiro}, {Bartlett},
  {Benabed}, {Benoit-L{\'e}vy}, {Bernard}, {Bersanelli}, {Bielewicz}, {Bobin},
  {Bock}, {Bond}, {Borrill}, {Bouchet}, {Bridges}, {Burigana}, {Butler},
  {Cardoso}, {Catalano}, {Challinor}, {Chamballu}, {Chiang}, {Chiang},
  {Christensen}, {Clements}, {Colombo}, {Couchot}, {Crill}, {Curto}, {Cuttaia},
  {Danese}, {Davies}, {Davis}, {de Bernardis}, {de Rosa}, {de Zotti},
  {Delabrouille}, {Diego}, {Donzelli}, {Dor{\'e}}, {Dupac}, {Efstathiou},
  {En{\ss}lin}, {Eriksen}, {Finelli}, {Forni}, {Frailis}, {Franceschi},
  {Galeotta}, {Ganga}, {Giard}, {Giardino}, {Gonz{\'a}lez-Nuevo}, {G{\'o}rski},
  {Gratton}, {Gregorio}, {Gruppuso}, {Hansen}, {Hanson}, {Harrison}, {Helou},
  {Hildebrandt}, {Hivon}, {Hobson}, {Holmes}, {Hovest}, {Huffenberger},
  {Jones}, {Juvela}, {Keih{\"a}nen}, {Keskitalo}, {Kisner}, {Knoche}, {Knox},
  {Kunz}, {Kurki-Suonio}, {L{\"a}hteenm{\"a}ki}, {Lamarre}, {Lasenby},
  {Laureijs}, {Lawrence}, {Leonardi}, {Lewis}, {Liguori}, {Lilje},
  {Linden-V{\o}rnle}, {L{\'o}pez-Caniego}, {Lubin}, {Mac{\'{\i}}as-P{\'e}rez},
  {Mandolesi}, {Maris}, {Marshall}, {Martin}, {Mart{\'{\i}}nez-Gonz{\'a}lez},
  {Masi}, {Massardi}, {Matarrese}, {Mazzotta}, {Meinhold}, {Melchiorri},
  {Mendes}, {Migliaccio}, {Mitra}, {Moneti}, {Montier}, {Morgante}, {Mortlock},
  {Moss}, {Munshi}, {Naselsky}, {Nati}, {Natoli}, {N{\o}rgaard-Nielsen},
  {Noviello}, {Novikov}, {Novikov}, {Osborne}, {Oxborrow}, {Pagano}, {Pajot},
  {Paoletti}, {Pasian}, {Patanchon}, {Perdereau}, {Perrotta}, {Piacentini},
  {Pierpaoli}, {Pietrobon}, {Plaszczynski}, {Pointecouteau}, {Polenta},
  {Ponthieu}, {Popa}, {Pratt}, {Pr{\'e}zeau}, {Puget}, {Rachen}, {Reach},
  {Reinecke}, {Ricciardi}, {Riller}, {Ristorcelli}, {Rocha}, {Rosset},
  {Rubi{\~n}o-Mart{\'{\i}}n}, {Rusholme}, {Santos}, {Savini}, {Scott},
  {Seiffert}, {Shellard}, {Spencer}, {Sunyaev}, {Sureau}, {Suur-Uski},
  {Sygnet}, {Tauber}, {Tavagnacco}, {Terenzi}, {Toffolatti}, {Tomasi},
  {Tristram}, {Tucci}, {T{\"u}rler}, {Valenziano}, {Valiviita}, {Van Tent},
  {Vielva}, {Villa}, {Vittorio}, {Wade}, {Wandelt}, {White}, {Yvon}, {Zacchei},
  {Zibin}, \& {Zonca}}]{Planck2013abber}
{Planck Collaboration} {et~al.}, 2014{\natexlab{b}}, \aap, 571, A27

\bibitem[{{Planck Collaboration} {et~al}\mbox{.}(2016{\natexlab{c}}){Planck
  Collaboration}, {Aghanim}, {Ashdown}, {Aumont}, {Baccigalupi}, {Ballardini},
  {Banday}, {Barreiro}, {Bartolo}, {Basak}, {Benabed}, {Bernard}, {Bersanelli},
  {Bielewicz}, {Bonavera}, {Bond}, {Borrill}, {Bouchet}, {Boulanger},
  {Burigana}, {Calabrese}, {Cardoso}, {Carron}, {Chiang}, {Colombo}, {Comis},
  {Couchot}, {Coulais}, {Crill}, {Curto}, {Cuttaia}, {de Bernardis}, {de
  Zotti}, {Delabrouille}, {Di Valentino}, {Dickinson}, {Diego}, {Dor{\'e}},
  {Douspis}, {Ducout}, {Dupac}, {Dusini}, {Elsner}, {En{\ss}lin}, {Eriksen},
  {Falgarone}, {Fantaye}, {Finelli}, {Forastieri}, {Frailis}, {Fraisse},
  {Franceschi}, {Frolov}, {Galeotta}, {Galli}, {Ganga}, {G{\'e}nova-Santos},
  {Gerbino}, {Ghosh}, {Giraud-H{\'e}raud}, {Gonz{\'a}lez-Nuevo}, {G{\'o}rski},
  {Gruppuso}, {Gudmundsson}, {Hansen}, {Helou}, {Henrot-Versill{\'e}},
  {Herranz}, {Hivon}, {Huang}, {Jaffe}, {Jones}, {Keih{\"a}nen}, {Keskitalo},
  {Kiiveri}, {Kisner}, {Krachmalnicoff}, {Kunz}, {Kurki-Suonio}, {Lamarre},
  {Langer}, {Lasenby}, {Lattanzi}, {Lawrence}, {Le Jeune}, {Levrier}, {Lilje},
  {Lilley}, {Lindholm}, {L{\'o}pez-Caniego}, {Ma}, {Mac{\'{\i}}as-P{\'e}rez},
  {Maggio}, {Maino}, {Mandolesi}, {Mangilli}, {Maris}, {Martin},
  {Mart{\'{\i}}nez-Gonz{\'a}lez}, {Matarrese}, {Mauri}, {McEwen}, {Melchiorri},
  {Mennella}, {Migliaccio}, {Miville-Desch{\^e}nes}, {Molinari}, {Moneti},
  {Montier}, {Morgante}, {Moss}, {Natoli}, {Oxborrow}, {Pagano}, {Paoletti},
  {Patanchon}, {Perdereau}, {Perotto}, {Pettorino}, {Piacentini},
  {Plaszczynski}, {Polastri}, {Polenta}, {Puget}, {Rachen}, {Racine},
  {Reinecke}, {Remazeilles}, {Renzi}, {Rocha}, {Rosset}, {Rossetti}, {Roudier},
  {Rubi{\~n}o-Mart{\'{\i}}n}, {Ruiz-Granados}, {Salvati}, {Sandri},
  {Savelainen}, {Scott}, {Sirignano}, {Sirri}, {Soler}, {Spencer}, {Suur-Uski},
  {Tauber}, {Tavagnacco}, {Tenti}, {Toffolatti}, {Tomasi}, {Tristram},
  {Trombetti}, {Valiviita}, {Van Tent}, {Vielva}, {Villa}, {Vittorio},
  {Wandelt}, {Wehus}, {Zacchei}, \& {Zonca}}]{Planck2016GNILC}
{Planck Collaboration} {et~al.}, 2016{\natexlab{c}}, \aap, 596, A109

\bibitem[{{Ponente} {et~al}\mbox{.}(2011){Ponente}, {Diego}, {Sheth},
  {Burigana}, {Knollmann}, \& {Ascasibar}}]{Ponente2011MNRAS}
{Ponente} P.~P., {Diego} J.~M., {Sheth} R.~K., {Burigana} C., {Knollmann}
  S.~R., {Ascasibar} Y., 2011, \mnras, 410, 2353

\bibitem[{{Pritchard} \& {Loeb}(2010)}]{Pritchard2010}
{Pritchard} J.~R., {Loeb} A., 2010, \prd, 82, 023006

\bibitem[{{Remazeilles} {et~al}\mbox{.}(2016){Remazeilles}, {Dickinson},
  {Eriksen}, \& {Wehus}}]{Remazeilles2016}
{Remazeilles} M., {Dickinson} C., {Eriksen} H.~K.~K., {Wehus} I.~K., 2016,
  \mnras, 458, 2032

\bibitem[{{Righi} {et~al}\mbox{.}(2008){Righi}, {Hern{\'a}ndez-Monteagudo}, \&
  {Sunyaev}}]{Righi2008b}
{Righi} M., {Hern{\'a}ndez-Monteagudo} C., {Sunyaev} R.~A., 2008, \aap, 489,
  489

\bibitem[{{Rybicki} \& {Lightman}(1979)}]{Rybicki1979}
{Rybicki} G.~B., {Lightman} A.~P., 1979, {Radiative processes in astrophysics}.
  New York, Wiley-Interscience, 1979.~393 p.

\bibitem[{{Sathyanarayana Rao} {et~al}\mbox{.}(2015){Sathyanarayana Rao},
  {Subrahmanyan}, {Udaya Shankar}, \& {Chluba}}]{Mayuri2015}
{Sathyanarayana Rao} M., {Subrahmanyan} R., {Udaya Shankar} N., {Chluba} J.,
  2015, \apj, 810, 3

\bibitem[{{Sazonov} \& {Sunyaev}(1998)}]{Sazonov1998}
{Sazonov} S.~Y., {Sunyaev} R.~A., 1998, \apj, 508, 1

\bibitem[{{Sunyaev} \& {Khatri}(2013)}]{Sunyaev2013}
{Sunyaev} R.~A., {Khatri} R., 2013, IJMPD, 22, 30014

\bibitem[{{Tashiro}(2014)}]{Tashiro2014}
{Tashiro} H., 2014, Prog. of Theo. and Exp. Physics, 2014, 060000

\bibitem[{{Thorne} {et~al}\mbox{.}(2016){Thorne}, {Dunkley}, {Alonso}, \&
  {Naess}}]{PySky}
{Thorne} B., {Dunkley} J., {Alonso} D., {Naess} S., 2016, ArXiv:1608.02841

\bibitem[{{Zeldovich} \& {Sunyaev}(1969)}]{Zeldovich1969}
{Zeldovich} Y.~B., {Sunyaev} R.~A., 1969, \apss, 4, 301

\end{thebibliography}

\begin{appendix}

\vspace{-5mm}
\section{Free-free emission in the optically thin regime}
\label{app:ff_derivation}
The radiative transfer equation for the photon occupation number, $n(\xe=h\nu/k\Te)$, under free-free emission and absorption in an isotropic medium takes the simple form \citep{Rybicki1979, Hu1995PhD}
\beal
\frac{\partial n(\xe)}{\partial \tau}=\frac{\kappa(\xe, \Te)\,\expf{-\xe}}{\xe^3}\left[1-n(\xe) (\expf{\xe}-1)\right].
\end{align}
Here, $\tau = \int \Ne \sigT  \id l$ is the Thomson optical depth along the photon path and $\kappa(\xe, \Te)$ describes the free-free emissivity of the plasma, which depends on the electron temperature\footnote{A single temperature plasma is assumed}, ionization degree, composition and weakly on frequency (through the Gaunt-factor). Assuming that the ambient radiation field is given by the CMB blackbody, $n_0\approx 1/(\expf{x}-1)$ with $x=h\nu/kT_0$, we have 
\beal
\delta n(\xe)\approx \delta \tau \,\frac{\kappa(\xe, \Te)\,\expf{-\xe}}{\xe^3}\frac{\expf{x}- \expf{\xe}}{\expf{x}-1}
=\delta \tau \,\frac{\kappa(\xe, \Te)}{\xe^3}\frac{\expf{x-\xe}- 1}{\expf{x}-1},
\end{align}
which implies a suppression factor $f(x, \xe)=(\expf{x-\xe}- 1)/(\expf{x}-1)$ from stimulated CMB emission and CMB absorption.

\section{Basis for free-free spectra}
\label{app:derivs_ff}
Here we give the first few functions, $G^{(l, m)}_{k}$, defined by Eq.~\eqref{eq:Gkm_ff}. They read
\beal
\label{eq:Gkm_ff_examples}
G^{(1, 0)}_{1}(\bar{\nu}_{\rm ff}, \bar\zeta, \nu)&=
\left(-\frac{2}{3}, 0, 0, 0\right)^{T}\cdot\vek{g}_{\rm ff}(\bar{\nu}_{\rm ff}, \bar\zeta, \nu)
\nonumber\\ 
G^{(2, 0)}_{1}(\bar{\nu}_{\rm ff}, \bar\zeta, \nu)&=
\left(\frac{2}{9}, -\frac{2}{3}, 0, 0\right)^{T}\cdot\vek{g}_{\rm ff}(\bar{\nu}_{\rm ff}, \bar\zeta, \nu)
\nonumber\\ 
G^{(3, 0)}_{1}(\bar{\nu}_{\rm ff}, \bar\zeta, \nu)&=
\left(-\frac{4}{81}, \frac{2}{9}, - \frac{1}{3}, 0\right)^{T}\cdot\vek{g}_{\rm ff}(\bar{\nu}_{\rm ff}, \bar\zeta, \nu)
\\ \nonumber
G^{(4, 0)}_{1}(\bar{\nu}_{\rm ff}, \bar\zeta, \nu)&=
\left(\frac{2}{243}, - \frac{4}{81}, \frac{1}{9}, -\frac{1}{9}\right)^{T}\cdot\vek{g}_{\rm ff}(\bar{\nu}_{\rm ff}, \bar\zeta, \nu)
\nonumber\\[2mm]
G^{(2, 0)}_{2}(\bar{\nu}_{\rm ff}, \bar\zeta, \nu)&=
\left(\frac{2}{9}, 0, 0, 0\right)^{T}\cdot\vek{g}_{\rm ff}(\bar{\nu}_{\rm ff}, \bar\zeta, \nu)
\nonumber\\ 
G^{(3, 0)}_{2}(\bar{\nu}_{\rm ff}, \bar\zeta, \nu)&=
\left(-\frac{4}{27}, \frac{2}{9}, 0, 0\right)^{T}\cdot\vek{g}_{\rm ff}(\bar{\nu}_{\rm ff}, \bar\zeta, \nu)
\nonumber \\ \nonumber
G^{(4, 0)}_{2}(\bar{\nu}_{\rm ff}, \bar\zeta, \nu)&=
\left(\frac{14}{243}, - \frac{4}{27}, \frac{1}{9}, 0\right)^{T}\cdot\vek{g}_{\rm ff}(\bar{\nu}_{\rm ff}, \bar\zeta, \nu)
\nonumber\\[2mm]
G^{(0, 1)}_{1}(\bar{\nu}_{\rm ff}, \bar\zeta, \nu)&=
\left(1, 0, 0, 0\right)^{T}\cdot\vek{g}_{\rm ff}(\bar{\nu}_{\rm ff}, \bar\zeta, \nu)
\nonumber\\ 
G^{(1, 1)}_{1}(\bar{\nu}_{\rm ff}, \bar\zeta, \nu)&=
\left(-\frac{2}{3}, 1, 0, 0\right)^{T}\cdot\vek{g}_{\rm ff}(\bar{\nu}_{\rm ff}, \bar\zeta, \nu)
\nonumber\\ 
G^{(2, 1)}_{1}(\bar{\nu}_{\rm ff}, \bar\zeta, \nu)&=
\left(\frac{2}{9}, -\frac{2}{3}, \frac{1}{2}, 0\right)^{T}\cdot\vek{g}_{\rm ff}(\bar{\nu}_{\rm ff}, \bar\zeta, \nu)
\nonumber\\ 
G^{(3, 1)}_{1}(\bar{\nu}_{\rm ff}, \bar\zeta, \nu)&=
\left(-\frac{4}{81}, \frac{2}{9}, - \frac{1}{3}, \frac{1}{6}\right)^{T}\cdot\vek{g}_{\rm ff}(\bar{\nu}_{\rm ff}, \bar\zeta, \nu)
\nonumber\\[2mm]
G^{(1, 1)}_{2}(\bar{\nu}_{\rm ff}, \bar\zeta, \nu)&=
\left(-\frac{2}{3}, 0, 0, 0\right)^{T}\cdot\vek{g}_{\rm ff}(\bar{\nu}_{\rm ff}, \bar\zeta, \nu)
\nonumber\\ 
G^{(2, 1)}_{2}(\bar{\nu}_{\rm ff}, \bar\zeta, \nu)&=
\left(\frac{2}{3}, -\frac{2}{3}, 0, 0\right)^{T}\cdot\vek{g}_{\rm ff}(\bar{\nu}_{\rm ff}, \bar\zeta, \nu)
\nonumber \\ \nonumber
G^{(3, 1)}_{2}(\bar{\nu}_{\rm ff}, \bar\zeta, \nu)&=
\left(-\frac{28}{81}, \frac{2}{3}, - \frac{1}{3}, 0\right)^{T}\cdot\vek{g}_{\rm ff}(\bar{\nu}_{\rm ff}, \bar\zeta, \nu)
\nonumber\\[2mm]
G^{(0, 2)}_{2}(\bar{\nu}_{\rm ff}, \bar\zeta, \nu)&=
\left(\frac{1}{2}, 0, 0, 0\right)^{T}\cdot\vek{g}_{\rm ff}(\bar{\nu}_{\rm ff}, \bar\zeta, \nu)
\nonumber\\ 
G^{(1, 2)}_{2}(\bar{\nu}_{\rm ff}, \bar\zeta, \nu)&=
\left(-\frac{2}{3}, \frac{1}{2}, 0, 0\right)^{T}\cdot\vek{g}_{\rm ff}(\bar{\nu}_{\rm ff}, \bar\zeta, \nu)
\nonumber \\ \nonumber
G^{(2, 2)}_{2}(\bar{\nu}_{\rm ff}, \bar\zeta, \nu)&=
\left(\frac{4}{9}, -\frac{2}{3}, \frac{1}{4}, 0\right)^{T}\cdot\vek{g}_{\rm ff}(\bar{\nu}_{\rm ff}, \bar\zeta, \nu),
\end{align}
where we defined $\vek{g}_{\rm ff}=\left(g_{\rm ff}, g^{(1)}_{\rm ff}, g^{(2)}_{\rm ff}, g^{(3)}_{\rm ff}\right)$. These are needed for a fourth order moment expansion. We also have $G^{(0, 0)}_{1}=G^{(0, 0)}_{2}=G^{(1, 0)}_{2}=G^{(0, 1)}_{2}=0$.

\section{Derivatives of Planckian}
\label{app:derivs}
In this section we give a closed form for the derivatives of a Planckian, $\nbb(\nu, \beta)=(\expf{h\nu\beta/k}-1)^{-1}$ with respect to $\beta=1/T$. Since for $x=h\nu\beta/k$ we have $\partial_\beta x = h\nu/k$, we can simply write
\beal
\beta^j \partial_\beta^j \nbb(\nu, \beta)=x^j \partial_x^j \nbb(x)
=\nbb(\nu, \beta)\,\frac{(-x)^j}{(1-\expf{-x})^{j}}\sum_{m=0}^{j-1}
\left<\!\!
\begin{array}{c}
j
\\
m
\end{array}
\!\!\right>\expf{-mx}
\end{align}
where we used the result of \citet{Chluba2012SZpack} for $x^j \partial_x^j \nbb(x)$. 
Here $\left<\!\!
\begin{array}{c}
j
\\
m
\end{array}
\!\!\right>$ denotes the Eulerian numbers, which can be computed using recursion relations \citep[compare][]{Chluba2012SZpack}.

\end{appendix}

\end{document}